\documentclass[twocolumn]{aastex6}
\usepackage{natbib}
\usepackage{threeparttablex} 
\usepackage{booktabs} 
\usepackage{longtable}
\usepackage{graphicx}
\usepackage{amssymb, amsmath, amsbsy}

\fullcollaborationName{The Friends of AASTeX Collaboration}

\shorttitle{Variable stars in ISLAndS galaxies}
\shortauthors{Mart\'inez-V\'azquez et al.}

\begin{document}


\title{The ISLAnds Project III: Variable Stars in Six Andromeda Dwarf Spheroidal 
Galaxies$^\star$}\thanks{$^\star$Based on observations made with the NASA/ESA 
Hubble Space Telescope, obtained at the Space Telescope Science Institute, which 
is operated by the Association of Universities for Research in Astronomy, Inc., 
under NASA contract NAS 5-26555. These observations are associated with programs 
$\#$13028 and $\#$13739.}


\author{
Clara E. Mart{\'i}nez-V{\'a}zquez\altaffilmark{1,2,3},
Matteo Monelli\altaffilmark{1,2},
Edouard J. Bernard\altaffilmark{4},
Carme Gallart\altaffilmark{1,2},
Peter B. Stetson\altaffilmark{5},
Evan D. Skillman\altaffilmark{6},
Giuseppe Bono\altaffilmark{7,8},
Santi Cassisi\altaffilmark{9},
Giuliana Fiorentino\altaffilmark{3},
Kristen B. W. McQuinn\altaffilmark{10},
Andrew A. Cole\altaffilmark{11},
Alan W. McConnachie\altaffilmark{5},
Nicolas F. Martin\altaffilmark{12,13},
Andrew E. Dolphin\altaffilmark{14},
Michael Boylan-Kolchin\altaffilmark{10},
Antonio Aparicio\altaffilmark{1,2},
Sebastian L. Hidalgo\altaffilmark{1,2},
Daniel R. Weisz\altaffilmark{15}}

\affil{}
\and
\email{clara.marvaz@gmail.com}
\altaffiltext{1}{IAC-Instituto de Astrof\'isica de Canarias, Calle V\'ia Lactea s/n, E-38205 La Laguna, Tenerife, Spain}
\altaffiltext{2}{Departmento de Astrof\'isica, Universidad de La Laguna, E-38206 La Laguna, Tenerife, Spain}
\altaffiltext{3}{INAF-Osservatorio Astronomico di Bologna, Via Gobetti 93/3, I-40129 Bologna, Italy}
\altaffiltext{4}{Universit\'e C\^ote d'Azur, OCA, CNRS, Lagrange, France}
\altaffiltext{5}{Dominion Astrophysical Observatory, Herzberg Institute of
Astrophysics, National Research Council, 5071 West Saanich Road, Victoria,
British Columbia V9E 2E7, Canada}
\altaffiltext{6}{Minnesota Institute for Astrophysics, University of Minnesota, Minneapolis, MN, USA}
\altaffiltext{7}{Department of Physics, Universit\`a di Roma Tor Vergata, via della Ricerca Scientifica 1, I-00133 Roma, Italy}
\altaffiltext{8}{INAF-Osservatorio Astronomico di Roma, via Frascati 33, I-00040 Monte Porzio Catone, Italy}
\altaffiltext{9}{INAF-Osservatorio Astronomico di Teramo, Via M. Maggini, I-64100 Teramo}
\altaffiltext{10}{Department of Astronomy, The University of Texas at Austin, 2515 Speedway, Stop C1400, Austin, TX 78712-1205, USA}
\altaffiltext{11}{School of Physical Sciences, University of Tasmania, Hobart, Tasmania, Australia}
\altaffiltext{12}{Observatoire astronomique de Strasbourg, Universit\'e de Strasbourg, CNRS, UMR 7550, 11 rue de l'Universit\'e, F-67000 Strasbourg, France} 
\altaffiltext{13}{Max-Planck-Institut f\"ur Astronomie, K\"onigstuhl 17, D-69117 Heidelberg, Germany}
\altaffiltext{14}{Raytheon; 1151 E. Hermans Rd., Tucson, AZ 85706, USA}
\altaffiltext{15}{Department of Astronomy, University of California Berkeley, Berkeley, CA 94720, USA}

\begin{abstract}

We present a census of variable stars in six M31 dwarf spheroidal 
satellites observed with the Hubble Space Telescope. We detect 870 RR 
Lyrae (RRL) stars in the fields of And~I (296), II (251), III (111), XV 
(117), XVI (8), XXVIII (87). We also detect a total of 15 Anomalous 
Cepheids, three Eclipsing Binaries, and seven field RRL stars 
compatible with being members of the M31 halo or the Giant Stellar Stream. 
We derive robust and homogeneous distances to the six galaxies using 
different methods based on the properties of the RRL stars. Working with the 
up-to-date set of Period-Wesenheit ($I$, $B$ -- $I$) relations published by 
Marconi et al., we obtain distance moduli of 
$\mu_0$ = [24.49, 24.16, 24.36, 24.42, 23.70, 24.43] mag (respectively), with 
systematic uncertainties of 0.08 mag and statistical uncertainties $<$ 0.11 mag.
We have considered an enlarged sample of 
sixteen M31 satellites with published variability studies, and compared
their pulsational observables (e.g., periods, amplitudes), with those of 
fifteen Milky Way satellites for which similar data are available. 
The properties of the (strictly old) RRL in both satellite systems do not 
show any significant difference. In particular, we found a strikingly similar 
correlation between the mean period distribution of the fundamental RRL 
pulsators (RRab) and the mean metallicities of the galaxies. This indicates 
that the old RRL progenitors were similar at the early stage in the two 
environments, suggesting very similar characteristics for the earliest 
stages of evolution of both satellite systems. 

\end{abstract}




\section{Introduction} \label{sec:intro}

RR Lyrae variable stars (RRLs) are unambiguous stellar tracers of an old  
($>$10 Gyr) stellar population. As such, they are a fossil record of the 
early stages of galaxy evolution. Their pulsational properties and their 
position in the color-magnitude diagram (CMD) -- on the horizontal branch 
(HB), $\sim$3 mag above the old main sequence turn-off (oMSTO)-- make RRLs 
easily identifiable objects even beyond the Local Group \citep[LG;][]{DaCosta2010}. 
They are excellent distance indicators, and powerful tools to investigate the early 
evolution of the host stellar system, since their metallicity can be inferred from 
their pulsational properties 
\citep[see e.g.,][and references therein]{Jeffery2011,Nemec2013,MartinezVazquez2016a}. 
Thus, RRL can provide valuable information on the nature of the building-blocks 
of large galaxies such as the Milky Way (MW) or M31 (see, e.g., \citealt{Fiorentino2015a,Monelli2017}). 
Indeed, in the last few years, the study of the populations of 
RRL in galaxies has become increasingly relevant for research on galaxy formation 
and evolution in addition to the more classical field of stellar astrophysics.

Basically, RRL have been detected in all the dwarf galaxies where they have 
been searched for. At least one RRL has been found in all very low mass 
(--8$\lesssim M_V \lesssim$--1.5;) dwarf spheroidal (dSph) galaxies (see, e.g., 
\citealt{Baker2015,Vivas2016a} and references therein). 
In many brighter dSph galaxies (--13$\lesssim M_V \lesssim$--9), both satellites and 
isolated, the number of RRL is greater than $\approx$100. In this way, they are
statistically sufficient to study in detail, for example, possible radial 
gradients in the old stellar populations of their host galaxies 
\citep[e.g.,][]{Bernard2008,MartinezVazquez2015,MartinezVazquez2016a}.
The great advance in observational studies of RRLs 
in nearby dwarf galaxies (see discussion in \S~\ref{sec:discussion})
has led to a much better understanding of their relative distributions in 
dwarf galaxies of different morphological type.
The study of variable stars in satellites of the Andromeda galaxy (And, M31) is
largely incomplete. This has been long due to two
main reasons: {\em i)} their (relatively) faint apparent magnitude ($V\sim$25
mag), and {\em ii)} the stellar crowding. The first successful attempt to identify RRL
stars in the M31 halo was achieved by \citet{pritchet87a}, using Canada-France-Hawaii telescope
data. \citet{saha90a} and \citet{saha90c} detected candidate
RRL stars in the dwarf elliptical M31 satellites NGC185 and NGC147.
Nevertheless, with the advent of the {\it Hubble Space Telescope} (HST) it was 
possible to reach well below the HB. This allowed the first
determination of the properties of RRL stars in the M31 field and its
satellites. Based on WFPC2 data, the discovery of RRL stars was reported in
And~I \citep{DaCosta1996}, And~II \citep{DaCosta2000} and And~III
\citep{DaCosta2002}. The population of variable stars detected in the three
galaxies were later analyzed in detail by \citet[][And~II]{Pritzl2004} and
\citet[][And~I and And~III]{Pritzl2005}. Additionally, And~VI was studied by
\citet{Pritzl2002} on the basis of data of comparable quality. Since then,
the number of known satellites of M31 has increased dramatically, primarily due to the PAndAS
survey \citep{mcconnachie09}. With a few exceptions (And~XI, And~XIII, \citealt{Yang2012};
And~XIX; \citealt{Cusano2013}; And~XXI; \citealt{Cusano2015};  And~XXV; \citealt{Cusano2016})
most of them have not been investigated for stellar variability.
Moreover, the knowledge of the properties of RRL stars in M31 itself and in the largest
satellites (M32, M33) is limited to a few ACS fields and is far from being complete.

Under the ISLAndS\footnote{Initial Star formation and Lifetimes of Andromeda 
Satellites} project (based on very deep, multi-epoch 
HST ACS and WFC3 data), six M31 dSph satellite 
companion galaxies were observed: And~I, And~II, And~III, And~XV, And~XVI and 
And~XXVIII. The main goal of this project is to determine whether the star formation 
histories (SFHs) of the M31 dSph satellites show notable differences from those of 
the MW. The project is described in more detail in the project presentation paper \citep{Skillman2017} 
while the first results concerning the SFH of And~II and And~XVI were presented 
in \citet{Weisz2014a} and \citet{Monelli2016}.

In order to complement these previous studies, this paper focuses on the study of 
variable stars --mainly RRLs, but also Anomalous Cepheids (ACs)-- present in the 
six ISLAndS galaxies. The data obtained within the framework of this project 
have allowed us to increase by a factor ~2--3.4 the number of known variable stars 
and the quality of the light curves in And~I, And~II, and And~III compared
to previous studies \citep{Pritzl2004,Pritzl2005}. On the other hand, this project 
provides the first discoveries of variable stars in And~XV, And~XVI, and And~XXVIII,
although an analysis of the RRL in AndXVI within the context of its SFH has been
presented in \citet{Monelli2016}; for homogeneity with the rest of the
observed ISLAndS galaxies, in this work we reanalyzed the And~XVI variable stars 
from scratch obtaining slightly refined values.

This paper is structured as follows. 
In \S~\ref{sec:observations} we present a summary of the 
observations and data reduction. In \S~\ref{sec:variables} we describe the variable 
star detection and classification, while \S~\ref{sec:rrl} focuses on the properties 
of RRL stars. The properties of the HB and of RRL stars of M31 satellites are compared 
to those of MW dwarfs in \S~\ref{sec:discussion}. RRL stars are used in \S~\ref{sec:distances} 
to derive new, homogeneous distances to the six galaxies. Furthermore, distance 
estimations based on the tip of the red giant branch (TRGB) are provided for the three most 
massive galaxies (And~I, II, and III). Finally, ACs and eclipsing binary (EB) candidates are 
also presented in \S~\ref{sec:ac} and \S~\ref{sec:eb}, respectively.
We note that in the on-line version of the paper we provide full details on all 
the variable stars discussed: time-series photometry, light curves (LCs), and 
mean photometric and pulsational properties.

\section{Observations and data reduction} \label{sec:observations}

\begin{table*}
\renewcommand{\thetable}{\arabic{table}}
\centering
\caption{Positions and structural parameters for the ISLAndS galaxies.}
\label{tab:parameters}
\begin{scriptsize}
\begin{tabular}{lccccccccc} 
\hline
\hline
Galaxy & RA & Dec & M$_V$ & E(B--V) & $\epsilon=1-b/a$ & PA & r$_h$ & r$_t$ & References\\
(name) & (hh mm ss) & ($^o$ $\arcmin$ $\arcsec$) & (mag) & (1-b/a) & ($^o$) & ($\arcmin$) & ($\arcmin$) & \\
\hline
   And~I   & 00:45:39.7 & 38:02:15.0 & --11.2$\pm$0.2& 0.047 & 0.28$\pm$0.03 &  30$\pm$4  &  3.9$\pm$0.1   & 10.4$\pm$0.9 & 1, 2\\
   And~II  & 01:16:26.8 & 33:26:07.0 & --11.6$\pm$0.2& 0.063 & 0.16$\pm$0.02 &  31$\pm$5  &  5.3$\pm$0.1   & 22.0$\pm$1.0 & 1, 2\\
   And~III & 00:35:30.9 & 36:29:56.0 & --9.5$\pm$0.3& 0.050 & 0.59$\pm$0.04 & 140$\pm$3  &  2.0$\pm$0.2   &  7.2$\pm$1.2 & 1, 2\\
   And~XV  & 01:14:18.3 & 38:07:11.0 & --8.0$^{+0.3}_{-0.4}$ & 0.041 & 0.24$\pm$0.10 &  38$\pm$15 &  1.3$\pm$0.1   &  $\sim$5.7 & 1, 3 \\
   And~XVI & 00:59:30.3 & 32:22:34.0 & --7.3$\pm$0.3& 0.066 & 0.29$\pm$0.08 &  98$\pm$9  &  1.0$\pm$0.1   &  $\sim$4.3 & 1, 3 \\
And~XXVIII & 22:32:41.5 & 31:13:03.7 & --8.7$\pm$0.4& 0.080 & 0.43$\pm$0.02 &  34$\pm$1  &  1.20$\pm$0.01 &  $\sim$18.0  & 4, 5 \\
\hline
\hline
\end{tabular}
\end{scriptsize}
\begin{tablenotes}
\begin{scriptsize}
\item References:
(1) \citet{Martin2016},
(2) \citet{McConnachie2006},
(3) \citet{Ibata2007},
(4) \citet{Slater2015},
(5) \citet{Tollerud2013}.
\end{scriptsize}
\end{tablenotes}
\end{table*}

Table~\ref{tab:parameters} presents a compilation of updated values for the position 
of the center (RA and Dec, column 2 and 3, respectively), absolute M$_V$ magnitude
(column 4), reddening (E($B$-$V$), column 5) and structural parameters --ellipticity 
($\epsilon$, column 6), position angle (PA, column 7), half-light radius (r$_h$, column 8) 
and tidal radius (r$_t$, column 9)-- for each of the six observed galaxies under the
ISLAndS project (hereafter called \textit{ISLAndS galaxies}).

The data for these six ISLAndS galaxies have been obtained under proposals GO-13028 and 
GO-13739, for a total of 111 HST orbits. They consist of one ACS pointing on 
the central region and a WFC3 parallel field (at 6$\arcmin$ from the ACS center) 
for each galaxy. For further details about the ACS and WFC3 field location, the reader 
is referred to Figure 4 by \citet{Skillman2017}, where the strategy and the description
of the ISLAndS project is explained in depth.

For both cameras, the $F475W$ and $F814W$ passbands were chosen. The observing strategy 
was designed in order to optimize the phase coverage of short period variables (between 0.3 and 1.2 d), 
specifically RRL and AC stars. In particular, the observations were spread over a few
days (from two to five), and the visits were planned to avoid accumulation of data around
the same time of day, in order to avoid aliasing problems around 0.5 or 1 day periods.
An overview of the observing runs is provided in Table~\ref{tab:runs}, which specifies, 
for each galaxy (column 1), the beginning and ending dates (column 2), and the 
number of orbits obtained (column 3). For an optimal sampling of the light curves, 
each orbit was split into one $F475W$ and one $F814W$ exposures, yielding the same number 
of epochs in each band for each galaxy.
Detailed observing logs are presented in the Appendix~\ref{sec:observing_logs} 
(Tables~\ref{tab:observing_log_andi}, \ref{tab:observing_log_andii}, 
\ref{tab:observing_log_andiii}, \ref{tab:observing_log_andxv}, 
\ref{tab:observing_log_andxvi}, and \ref{tab:observing_log_andxxviii}). 

\begin{table}
\renewcommand{\thetable}{\arabic{table}}
\centering
\caption{Summary of HST observation.}
\label{tab:runs}
\begin{scriptsize}
\begin{tabular}{llccc}
\hline 
\hline
Galaxy           & Obs. Dates       & Orbits \\
\hline
And~I       & September 1-6, 2015   & 22  \\
And~II      & October 4-6, 2013     & 17  \\
And~III     & November 24-28, 2014  & 22  \\
And~XV      & September 17-20, 2014 & 17  \\
And~XVI     & November 20-22, 2013  & 13  \\
And~XXVIII  & January 20-25, 2015   & 20  \\
\hline
\hline
\end{tabular}
\end{scriptsize}
\end{table}

The photometry has been homogeneously performed with the DAOPHOT/ALLFRAME suite
of routines, following the prescriptions described in \citet{Monelli2010b}, for 
both the ACS and parallel WFC3 fields. The photometric catalogs have been 
calibrated to the VEGAMAG photometric systems adopting the updated zero points 
from the instrument web page.

\section{Variable stars identification}\label{sec:variables}

\begin{table*}
\renewcommand{\thetable}{\arabic{table}}
\centering
\caption{Variable star detections.}
\label{tab:variables}
\begin{scriptsize}
\begin{tabular}{lrccccccc}
\cline{3-9}
      &        & And~I & And~II & And~III & And~XV & And~XVI & And~XXVIII & Total \\
\hline
      &  ACS   &       261 &       217 &       108 &       117 &       8 &      87$^{a}$ &  798 \\
RRL   &  WFC3  &        35 &        34 &         3 &         0 &       0 &       0 &         72 \\
\cline{3-9}
      &  total & {\bf 296} & {\bf 251} & {\bf 111} & {\bf 117} & {\bf 8} & {\bf 87} & {\bf 870} \\
\hline
      &  ACS   &         0 &         4 &         4 &         4 &       0 &        3 &        15 \\
AC    &  WFC3  &         0 &         0 &         0 &         0 &       0 &        0 &         0 \\
\cline{3-9}
      &  total &   {\bf 0} &   {\bf 4} &   {\bf 4} &   {\bf 4} & {\bf 0} &  {\bf 3} &  {\bf 15} \\
\hline								
      &  ACS   &         1 &         1 &         0 &         0 &       0 &        0 &         2 \\
EB    &  WFC3  &         0 &         1 &         0 &         0 &       0 &        0 &         1 \\
\cline{3-9}
      &  total &   {\bf 1} &   {\bf 2} &   {\bf 0} &   {\bf 0} & {\bf 0} &  {\bf 0} &    {\bf 3} \\					
\hline								
      &  ACS   &   5$^{b}$ &         0 &   1$^{c}$ &         0 & 1$^{c}$ &        0 &          7 \\
Field &  WFC3  &         0 &         0 &         0 &         0 &       0 &        0 &          0 \\
\cline{3-9}
      &  total &   {\bf 5} &   {\bf 0} &   {\bf 0} &   {\bf 0} & {\bf 1} &  {\bf 0} &     {\bf 7} \\					
\hline
\hline
      & TOTAL$_{ACS}$  &  267 &   222 &   113 &   121 &   9 &  90 & 822 \\
      & TOTAL$_{WFC3}$ &   35 &    35 &     3 &     0 &   0 &   0 &  73 \\
\cline{3-9}			
      & TOTAL & {\bf 302} & {\bf 257} & {\bf 116} & {\bf 121} & {\bf 9} & {\bf 90} & {\bf 895} \\					
\hline
\hline
\end{tabular}
\end{scriptsize}
\begin{tablenotes}
\begin{scriptsize}
\item $^{a}$ Includes two stars with rather noisy light curves. Based on their position in the
CMD, we assume they are RRL stars. 
\item $^{b}$ RRL (3 RRab + 2 RRc) stars compatible with being field stars of
the giant stellar stream (GSS) of M31.
\item $^{c}$ RRab star compatible with a candidate field star from M31. 
\end{scriptsize}
\end{tablenotes}
\end{table*}

\begin{figure*}
    \hspace{2cm}
	\includegraphics[scale=0.7]{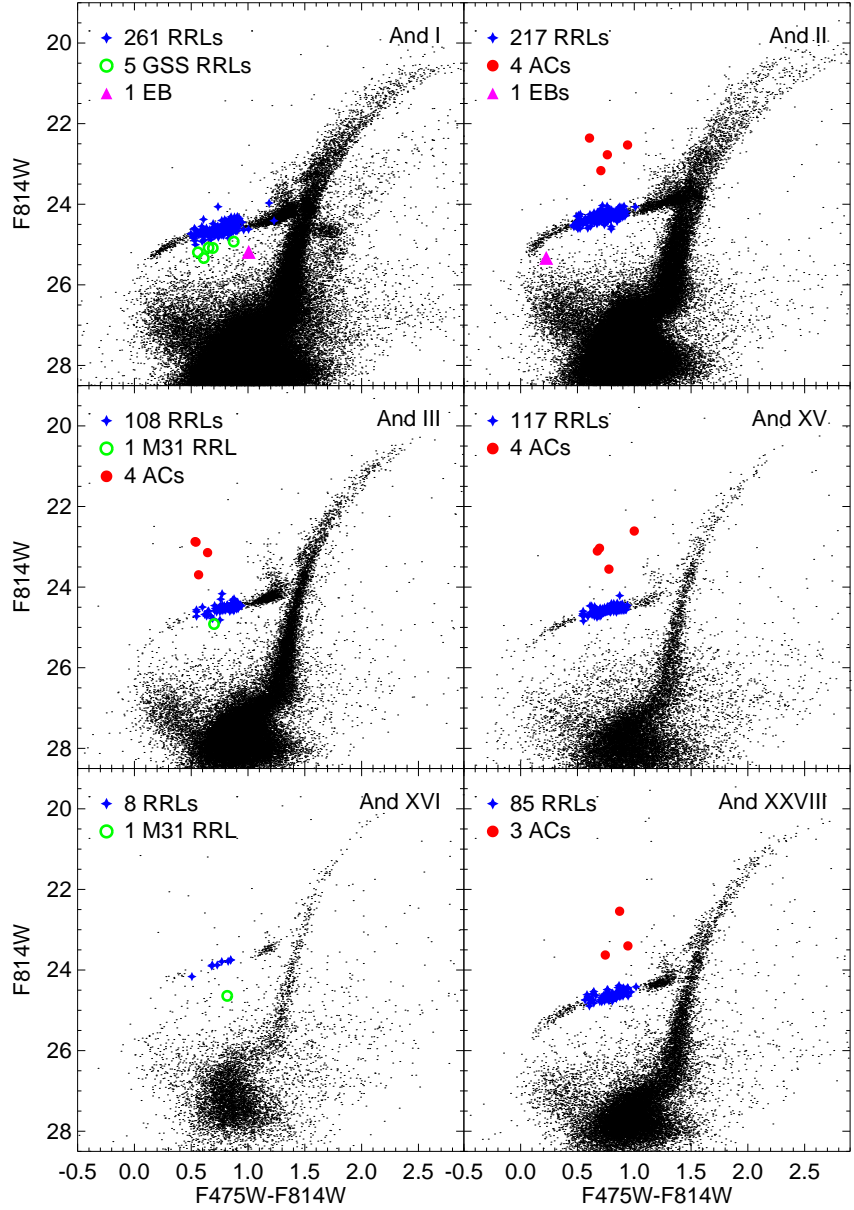}
	\caption{CMDs of the ACS fields for each ISLAndS galaxy. 
		The And~I CMD shows a significant contamination from M31 Giant Stellar Stream 
	  \citep{Ibata2001,Ferguson2002,McConnachie2003}.
	    Variable stars are overplotted.
        Blue stars represent the RRL stars.
		Red circles are the ACs. Green open circles are RRL stars tentatively
		associated with the field of M31. Magenta triangles are the probable
		eclipsing binaries.}
	\label{fig:cmds_acs}
\end{figure*}

\begin{figure*}
    \hspace{-1.5cm}
	\includegraphics[scale=0.75]{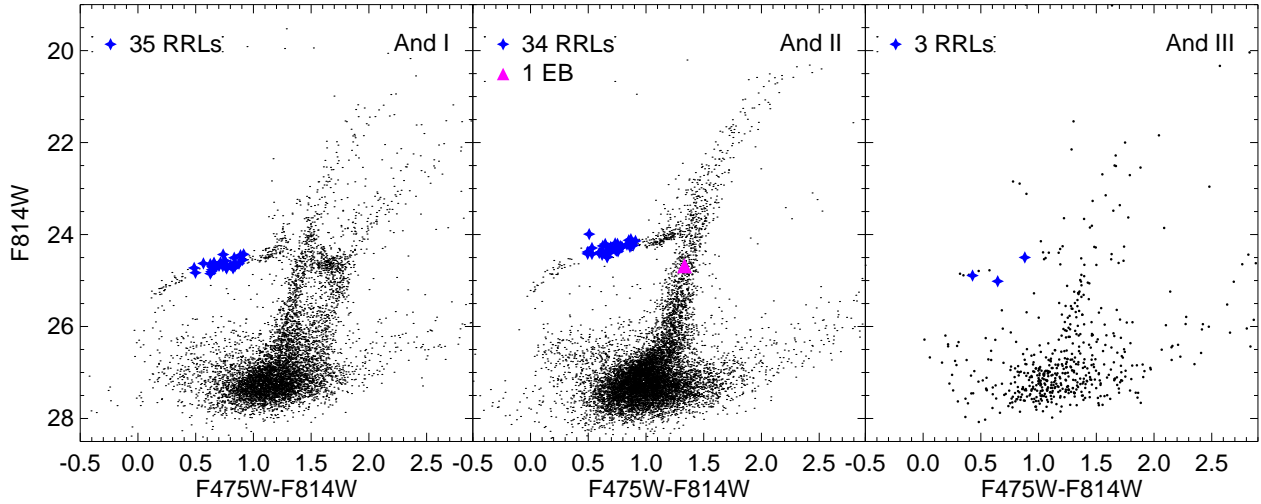}
	\caption{CMDs of the parallel WFC3 fields for the 
        three ISLAndS galaxy where there is still a relevant stellar
		population. Variable stars are overplotted. As in Figure~\ref{fig:cmds_acs},
		blue stars represent the RRL stars, and magenta triangles are the probable
		eclipsing binaries. In the case of the And~I CMD, the contamination 
        from M31 Giant Stellar Stream \citep{Ibata2001,Ferguson2002,McConnachie2003}
		is still present.}
	\label{fig:cmds_wfc3}
\end{figure*}

\begin{figure}
	\includegraphics[width=0.48\textwidth]{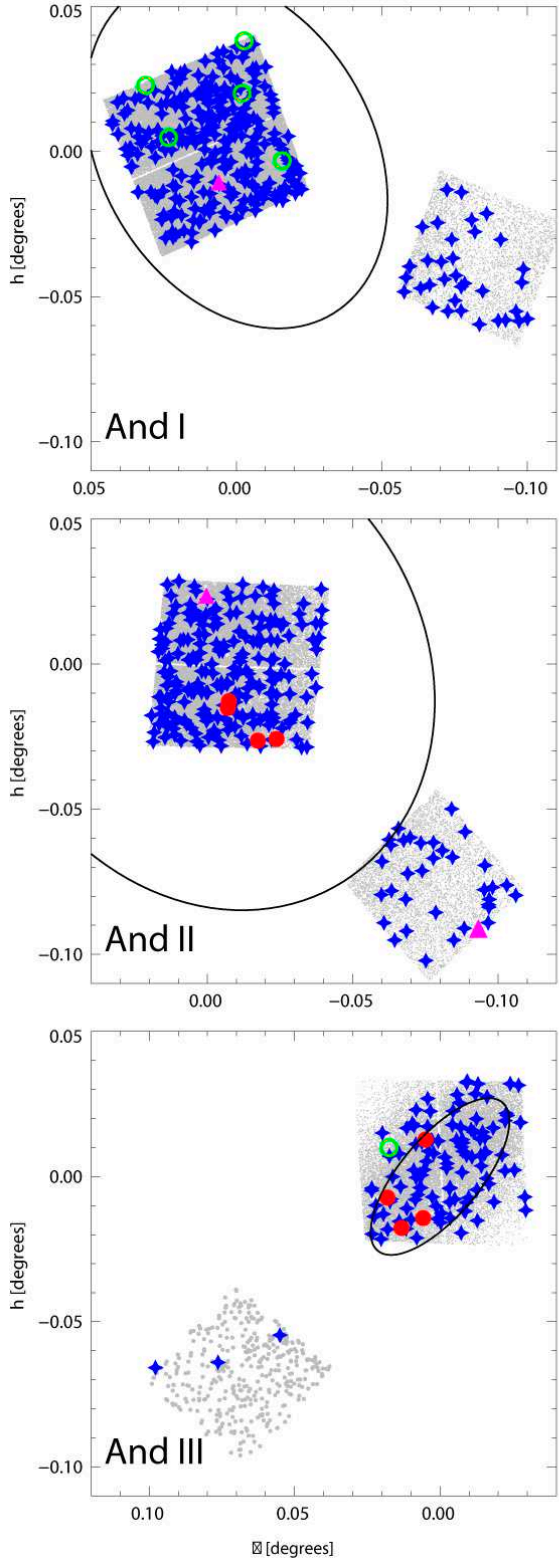}
	\caption{Spatial distribution of the variable stars found in 
	the observed ACS+WFC3 fields for And~I, II and III. Non-variable stars 
	are represented by gray dots. Variables are shown with the same symbol 
    and color code as in Figure~\ref{fig:cmds_acs}. The black ellipses represent
    the half-light radius (r$_h$) for each galaxy (column 6 in Table~\ref{tab:parameters}).}
	\label{fig:spatial_dist_123}
\end{figure}

\begin{figure*}
	\hspace{-1.2cm}
	\includegraphics[scale=0.67]{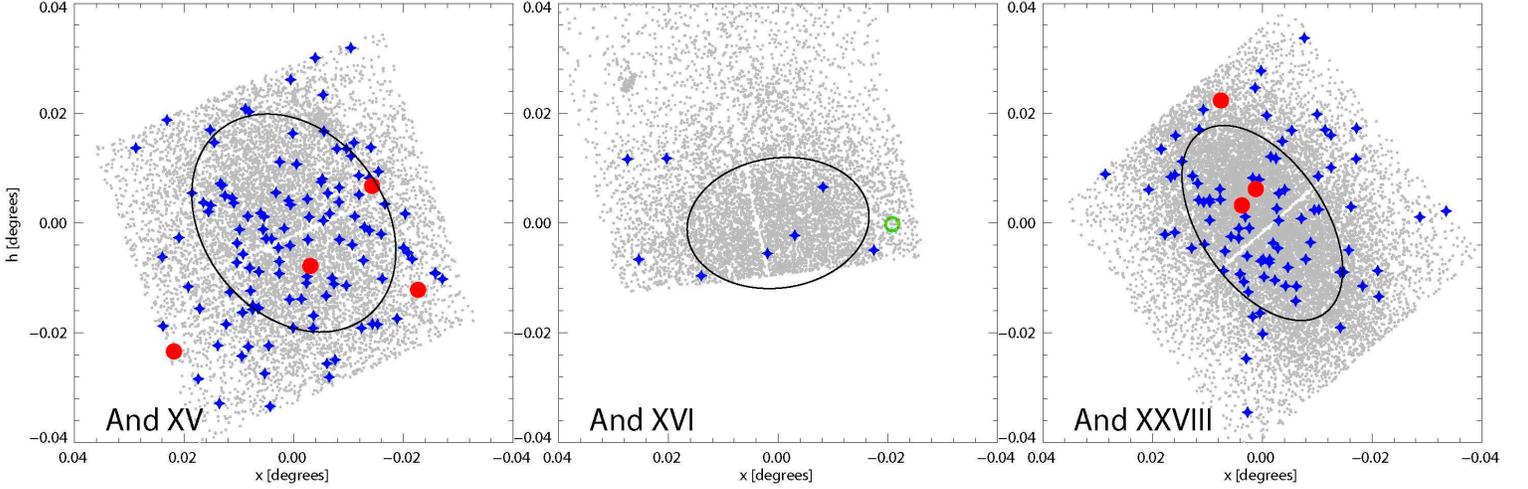}
	\caption{Spatial distribution of the variable stars found in 
	the observed ACS fields for And~XV, XVI and XXVIII. Non-variable stars 
	are represented by gray dots. Variables are shown with the same symbol 
    and color code as in Figure~\ref{fig:cmds_acs}. The black ellipses represent
    the half-light radius (r$_h$) for each galaxy (column 6 in 
    Table~\ref{tab:parameters}). The WFC3 fields are not shown here because the CMDs of 
    these three fields do not have any evidence of a satellite stellar population.}
	\label{fig:spatial_dist_151628}
\end{figure*}

\begin{figure}
	\hspace{-1.2cm}
	\includegraphics[scale=0.50]{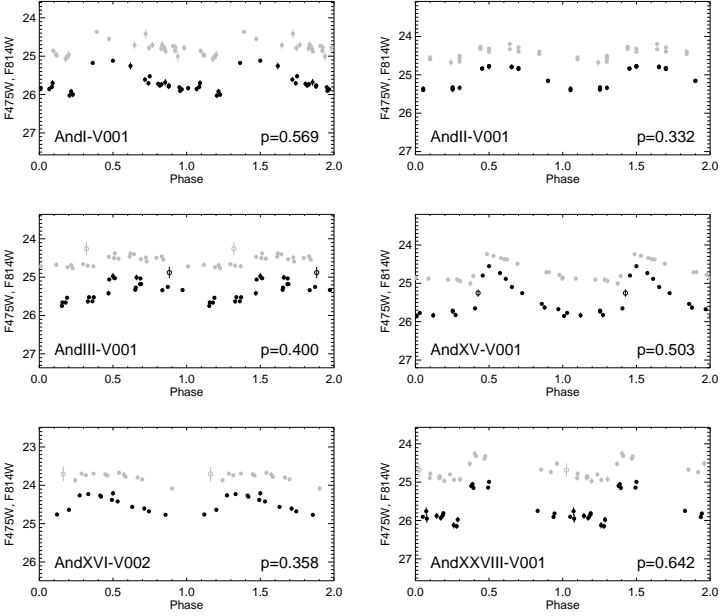}
	\caption{Examples of light curves of member RRL stars from each of the six ISLAndS galaxies
	in the $F475W$ (black) and $F814W$ (gray) bands. Periods (in days) are given
	in the lower-right corner, while the name of the variable is displayed
	in the left-hand side of each panel. Open symbols show the data-points for which the 
	uncertainties are larger than 3-$\sigma$ above the mean error of a given star; these data
	were not used in periods and mean magnitudes calculations.  
	All RRL light curves are available in the electronic edition of
 {\it The Astrophysical Journal}.}
	\label{fig:rrl_lcv}
\end{figure}

\begin{figure}
    \hspace{-1cm}
	\includegraphics[scale=0.52]{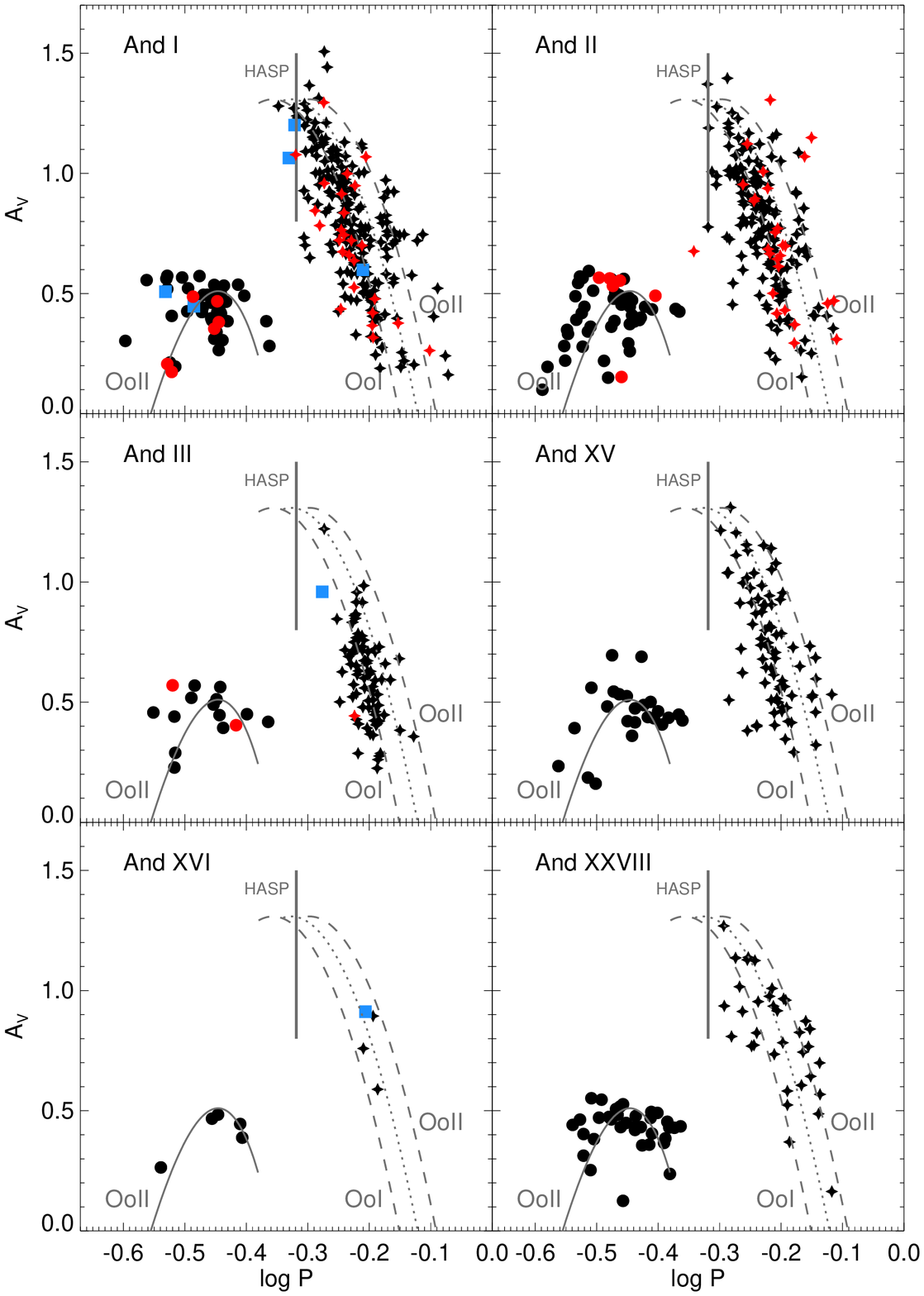}
	\caption{Period-amplitude or Bailey diagrams for the RRL samples. Stars and
	circles represent RRab and RRc stars (respectively) found in the ACS field (black) and in the
	WFC3 (red). Blue squares display the five RRLs which are probable M31 field stars. 
	The dashed gray lines are the relations for RRab stars in Oo-I and Oo-II clusters 
	obtained by \citet{Cacciari2005} while the dotted gray line delimits the 
    middle position between the last two. The gray solid curve is derived from the M22 (Oo-II cluster)
	RRc stars by \citet{Kunder2013}. Grey vertical lines mark the HASP limit defined by
	\citet{Fiorentino2015a} (see text for further details). For the sake of clarity, RRd stars are not plotted.}
	\label{fig:bailey}
\end{figure}

Candidate RRL stars and ACs were searched for in a rectangular region of the CMD 
with a width that covers the full color range of the HB, and with height between
$\sim$1.5 mag fainter than the HB to the magnitude of the TRGB, i.e., enclosing 
the instability strip (IS) where RRL stars and ACs are located\footnote{Other types 
of variable stars, such as long period RGB/AGB stars or very short period such as 
$\delta$-Scuti could not be detected nor properly characterized with the current 
data set, so we focus on the core helium-burning ones only.}. We visually 
inspected the LCs of all the stars in this region, without any cut on a 
variability index. The number of candidates ranged from 201 in And~XVI to 7414 
in And~I. The periodogram was calculated between 0.2 and 10 days, a range which 
encompasses all the possible periods of RRL stars and ACs.
Pulsational parameters were derived 
for the confirmed variables sources. Using widget based software, we first estimate the 
period of candidate variables through the Fourier analysis of the time series, following 
the prescription of \citet{Horne1986}. The analysis is refined by visual inspection of the
LCs in both bands simultaneously in order to fine-tune the period. The intensity-averaged 
magnitudes and amplitudes of the mono-periodic light curves were obtained by fitting the 
LCs with a set of templates partly based on the set of \citet{Layden1999} following the 
procedure described in \citet{Bernard2009}. We expect that the completeness of both 
the RRL star and AC samples are 100\% within each pointing for the following reasons: 
\textit{i)} the search for candidates, described above, insures that any star showing 
brightness variations has been visual inspected; \textit{ii)} according to the artificial 
star tests presented in \citet{Skillman2017}, the photometric completeness at the magnitude 
of the HB (and above) is about 100\%; and \textit{iii)} the amplitude of the RRLs and ACs 
pulsations are significantly larger than the magnitude uncertainty in the region 
of the HB and above.

The classification of variable stars was based on their pulsational properties 
(period and amplitude), LCs, and positions on the CMD. Table~\ref{tab:variables} 
summarizes the total number of different types of variable stars detected. Most of 
them are RRL stars (870 in the dwarfs + 7 field stars), while a few are ACs (15) 
and EBs (3). Each variable type will be described in more detail in the subsequent 
sections.

The individual $F475W$ and $F814W$ measurements (time-series) for all the detected 
variables are available in Appendix~\ref{sec:time_series} (Tables~\ref{tab:photometry_andi}, 
\ref{tab:photometry_andii}, \ref{tab:photometry_andiii}, \ref{tab:photometry_andxv}, 
\ref{tab:photometry_andxvi}, and \ref{tab:photometry_andxxviii}). The typical 
photometric uncertainties on individual measurements are of the order of 0.07 mag 
and 0.06 mag in $F475W$ and $F814W$, respectively, for the most distant galaxy 
(And~I), while for the nearest galaxy (And~XVI) it is of the order of 0.04 mag in 
both passbands. The variable stars were named with a prefix which refers 
to the galaxy,  followed by ``V'', indicating that the star is a variable (e.g., ``AndI-V'') 
and a number which increases with increasing right ascension. Interestingly, we note that 
no variable stars were detected in the parallel fields (WFC3) of And~XV, And~XVI, and 
And~XXVIII, in  agreement with the visual appearance of the CMD that does not show any 
obvious evolutionary sequence (HB, RGB, nor the more populous main sequence turn-off). 
We also note that the RRL stars of And~XVI were already presented
in \citet{Monelli2016}, but are included in this work as well for completeness.
As some of our target galaxies have been previously investigated for variability
(And~I: \citealt{Pritzl2005}; And~II: \citealt{Pritzl2004}; And~III:
\citealt{Pritzl2005}), a detailed comparison is presented in Appendix~\ref{sec:comp_pritzl}.

The derived values of the pulsational properties (period, amplitudes, mean magnitudes) 
for the variable stars detected in the different galaxies are presented in Appendix~\ref{sec:pulsation_properties} 
(Tables~\ref{tab:variables_andi}, 
\ref{tab:variables_andii}, \ref{tab:variables_andiii}, \ref{tab:variables_andxv}, 
\ref{tab:variables_andxvi}, and \ref{tab:variables_andxxviii}). These tables include the star name,
position (RA, Dec), period, mean magnitude and amplitude in the $F475W$, $F814W$, $B$, $V$ and 
$I$ passbands, and the classification. We note that the {\it HST} magnitudes in
the VEGAMAG system were transformed to the Johnson system using the calibration
provided by \citet{Bernard2009}.
The main purpose of this conversion from $F475W$ and $F814W$ magnitudes to Johnson $BVI$ 
is not only to allow comparison with observations of variable stars in globular clusters (GCs) 
and other galaxies reported in the literature (see \S~\ref{sec:discussion}) but also for 
using the period-luminosity relations (for example to obtain distances, as
we do in \S~\ref{sec:distances}) or the Bailey (period-amplitude) diagram 
(see \S~\ref{sec:rrl}) that are most commonly used in the $V$ band.

We display in Figure~\ref{fig:cmds_acs} the CMDs 
of the ACS fields of the six galaxies highlighting in them the different types of variable 
stars detected: RRL stars (blue stars symbols for those dSph members and green open circles
for field RRL stars), ACs (red circles) and EBs (magenta triangles). Table~\ref{tab:variables}
displays the number of detected variables of each type in the ACS fields. The CMD of And~I 
shows clearly the contamination of the M31 Giant Stellar Stream 
\citep[GSS,][]{Ibata2001,Ferguson2002,McConnachie2003} as shown by the presence of a second, 
redder RGB and red clump visible in the CMD. In particular, we have found 5 RRL stars with 
properties compatible with membership in the GSS (see \S~\ref{sec:rrl_gss}).

RRLs were detected in all six galaxies with as few as 8 (in And~XVI\footnote{Excluding 
the RRL star AndXVI-V001 (V0 in \citealt{Monelli2016}) as it is a candidate M31 halo field 
star not belonging to And~XVI.}) and as a many as 296 (in And~I). The striking difference
in the number of RRL between And~XVI and And~XV, despite having a similar mass, can be 
explained as a consequence of their different SFHs: the mass fraction already in place at 
old ages (10 Gyr ago) was only about 50\% in And~XVI, while it was 90\% in And~XV (see 
Figure 7 in \citealt{Skillman2017}).

A few (3-4) ACs are present in And~II, III, XV, and XXVIII, but none have been detected 
in And~I nor in And~XVI. This is not surprising in the case of the latter, due to its low 
mass\footnote{The initial estimate of its luminosity \citep[M$_V$ = --9.2 mag][]{Ibata2007} 
suggested a relatively bright object. However, more recent estimates \citep{Martin2016} 
revised this value to a significantly fainter value (see Table~\ref{tab:parameters})}. 
The lack of ACs is however remarkable in the case of And~I, as no other massive dSph 
presents such a dearth of ACs (see \S~\ref{sec:ac}). Nevertheless, the high mean 
metallicity \citep{Kalirai2010} may explain such occurrence.

Figure~\ref{fig:cmds_wfc3} presents the CMDs for the parallel WFC3 fields of And~I, II, 
and III, where variable stars have been detected. The symbols are the same
as in Figure~\ref{fig:cmds_acs}. The presence of the GSS is also noticeable in the CMD 
of the parallel WFC3 field of And~I. For the cases of And~XV, XVI, and XXVIII, the 
parallel WFC3 field do not show a significant component of the galaxy; in fact, no 
variable stars have been detected.

Figures~\ref{fig:spatial_dist_123} and \ref{fig:spatial_dist_151628} present the spatial 
distribution of variable stars in the six galaxies, as detected by the two cameras. 
The black ellipses represent the half-light radius (column 6 in Table~\ref{tab:parameters}). 
These two plots show that the area covered for the six galaxies is far from being complete. 
Nevertheless, for four of the six galaxies we cover beyond the half-light radius, 
thus implying that the large majority of RRL stars have been detected. Wide-field, 
ground based photometric follow-up would be valuable to complete the census, 
especially in the case of the largest galaxies.


\section{RR Lyrae stars} \label{sec:rrl}

\subsection{Mean properties and Bailey diagrams}

\begin{table*}
\renewcommand{\thetable}{\arabic{table}}
\centering
\caption{RRL star subgroups.}
\label{tab:rrl}
\begin{scriptsize}
\begin{tabular}{lrccccccc}
\cline{3-9}
      &        & And~I & And~II & And~III & And~XV & And~XVI & And~XXVIII & Total \\
\hline
      &  ACS   &       203 &       160 &      83  &       80 &       3 &      35 &       562 \\
RRab  &  WFC3  &        26 &        27 &       1  &        0 &       0 &       0 &        53 \\
\cline{3-9}
      &  total & {\bf 229} & {\bf 187} & {\bf 84} & {\bf 80} & {\bf 3} & {\bf 35} & {\bf 615} \\
\hline
      &  ACS   &        42 &        42 &      13 &        24 &       5 &       35 &       158 \\
RRc   &  WFC3  &         6 &         6 &       2 &         0 &       0 &        0 &        14 \\
\cline{3-9}
      &  total &  {\bf 48} &  {\bf 48} & {\bf 15} & {\bf 24} & {\bf 5} & {\bf 34} & {\bf 172} \\
\hline								
      &  ACS   &        16 &        15 &       12 &       13 &       0 &        15 &       69 \\
RRd   &  WFC3  &         3 &         1 &        0 &        0 &       0 &         0 &        4 \\
\cline{3-9}
      &  total &   {\bf 19} & {\bf 16} & {\bf 12} & {\bf 13} & {\bf 0} &  {\bf 15} &  {\bf 73} \\					
\hline
\hline
      & TOTAL$_{ACS}$  &  261 &   217 &  108 &   117 &   8 &  85$^{a}$ & 797 \\
      & TOTAL$_{WFC3}$ &   35 &    34 &    3 &     0 &   0 &   0 &  72 \\
\cline{3-9}			
      & TOTAL & {\bf 296} & {\bf 251} & {\bf 111} & {\bf 117} & {\bf 8} & {\bf 85$^{a}$} & {\bf 869} \\					
\hline
\hline
\end{tabular}
\end{scriptsize}
\begin{tablenotes}
\begin{scriptsize}
\item $^{a}$ We have identified 2 additional RRL star candidates with noisy light curves. We do not include 
them here because of the uncertainty in their classification.
\end{scriptsize}
\end{tablenotes}
\end{table*}

\begin{table*}
\renewcommand{\thetable}{\arabic{table}}
\centering
\caption{Mean properties of the RRL stars.}
\label{tab:mean_periods}
\begin{scriptsize}
\begin{tabular}{lccccccc}
\hline 
\hline
Galaxy & $\langle$P$_{ab} \rangle$ & $\langle$P$_{c} \rangle$ & f$_c$ & f$_{cd}$ & \% Oo-I & \% Oo-II  & $\langle$m$_V\rangle$\\   
\hline
And~I       & 0.597$\pm$0.004 ($\sigma$=0.07) & 0.343$\pm$0.005 ($\sigma$=0.03) & 0.17 & 0.23 & 80 & 20 & 25.13 \\ 
And~II      & 0.601$\pm$0.005 ($\sigma$=0.07) & 0.332$\pm$0.006 ($\sigma$=0.04) & 0.20 & 0.25 & 80 & 20 & 24.78 \\ 
And~III     & 0.622$\pm$0.004 ($\sigma$=0.03) & 0.344$\pm$0.011 ($\sigma$=0.04) & 0.15 & 0.24 & 89 & 11 & 25.04 \\ 
And~XV      & 0.608$\pm$0.006 ($\sigma$=0.05) & 0.360$\pm$0.009 ($\sigma$=0.04) & 0.23 & 0.32 & 78 & 22 & 25.07 \\ 
And~XVI     & 0.636$\pm$0.010 ($\sigma$=0.02) & 0.356$\pm$0.019 ($\sigma$=0.04) &  --- & --- & 67 & 33 & 24.34 \\ 
And~XXVIII  & 0.624$\pm$0.012 ($\sigma$=0.07) & 0.359$\pm$0.007 ($\sigma$=0.04) & 0.50 &  0.59 & 49 & 51 & 25.14 \\ 
\hline
\hline
\end{tabular}
\end{scriptsize}
\begin{tablenotes}
\begin{scriptsize}
\item Notes.- 
\item Mean periods are given in days.
\item The definition of f$_c$ is $\frac{Nc}{Nab+Nc}$ while f$_{cd}$ is defined as $\frac{Nc+Nd}{Nab+Nc+Nd}$
\end{scriptsize}
\end{tablenotes}
\end{table*}

RRL stars are low-mass ($\sim$ 0.6 - 0.8M$_{\odot}$) and radially pulsating 
variable stars with periods ranging from 0.2 to 1.0 d and $V$ amplitudes from 
0.2 to $\lesssim$ 2 mag. They are found in stellar systems which host an old 
(t $>$ 10 Gyr) stellar population \citep{Walker1989,Smith1995,Catelan2015}.
A total of 870 RRL stars were detected and characterized in the six ISLAndS galaxies.
Table~\ref{tab:rrl} summarizes, for each galaxy, the number of fundamental (RRab), 
first overtone (RRc) and double-mode (RRd) pulsators in both the ACS and WFC3 fields of view.
Different types of RRL stars are usually easy to classify on the basis of a visual inspection of the 
light curve and the period. RRab stars are characterized by longer periods 
($\sim$0.45-1.0 days) and saw-tooth light curves, with a steep rise up to the maximum and 
a less steep fall to the minimum. RRc have shorter periods ($\sim$0.2-0.45 days), 
lower amplitudes (A$_V \lesssim$ 0.8) and almost sinusoidal light variations. 
Conversely, RRd stars have usually periods around 0.4 d and their light curves are 
particularly noisy due to simultaneous pulsation in the fundamental mode and first overtone. 
Figure~\ref{fig:rrl_lcv} presents an example RRL light curve for each galaxy. Black and 
gray points are used for data in the $F475W$ and $F814W$ passbands, respectively. Open 
symbols are used to indicate outlier measurements that have not been taken into 
account in deriving the pulsational properties. We emphasize that the whole set of light
curves is available in the electronic edition of this paper.
Additionally, the properties of the individual variable stars can be found in 
Appendix~\ref{sec:pulsation_properties}. 

Figure~\ref{fig:bailey} presents the period-amplitude (Bailey) diagram for the six
galaxies (see caption for details). The plot shows the two different relations for the 
Oosterhoff types, represented in the plot by the dashed lines 
\citep[Oosterhoff I and II, or Oo-I and Oo-II][]{Cacciari2005}. As long known 
\citep{Oosterhoff1939,Oosterhoff1944}, the properties of RRab stars divide Galactic GCs 
into two groups, called Oosterhoff I and Oosteroff II. The mean period of fundamental
pulsators of the former group is shorter (P$\sim$0.55 d) than the latter (P$\sim$0.65).
Although the origin of such behavior has not been fully explained, the Oosterhoff 
dichotomy appears to be related to the metallicity of the clusters, being the Oo-II 
stars more metal-poor, on average \citep[e.g., see the review by][]{Catelan2009}. 
On the other hand, dwarf galaxies do not show similar behavior, as the mean 
period of their RRab stars typically locates them in the \textit{Oosterhoff gap}
between the two Oostherhoof groups. For this reason, they have been often
considered as \textit{Oosterhoff-intermediate} types 
\citep[see e.g.,][]{Kuehn2008, Bernard2009, Bernard2010, Garofalo2013, Stetson2014,Cusano2015, Ordonez2016}. 

Table~\ref{tab:mean_periods} summarizes the mean pulsational properties for the galaxies in 
our sample: the mean periods of RRab ($<P_{ab}>$) and RRc ($<P_{c}>$) type stars, the fraction 
of RRc ($f_c$) and of RRc+RRd ($f_{cd}$) stars, the fraction of Oo-I-like and Oo-II-like stars 
(defined below in this section), and the apparent mean magnitude in $V$-band (which will be 
used in \S~\ref{sec:distances} for determining the distance to the galaxy). From the information  
in the Table, the six ISLAndS galaxies could also be considered \textit{Oosterhoff-intermediate}, 
since they have $<P_{ab}>$ $\sim$ 0.6 d. In this respect, the ISLAndS galaxies are similar to the 
MW dSph satellites. However, an intermediate mean period does not mean that the stars are 
distributed in the Bailey diagram {\it between} the two typical Oosterhoff lines. 
Figure~\ref{fig:bailey} clearly shows that stars tend to clump around each Oosterhoff group 
locus, and with a predominance of \textit{Oo-I like} stars. In fact, if we split the sample 
using the dotted, intermediate line, and classify stars as \textit{Oo-I like} or \textit{Oo-II like} 
according to their relative position with respect of this separator, four galaxies (And~I, II, III, 
and XV) present a majority ($\sim$80\%) of \textit{Oo-I like} stars (see Table~\ref{tab:oo}). In the 
case of And~I and II, the same result was found for the variable stars in the parallel WFC3.

And~XXVIII is the exception, with a fraction of \textit{Oo-I like} stars close to 
50\%. Moreover, the distribution of RRLs in the Bailey diagram 
is also different from the other And dSphs; the RRab stars show a broad
spread and are not concentrated on either Oosterhoff line. And~XXVIII is also peculiar 
for the large fraction of RRcd type stars, which represent $\sim$58\% of the 
total. In the LG, if we exclude low-mass galaxies with very small 
samples of RRLs ($<$15, such as e.g., Bootes I and And~XVI, see \S~\ref{sec:global_rrl}), 
And~XXVIII is the only galaxy with more RRcd than RRab type stars. Similar to
And~XXVIII, the galaxies with particularly large fraction of RRcd (Ursa Minor: 
43\%, \citealt{Nemec1988}; Sculptor: 46\%, \citealt{MartinezVazquez2016b}; Tucana: 40\%, 
\citealt{Bernard2009}) are all also characterized by the presence 
of a strong blue HB component. This may be connected to a sizable population
of very metal-poor stars.

The black vertical line in Figure~\ref{fig:bailey} marks the limit of 
the High Amplitude Short Period (HASP) region, defined by \citet{Fiorentino2015a}
as those RRab stars with periods shorter than 0.48 d and amplitudes in the $V$
band larger than 0.75 mag. These stars are interpreted as the metal-rich
tail of the metallicity distribution of RRL stars ([Fe/H]$>$--1.5), and have
been found only in systems that were dense or massive enough to enrich to this
metallicity before 10 Gyr ago \citep{Fiorentino2017}.
We confirm this trend with the six ISLAndS galaxies, as HASPs have only been detected
in the two most massive satellite galaxies: And~I (3\footnote{The other two most 
likely belong to the M31 GSS, see \S~\ref{sec:rrl_gss} for further details.}) and And~II (2). 
A detailed analysis of the chemical properties of RRL stars will be discussed 
in a forthcoming paper.

It is worth noting that a few stars with HASP properties were already identified in the 
catalogs by \citet{Pritzl2004} and \citet{Pritzl2005} for And~II and And~I, 
respectively. In the case of And~I, we confirm the HASP nature of 3 out of the 
7 stars, while the period was likely underestimated for the other 4, possibly due to aliasing 
(see Appendix~\ref{sec:comp_pritzl}).
However, we do not confirm any of the 8 HASP stars in And~II (see the Appendix~\ref{sec:comp_pritzl} for a
detailed comparison with literature values). Nevertheless, we discovered 2 new HASPs in
And~I and 2 in And~II, which are all located outside the WFPC2 field studied by 
\citet{Pritzl2004,Pritzl2005}.

	\subsection{Five detected RR Lyrae stars from M31 GSS} \label{sec:rrl_gss}

Five RRLs in And~I have mean magnitudes that are a few tenths of a magnitude fainter 
than the HB (three RRab: AndI-V053, AndI-V110 
and AndI-V113; and two RRc: AndI-V257 and AndI-V280). We exclude the possibility that
sampling problems of the light curve may be causing a bias toward fainter magnitudes.
Possible explanations are: {\em i)} a significantly higher metal content, or 
{\em ii)} a distance effect. Assuming they are at the distance of And I, in order 
to explain such faint luminosity (0.45 mag fainter) a super solar metallicity is required.
This value appears to be unlikely given the morphology of the CMD and the star
formation history \citep{Skillman2017}. 

On the other hand, as indicated in the previous section, the CMD of And~I shows 
that a significant contamination by the 
GSS is present along the line of sight to And~I. In particular, And~I 
is projected on the GSS ``Field 3'' studied by \citet{McConnachie2003}, which is 
located at 860$\pm$20 kpc according to the TRGB determination. To verify whether
the faint RRL stars can be associated to the GSS, we first note that two of the 
three RRab are HASP RRL stars. This suggests that their metallicity is likely to be
higher than --1.5 dex. Assuming [Fe/H]=--1.5 and using the period-Wesenheit relation 
described in \S~\ref{sec:wese}, we obtain a mean distance modulus of $\mu_0$=24.86 mag 
(sys=0.08; rand=0.11), for the five stars, corresponding to 937 kpc (sys=34; rand=47). 
This means that they are likely
located $\sim$140 kpc beyond And~I (d$_{\odot}\sim$800 kpc, see \S~\ref{sec:distances}).
Given the error bars, we conclude that the five faint RRL stars are compatible with 
being connected to the metal-poor component of the GSS \citep{Gilbert2009} 
rather than members of And~I.

\section{Properties of the old population in the M31 and MW satellites system}\label{sec:discussion}

\begin{figure}
\hspace{-0.5cm}
\includegraphics[scale=0.58]{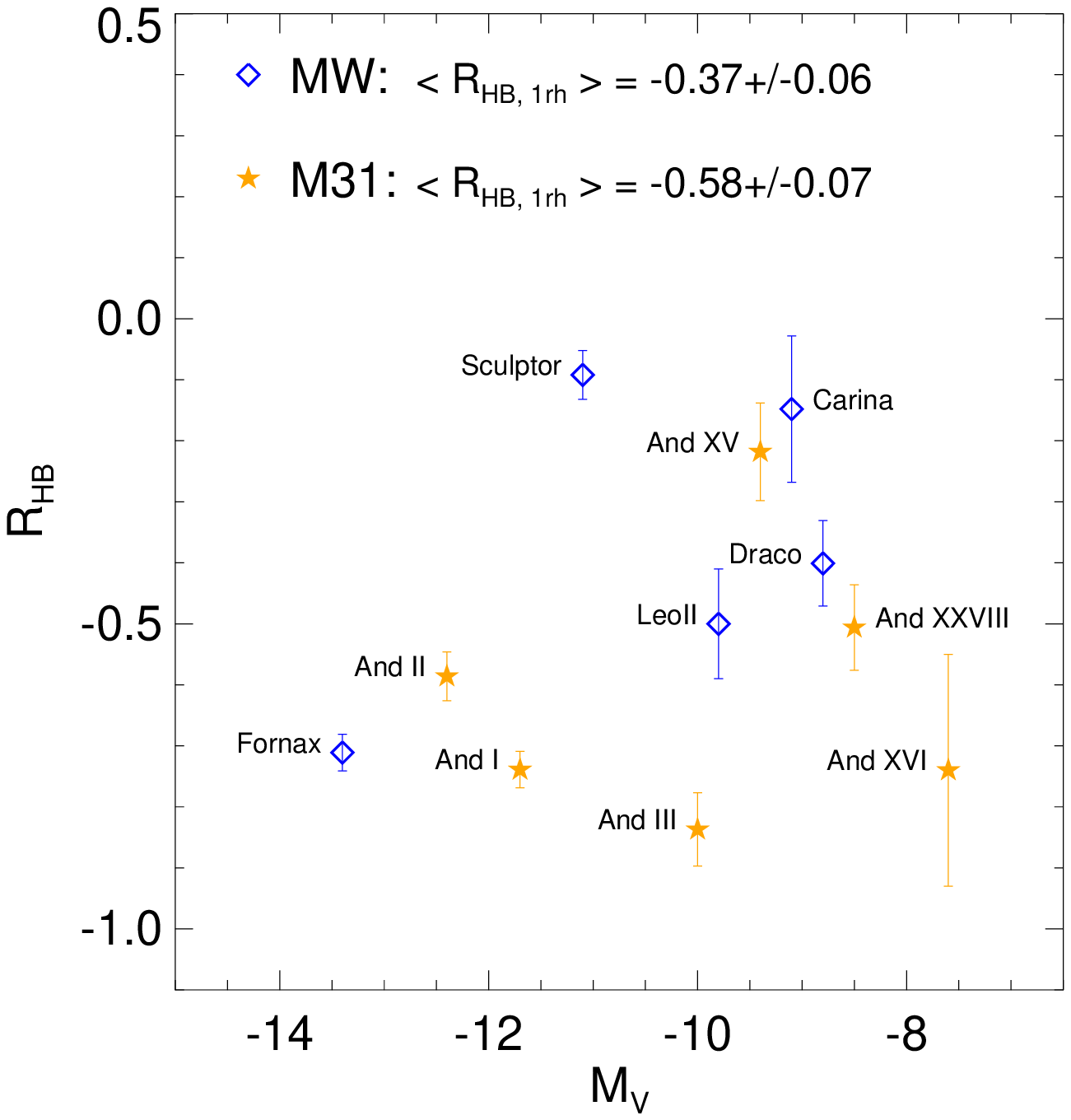}
	\caption{R$_{HB}$ index versus the luminosity of the host
    galaxy, M$_V$, for the ISLAndS targets (orange filled stars) and a 
    sample of MW satellites (blue open diamonds). The values have been 
    calculated within 1 r$_h$, except for And~I and II (red stars) 
    since the field of view of the ACS is not large enough. The mean 
    value for M31 satellites support redder HB morphology than for 
    MW satellites.}
	\label{fig:hbr}
\end{figure}

\begin{figure}
\hspace{-1.2cm}
	\includegraphics[scale=0.45]{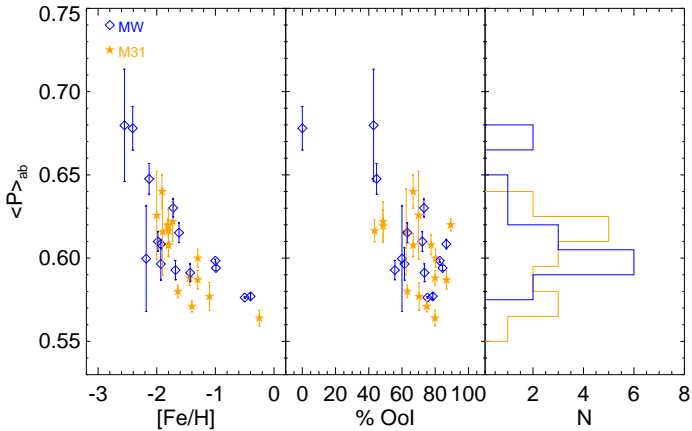}
	\caption{{\em Left and Middle - }Mean period of the RRab stars of the 
    sample of MW (blue) and M31 (orange) satellites versus the mean 
    metallicity and the percentage of Oo-I type stars in the system. 
	There is no obvious between the two subgroups.
    {\em Right - }Mean period distribution of the sample of MW (blue histogram) 
    and M31 dwarf galaxies (orange histogram). The peaks of the two 
    distribution are very close to each other.}
	\label{fig:mvp}
\end{figure}

\begin{figure}
\hspace{-1.3cm}
	\includegraphics[scale=0.45]{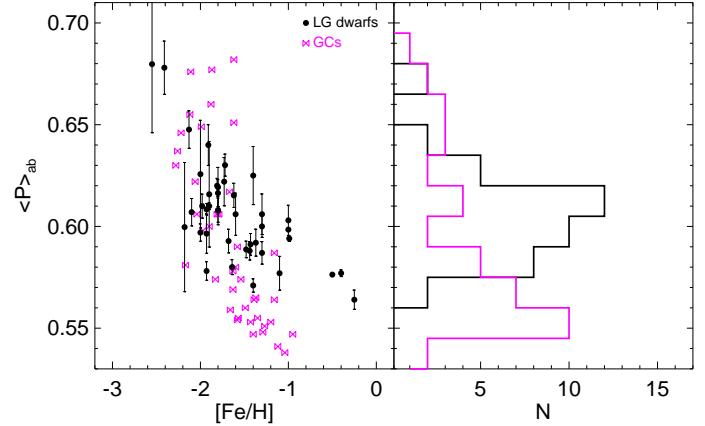}
	\caption{{\em Left -} $<P_{ab}>$ for a sample of 41 dwarf galaxies
    reported in Table~\ref{tab:oo} (black dots) as a function of [Fe/H], 
    compared to that of GCs (purple bowties).
    {\em Right -} Period distribution of the sample of dwarf galaxies 
    and GCs. The peak of the former occurs at a period typical of the 
    Oo-intermediate system, while the latter peaks in the short period 
    regime, populated by Oo-I systems, which is devoid of galaxies.}
	\label{fig:histo_galgc}
\end{figure}

\begin{table*}
\renewcommand{\thetable}{\arabic{table}}
\caption{Properties of the set of RRL stars in a sample of 41 LG dwarf
galaxies of different morphological type within $\sim$2 Mpc, with at least five RRab and with data available in literature.}
\label{tab:oo}
\begin{scriptsize}
\setlength\tabcolsep{3pt}
\hspace{-2.3cm}
\begin{tabular}{lccccccccccc} 
\hline
\hline
& & &\multicolumn{2}{c}{RRab} & & & \multicolumn{2}{c}{$\langle$P$_{ab}\rangle$} & \multicolumn{2}{c}{$\langle$P$_{cd}\rangle$} & \\ 
\cmidrule(lr){4-5}
\cmidrule(lr){8-9}
\cmidrule(lr){10-11}
Galaxy & $\langle$[Fe/H]$\rangle^{*}$ & RRab & \%OoI & \%OoII & RRcd & f$_{cd}$ & Median & Mean & Median & Mean & References \\
\hline
\multicolumn{11}{c}{MW dwarf satellites} \\ 
\hline
        Ursa Major I  &  -2.18  &     5  &  60  &  40  &     2  &  0.29  &  0.600  &  0.628$\pm$0.031(0.07)  &  0.407  &  0.402$\pm$0.005(0.008) & \citet{Garofalo2013} \\
            Bootes I  &  -2.55  &     7  &  43  &  57  &     8  &  0.53  &  0.680  &  0.691$\pm$0.034(0.09)  &  0.386  &  0.364$\pm$0.016(0.04) & \citet{Siegel2006} \\
            Hercules  &  -2.41  &     6  &   0  & 100  &     3  &  0.33  &  0.678  &  0.678$\pm$0.013(0.03)  &  0.400  &  0.399$\pm$0.002(0.003) & \citet{Musella2012} \\
    Canes Venatici I  &  -1.98  &    18  &  72  &  28  &     5  &  0.22  &  0.610  &  0.604$\pm$0.006(0.03)  &  0.390  &  0.378$\pm$0.012(0.03) & \citet{Kuehn2008} \\
               Draco  &  -1.93  &   211  &  87  &  13  &    56  &  0.21  &  0.608  &  0.615$\pm$0.003(0.04)  &  0.401  &  0.389$\pm$0.004(0.03) & \citet{Kinemuchi2008} \\
          Ursa Minor  &  -2.13  &    47  &  45  &  55  &    35  &  0.43  &  0.648  &  0.638$\pm$0.009(0.06)  &  0.383  &  0.375$\pm$0.011(0.07) & \citet{Nemec1988} \\
              Carina  &  -1.72  &    71  &  73  &  27  &    12  &  0.15  &  0.630  &  0.634$\pm$0.005(0.05)  &  0.364  &  0.350$\pm$0.013(0.04) & \citet{Coppola2015} \\
             Sextans  &  -1.93  &    26  &  62  &  38  &    10  &  0.28  &  0.596  &  0.606$\pm$0.010(0.05)  &  0.352  &  0.355$\pm$0.019(0.06) & \citet{Mateo1995} \\
              Leo II  &  -1.62  &   106  &  63  &  37  &    34  &  0.24  &  0.615  &  0.619$\pm$0.006(0.06)  &  0.370  &  0.363$\pm$0.008(0.05) & \citet{Siegel2000} \\
            Sculptor  &  -1.68  &   289  &  56  &  44  &   247  &  0.46  &  0.593  &  0.610$\pm$0.006(0.10)  &  0.355  &  0.346$\pm$0.002(0.04) & \citet{MartinezVazquez2016b} \\
               Leo I  &  -1.43  &   136  &  74  &  26  &    28  &  0.17  &  0.591  &  0.599$\pm$0.005(0.06)  &  0.367  &  0.352$\pm$0.007(0.04) & \citet{Stetson2014} \\
              Fornax  &  -0.99  &   998  &  84  &  16  &   445  &  0.31  &  0.594  &  0.595$\pm$0.001(0.05)  &  0.380  &  0.379$\pm$0.001(0.07) & \citet{Fiorentino2017} \\
         Sagittarius  &  -0.40  &  1636  &  79  &  21  &   409  &  0.20  &  0.576  &  0.575$\pm$0.002(0.07)  &  0.322  &  0.319$\pm$0.002(0.04) & \citet{Soszynski2014}  \\
                 SMC  &  -1.00  &  4961  &  83  &  17  &  1407  &  0.22  &  0.598  &  0.598$\pm$0.0008(0.06) &  0.366  &  0.360$\pm$0.001(0.04) & \citet{Soszynski2016} \\
                 LMC  &  -0.50  & 27620  &  75  &  25  & 11461  &  0.29  &  0.576  &  0.580$\pm$0.0004(0.07) &  0.339  &  0.333$\pm$0.000(0.04) & \citet{Soszynski2016} \\
\hline
\multicolumn{11}{c}{M31 dwarf satellites} \\ 
\hline
            And XIII  &  -1.90  &     8  &  63  &  37  &     1  &  0.11  &  0.616  &  0.648$\pm$0.026(0.07)  &  0.4287 &  0.4287                & \citet{Yang2012} \\
              And XI  &  -2.00  &    10  &  70  &  30  &     5  &  0.33  &  0.626  &  0.621$\pm$0.026(0.08)  &  0.428  &  0.423$\pm$0.013(0.03) & \citet{Yang2012} \\
          And XXVIII  &  -1.73  &    35  &  49  &  51  &    50  &  0.59  &  0.622  &  0.624$\pm$0.012(0.07)  &  0.366  &  0.361$\pm$0.005(0.04) & This work  \\
             And XVI  &  -1.91  &     3  &  67  &  33  &     5  &  0.63  &  0.640  &  0.636$\pm$0.010(0.02)  &  0.358  &  0.356$\pm$0.019(0.04) & This work  \\
             And XIX  &  -1.80  &    23  &  44  &  56  &     8  &  0.26  &  0.616  &  0.618$\pm$0.007(0.03)  &  0.401  &  0.392$\pm$0.010(0.03) & \citet{Cusano2013} \\
              And XV  &  -1.80  &    80  &  76  &  22  &    37  &  0.32  &  0.608  &  0.608$\pm$0.006(0.05)  &  0.366  &  0.364$\pm$0.006(0.04) & This work  \\
             And XXV  &  -1.80  &    45  &  67  &  33  &    11  &  0.20  &  0.608  &  0.607$\pm$0.007(0.05)  &  0.370  &  0.363$\pm$0.010(0.03) & \citet{Cusano2016} \\
             And XXI  &  -1.80  &    37  &  49  &  51  &     4  &  0.10  &  0.619  &  0.638$\pm$0.010(0.06)  &  0.387  &  0.343$\pm$0.028(0.06) & \citet{Cusano2015} \\
             And III  &  -1.81  &    84  &  89  &  11  &    27  &  0.24  &  0.620  &  0.623$\pm$0.004(0.03)  &  0.375  &  0.375$\pm$0.012(0.06) & This work  \\
              And VI  &  -1.30  &    91  &  87  &  13  &    20  &  0.18  &  0.587  &  0.588$\pm$0.006(0.05)  &  0.386  &  0.382$\pm$0.009(0.04) & \citet{Pritzl2002} \\
               And I  &  -1.44  &   229  &  80  &  20  &    67  &  0.23  &  0.588  &  0.597$\pm$0.004(0.07)  &  0.353  &  0.349$\pm$0.004(0.03) & This work  \\
              And II  &  -1.30  &   187  &  80  &  20  &    64  &  0.26  &  0.600  &  0.601$\pm$0.005(0.07)  &  0.350  &  0.341$\pm$0.005(0.04) & This work  \\
             And VII  &  -1.40  &   386  &  75  &  25  &   187  &  0.33  &  0.571  &  0.578$\pm$0.003(0.06)  &  0.342  &  0.338$\pm$0.003(0.04) & \citet{Monelli2017} \\
             NGC 147  &  -1.10  &   118  &  70  &  30  &    59  &  0.33  &  0.577  &  0.589$\pm$0.008(0.09)  &  0.331  &  0.325$\pm$0.006(0.05) & \citet{Monelli2017} \\
              NGC185  &  -1.64  &   544  &  63  &  37  &   276  &  0.34  &  0.580  &  0.587$\pm$0.004(0.09)  &  0.325  &  0.322$\pm$0.003(0.04) & \citet{Monelli2017} \\
                 M32  &  -0.25  &   314  &  80  &  20  &   102  &  0.25  &  0.564  &  0.569$\pm$0.005(0.08)  &  0.324  &  0.323$\pm$0.004(0.04) & \citet{Fiorentino2012c} \\
\hline
\multicolumn{11}{c}{Isolated dwarf galaxies} \\ 
\hline 
              Tucana  &  -2.00  &  216  &  68  &  32  &   142  &  0.40  &  0.597  &  0.604$\pm$0.004(0.06)  &  0.370  &  0.367$\pm$0.003(0.03) & \citet{Bernard2009} \\
             Phoenix  &  -1.37  &   95  &  70  &  30  &    26  &  0.21  &  0.592  &  0.602$\pm$0.007(0.06)  &  0.360  &  0.363$\pm$0.014(0.07) & \citet{Ordonez2014} \\
                LGS3  &  -2.10  &  109  &  69  &  31  &    24  &  0.18  &  0.607  &  0.616$\pm$0.007(0.07)  &  0.372  &  0.360$\pm$0.011(0.05) & Mart\'inez-V\'azquez et al. in prep. \\
             DDO 210  &  -1.30  &   24  &  92  &   8  &     8  &  0.25  &  0.606  &  0.609$\pm$0.010(0.05)  &  0.374  &  0.359$\pm$0.027(0.08) & \citet{Ordonez2016} \\
               Cetus  &  -1.90  &  506  &  83  &  17  &   124  &  0.20  &  0.610  &  0.613$\pm$0.002(0.04)  &  0.389  &  0.381$\pm$0.003(0.04) & \citet{Monelli2012} \\
               Leo A  &  -1.40  &    7  &  71  &  29  &     3  &  0.30  &  0.625  &  0.637$\pm$0.014(0.04)  &  0.372  &  0.366$\pm$0.017(0.03) & \citet{Bernard2013} \\
              IC1613  &  -1.60  &   61  &  64  &  36  &    29  &  0.32  &  0.606  &  0.611$\pm$0.010(0.08)  &  0.349  &  0.339$\pm$0.006(0.03) & \citet{Bernard2010} \\
             NGC6822  &  -1.00  &   24  &  83  &  17  &     2  &  0.08  &  0.603  &  0.605$\pm$0.007(0.04)  &  0.406  &  0.388$\pm$0.019(0.03) & \citet{Baldacci2005} \\
\hline
\multicolumn{11}{c}{Sculptor Group dwarf galaxies} \\ 
\hline 
         ESO410-G005  &  -1.93  &  224  &  66  &  34  &    44  &  0.16  &  0.578  &  0.589$\pm$0.005(0.07)  &  0.327  &  0.317$\pm$0.010(0.06) & \citet{Yang2014} \\
         ESO294-G010  &  -1.48  &  219  &  62  &  38  &    13  &  0.06  &  0.589  &  0.593$\pm$0.004(0.06)  &  0.345  &  0.330$\pm$0.017(0.06) & \citet{Yang2014} \\

\hline
\hline
\end{tabular}
\end{scriptsize}
\begin{tablenotes}
\begin{scriptsize}
\item $^{*}$Mean metallicity for each galaxy obtained from \citet{McConnachie2012}.
\end{scriptsize}
\end{tablenotes}
\end{table*}

	\subsection{Comparing the HB morphologies of the MW and M31 satellites}
    
Pioneering works by \citet{DaCosta1996,DaCosta2000,DaCosta2002} based on 
shallower WFPC2 data disclosed the first hint that the M31 satellites are 
characterized by redder HB morphology with respect to MW dwarfs. A 
similar conclusion was reached by \citet{Martin2017}, based on ACS data 
for 20 M31 galaxies. The analysis was based on a morphological index 
accounting for the number of blue and red HB stars. In this section we 
apply a similar approach, and taking advantage of the known number
of RRL stars, we can compare the morphology index 
R$_{HB}$\footnote{R$_{HB}$=(B-R)/(B+V+R) where B and R are the numbers 
of HB stars bluer and redder than the IS, respectively, and V is the 
number of RRL stars \citep{Lee1990}.} of the six ISLAndS galaxies and of
a sample of MW satellites. The latter consists of revised data for 
Carina, Fornax, Sculptor, Draco, and Leo II from the updated 
catalogs available in P. B. Stetson's database (Stetson 2017, priv. comm.). 
These studies are part of an ongoing series of papers on variable stars 
in globular clusters and dwarf galaxies by ourselves and our collaborators
\citep{Stetson2014,Braga2015,Coppola2015,MartinezVazquez2016b,Braga2016,
Fiorentino2017}.

The value of the R$_{HB}$ index was calculated in a homogeneous way 
considering only stars within 1 half-light radius, r$_h$. This was possible 
for all of the galaxies except for And~I and II, since the ACS only covers 
a fraction of such area (see Figure~\ref{fig:spatial_dist_123}). In the case
of the MW satellites, we estimated and subtracted the Galactic field-star 
contribution using a proper control field in the outskirts of each object.
The exact limits in color and magnitude for the selection of HB stars for the
R$_{HB}$ index were defined on a per-galaxy basis because of the variety of CMD 
morphology, filter bandpasses, and foreground contamination. However, these were 
carefully chosen to limit contamination from any RGB, AGB, RC, and blue straggler 
populations present, while also avoiding biases.

Figure~\ref{fig:hbr} shows, as a function of the host galaxy absolute
M$_V$ magnitude, the R$_{HB}$ index calculated inside 1 r$_h$
for the MW (open diamonds) satellites and for the ISLAndS (stars) 
galaxies. And~I and II are calculated over the full ACS area, which is
smaller than 1 $r_h$.
The plot suggests that, at least in the innermost regions of 
the available samples, the M31 satellites have slightly redder HBs than the MW dSph 
satellites although the difference is within 2-$\sigma$. In fact, within 1 r$_h$ 
the mean value of R$_{HB}$ is more negative in the case of M31 galaxies 
(R$_{HB,M31}$ = --0.58$\pm$0.07) than for the MW companions 
(R$_{HB,MW}$ = --0.37$\pm$0.06). However, we emphasize that the latter numbers 
may be biased due to the small subsample of satellites for which we have data in 
both MW and M31 systems.

It is worth mentioning that the present analysis presents several improvements
when compared with previous ones \citep{Harbeck2001,Martin2017}:
{\em i)} the better photometric precision at the HB level, and the filter 
combination providing better color discriminating power, allows us to clearly 
separate the red HB from the blue edge of the RGB, even in the case of And~I; 
{\em ii)} the larger field of view of ACS compared to WFPC2 provided a larger 
sample; 
{\em iii)} the up-to-date, wide field, homogeneous data available for the MW 
companions allowed us to perform the comparison in a more homogeneous manner;
{\em iv)} the better phase coverage allowed us to derive better defined mean 
colors of RRL stars.

The current data do not allow us to fully explore whether the HB morphology
presents significant variation as a function of galactocentric distance, 
i.e., distance from the center of each galaxy.
Nevertheless, when considering the parallel WFC3 field for And~I and And~II,
we derive larger values of the R$_{HB}$ index, and therefore an indication that
the HB morphology gets bluer when moving to an external region. This is in agreement 
with what was found for other LG galaxies \citep[e.g.,][]{Harbeck2001,Tolstoy2004,Cole2017}, and more 
in general with the populations gradients commonly found in dwarf galaxies 
\citep[][and references therein]{Hidalgo2013}. In fact, 
when considering the area within 2 r$_h$, the six galaxies tend to have bluer HB.
Unfortunately, a straight comparison between the two satellite systems is 
complicated by the fraction of area covered. This leaves open the question of 
whether the HB morphology remains different at larger galactocentric distances, 
or whether M31 and MW satellites tend to be more similar when their global 
properties are taken into account. More wide-field variability studies, 
particularly for the M31 satellites, would help solve this problem.

	\subsection{Global properties of RRL stars}\label{sec:global_rrl}

In \S~\ref{sec:rrl} we presented the Bailey diagram of the ISLAndS galaxies and
discussed their properties in terms of Oosterhoff classification. Despite the 
intermediate mean-period, stars in the Bailey diagram still tend to clump 
around the Oo-type lines, with predominance of Oo-I-like stars, rather than 
in between. Therefore, the period distribution provides a more detailed 
description than the mean period alone \citep{Fiorentino2015a,Fiorentino2017}. 
In the previous subsection we have presented the evidence that M31 and MW satellites 
present slightly different HB morphology. We now focus on the properties of 
the RRL stars only. 

Table~\ref{tab:oo} lists the properties of the RRL in a sample of 41 dwarf 
galaxies (39 LG dwarfs + 2 Sculptor group dwarfs) of different morphological 
type within 2 Mpc (column 1): the number of RRab stars (column 2), the percent of 
Oo-I type and Oo-II type RRab stars (column 3 and 4), the number of RRcd stars 
(column 5), the fraction of RRcd stars over the total of the RRL (column 6) 
and the median and mean period of the RRab and RRcd stars (column 7, 8, 9, 
and 10) derived from the literature (references in column 11).

The left panel of Figure~\ref{fig:mvp} shows the mean period of RRab type stars, 
$<P_{ab}>$ as a function of the mean metallicity of the host galaxy (left panel), 
for 16 satellites of M31 (filled orange stars) and 15 MW dwarfs (blue open diamonds). 
Galaxies with at least 5 known RRab stars have been included. The plot discloses 
that the mean period of RRab type stars decreases for increasing mean metallicity 
of the host system \citep{Sandage1981}, for both the MW and the M31 satellites. The trend presents 
some scatter, but interestingly a linear fit to the data provides very similar slope 
(0.040$\pm$0.008 and 0.038$\pm$0.008, respectively), thus suggesting an overall 
similar behavior in the two satellite systems.

The decreasing mean period for increasing metallicity can be related to the early
chemical evolution of the sample galaxies. On the one hand, the distribution of 
stars in the Bailey diagram suggests that galaxies tend to progressively populate 
the RRab short period range for increasing metallicity (and mass). This translates 
into a smaller mean period.
It may appear intriguing that a property of a purely old stellar tracer correlates
with the {\em present-day} mean metallicity of the host galaxy. This suggests 
that galaxies that today are more massive and more metal-rich on average also 
experienced faster early chemical evolution, which is imprinted in the properties 
of their RRL stars. This implies that the mass-metallicity relation \citep[e.g.,][]{Kirby2013}
was in place at early epoch \citep{MartinezVazquez2016b}.

The central panel of Figure~\ref{fig:mvp} shows the mean period as a function 
of the fraction of Oo-I type stars, as defined in \S~\ref{sec:rrl}. While there 
is no clear correlation for either satellite system, we find that the vast 
majority of galaxies host a larger fraction of Oo-I type stars, between 60 
and 90\% of the total amount of RRab stars. Nevertheless, the mean period of 
fundamental pulsators would classify them as Oo-intermediate system. Again, 
this suggests that the RRL stars in complex systems such as galaxies are not 
properly represented by a single parameter.

Finally, the right panel of Figure~\ref{fig:mvp} shows the mean period distribution 
for the RRab in MW satellites (blue) and in M31 satellites (orange). 
Apparently, both of them are similar and their peaks agree within 1-$\sigma$. 

The former analysis reveals that, if we limit the comparison to strictly old and well
defined populations
such that of RRL stars, there are no obvious differences between the RRL populations
of the satellite systems of M31 and the MW. 

Figure~\ref{fig:histo_galgc} shows the behavior of $\langle P_{ab}\rangle$ versus [Fe/H], but 
comparing a sample of 41 galaxies (black circles, including MW and M31 satellites,
isolated dwarfs and two galaxies in the Sculptor group) with GCs 
(magenta bowtie symbols). We use here the compilation from 
\citet{Catelan2009}, including all the GCs with more than 10 RRL stars. 
Galactic GCs, as well as clusters from the LMC and the Fornax dSph galaxy 
are shown. The plot shows that a few Oo-intermediate clusters overlap with 
galaxies in the Oosterhoff gap, but most off the Oo-I clusters 
(i.e., with P$_{ab}<0.58$) occupy a region of the parameter space where 
no galaxies are present -- this holds even if we restrict the GC sample 
to those with 30 RRL or more. This is even more evident in the right panel of 
Figure~\ref{fig:histo_galgc}, which shows the mean period distributions of the two 
samples. It clearly shows that the peak for the galaxy distribution occurs at 
a period typical of Oo-intermediate systems, while the peak of the GCs occurs 
in the Oo-I regime. 


\section{Distance moduli}\label{sec:distances}

In the following, we use four independent methods to derive 
the distances to the six ISLAndS galaxies, the first three based on the
properties of the RRL stars: {\em i)} the reddening-free period-Wesenheit
relations \citep[PWR,][]{Marconi2015}; {\em ii)} the luminosity-metallicity (M$_V$ versus 
[Fe/H]) relation \citep[LMR,][]{Bono2003,Clementini2003}; {\em iii)}  the first overtone 
blue edge (FOBE) relation \citep{Caputo2000}; these are supplemented by {\em iv)} the 
tip of the RGB (TRGB) method.

All the aforementioned relations require an assumption for the metal abundance.
In particular, in the case of the PWR, LMR, and FOBE relation, we need to 
assume a metallicity corresponding to the old population (representative of 
the RRL stars). On the other hand, the TRGB method uses the 
metallicity of the RGB stars to obtain the expected mean color value of the TRGB. 
In complex systems like dwarf galaxies, the metallicity of the global population
may range over $\sim$2 dex, and in many systems a mix of old and intermediate-age
populations is present. However, the metallicity adopted for the methods based on RRL 
stars must be representative of the {\it old} stellar population. In the next section, 
we discuss in detail the choice of the metallicity in order to determine the distance 
to the six galaxies.

	\subsection{The choice of the metallicity}\label{sec:metallicity}

The metallicity estimates available in the literature for the ISLAndS galaxies are 
all based on CaT spectroscopy of bright RGB stars\footnote{In the case of And~II, 
\citet{Kalirai2010} estimate both a photometric and a spectroscopic
metallicity, concluding that with the data at hand the former is less dependent
on the low S/N of the measurements.}.  
As the RGB can be populated by stars of any age larger than $\sim$1 Gyr,
the derived metallicity distribution may not be representative of the RRL 
stars, since relatively young and/or more metal-rich populations may exist on the
RGB but may not have counterparts among RRL stars \citep{MartinezVazquez2016a}. 
As a consequence, assuming a mean metallicity that may be too high by 1.0 dex for 
the RRL stars would introduce a systematic error in the distance modulus estimates, 
at the level of $\sim$0.2 mag.

Table~\ref{tab:metallicities} lists literature values for the mean metallicity 
(column 2) of ISLAndS galaxies, the $\sigma$ of the metallicity distribution 
(column 3), and the number of RGB stars (column 4) used in these studies 
(references in column 5). Relatively low values were found for And~III, 
And~XV, And~XVI, and And~XXVIII, on average close to [Fe/H]$\sim$--1.8 or 
lower. On the other hand, in the case of And~I and And~II, different authors 
\citep{Kalirai2010,Ho2012} agree on a much higher mean metallicity 
([Fe/H]$\sim$--1.4), and a relatively large metallicity spread 
($\sigma_{AndI}$=0.37 dex, $\sigma_{AndII}$=0.72 dex). 
Nevertheless, the small number of HASP stars (see \S~\ref{sec:rrl}) 
suggests that, even if the tail of the RRL metallicity distribution 
reaches such relatively high values, the bulk of the RRL stars must 
have a lower metallicity ([Fe/H]$<$--1.5, \citealt{Fiorentino2015a}). 
Therefore, as representative values of the metallicity for the RRL 
population, we adopted --in agreement with their SFHs
\citep{Skillman2017}-- [Fe/H]=--1.8 for And~I and And~II while,
for the rest of the galaxies, we assume that the metallicity
of the old population must to be quite similar to that obtained by 
the spectroscopic studies (see column 8). 

The adopted mean metallicities for each ISLAndS galaxy are summarized in the 
second to last column of Table~\ref{tab:metallicities}. We note that the values
have been homogenized to the scale of \citet{Carretta2009}. Column 2
reports the value in the original scale, which is specified in column 6. 
In those cases based on theoretical spectra, we applied a correction to 
take into account the different solar iron abundance (from 
$\log\epsilon_{Fe}$ = 7.45 to 7.54), which translates to a distance 
modulus change between 0.01 mag in the case of the FOBE and 0.04 in 
the case of the LMR.

\begin{table*}
\renewcommand{\thetable}{\arabic{table}}
\centering
\caption{Metallicity studies with the largest samples of RGB stars.}
\label{tab:metallicities}
\begin{scriptsize}
\begin{tabular}{rcccrccc} 
\hline
\hline
& \multicolumn{6}{c}{RGB stars} & RRL stars \\
 \cmidrule(lr){2-7}  \cmidrule(lr){8-8} 
Galaxy & $\langle$[Fe/H]$\rangle$ & $\sigma_{\langle[Fe/H]}\rangle$ & N$_{stars}$ & References & Metallicity scale$^{a}$ & $\langle$[Fe/H]$\rangle_{C09}^{b}$ & [Fe/H]$_{old~pop.}$\\
\hline
\textbf{And~I}      &  --1.45$\pm$0.04 & 0.37        &  80 & \citet{Kalirai2010} & ZW84 & --1.44 & --1.8 \\
\textbf{And~II}     &  --1.39$\pm$0.03 & 0.72        & 477 & \citet{Ho2012}      & G07  & --1.30 & --1.8 \\
\textbf{And~III}    &  --1.78$\pm$0.04 & 0.27        &  43 & \citet{Kalirai2010} & ZW84 & --1.81 & --1.8 \\
\textbf{And~XV}     &  --1.80$\pm$0.20 & $-^{c}$     &  13 & \citet{Letarte2009} & C09  & --1.80 & --1.8 \\
\textbf{And~XVI}    &  --2.00$\pm$0.10 & $-^{d}$     &  12 & \citet{Collins2015} & G07  & --1.91 & --2.0 \\
\textbf{And~XXVIII} &  --1.84$\pm$0.15 & 0.65$^{e}$  &  13 & \citet{Slater2015}  & C09  & --1.84 & --1.8 \\
\hline
\hline
\end{tabular}
\end{scriptsize}
\begin{tablenotes}
\begin{scriptsize}
\item $^{a}$Metallicity scales: ZW84 = \citet{ZinnWest1984}, G07 = \citet{Grevesse2007}, and C09 = \citet{Carretta2009}. \\
\item $^{b}$ We have either converted the metallicity to the C09 scale when it was possible, 
or shifted the metallicity value assuming the same Solar iron abundance (log$\epsilon$(Fe)=7.54). 
The C09 scale was chosen as the homogeneous scale for being the most up-to-date.
\item $^{c}$ \citet{Letarte2009} did not publish $\sigma_{[Fe/H]}$. Instead they provide an 
interquartile range of 0.08, with a median metallicity of [Fe/H]=--1.58 dex. \\
\item $^{d}$ \citet{Collins2015} did not publish $\sigma_{[Fe/H]}$. 
However, \citet{Letarte2009} published for And~XVI an interquartile range 
of 0.12, with a median of [Fe/H]=--2.23 dex. By stacking the spectra
of the member stars (8 in this case), they found [Fe/H]=--2.1 
with an uncertainty of $\sim$0.2 dex. 
This value agrees with that obtained by \citet{Collins2015} (shown in the Table). \\
\item $^{e}$ As this $\sigma$ is obtained from a small number of individual
measurements, it may not be representative of the actual distribution. \\
\end{scriptsize}
\end{tablenotes}
\end{table*}


	\subsection{The period-Wesenheit relations}\label{sec:wese}

PWRs are a powerful tool for distance determination, because they are 
reddening-free by construction and are only marginally metallicity 
dependent. They are theoretically described by:

\begin{equation}
W(X, X-Y) = \alpha + \beta  \log P + \gamma [Fe/H]
\end{equation}

where X and Y are magnitudes and W($X$, $X$--$Y$) denotes the reddening free 
Wesenheit magnitude \citep{Madore1982} obtained as 
W($X$, $X$--$Y$)=$X$--R($X$--$Y$), where R is the ratio of 
total-to-selective absorption, R=A$_X$/E($X$-$Y$).

An updated and very detailed analysis of the framework of the PWRs is 
provided by \citet{Marconi2015}. Their Tables 7 and 8 give a broad range
of optical, optical-NIR, and NIR PWRs, along with their corresponding uncertainties.
In particular, in this work, we use their PWR in the ($I$, $B-I$) filter 
combinations\footnote{According to the equations obtained by 
\citet{Bernard2009} to transform $F475W$ and $F814W$ to Johnson-Cousins $BVI$, 
both $B$ and $V$ are transformed from $F475W$. For this reason we cannot apply 
the \textit{metal-independent} PWR ($V$, $B-V$) published by \citet{Marconi2015}, 
because $B$ and $V$ are correlated.}:

\begin{eqnarray}
W(I, B-I)=-0.97(\pm0.01) +\nonumber\\
(-2.40\pm0.02)\log P + (0.11\pm0.01) [Fe/H]
\end{eqnarray}

which has an intrinsic dispersion of $\sigma$ = 0.04 mag. For this relation, 
a metallicity change of 0.2 dex translates into a change in the distance of 
order 0.03 mag.

The theoretical W($I$, $B$--$I$) was obtained from the individual stars assuming 
a metallicity for the old population (see column 8 in Table~\ref{tab:metallicities}
and discussion of \S~\ref{sec:metallicity}). We next calculated the individual
apparent Wesenheit magnitude as: w($I$, $B$--$I$)=$I$--0.78($B$--$I$). 
We report the distance moduli obtained by averaging individual estimates for the 
global sample (RRab + fundamentalized RRc: log P$_{fund}$ = log P$_{RRc}$ + 0.127;
\citealt{Bono2001}) in columnn 2 of Table~\ref{tab:distance_summary}. For comparison, 
if we use independently the sample of RRab and RRc, the values from
the different determinations agree on average within $\pm$0.04 mag. 
Column 2 of Table~\ref{tab:distance_summary} reports the true distance moduli 
obtained for each galaxy using this method.

	\subsection{The luminosity-metallicity relation}\label{sec:mvfeh} 

The LMR is another simple, widely used approach to determine distances, 
in this case using the mean $V$ magnitude of RRL stars. Despite the 
fact that both theoretical and empirical calibrations suggest that the 
relation is not linear (being steeper in the more metal-rich regime 
(see e.g., \citealt[][and references therein]
{Caputo2000, SandageTammann2006, Cassisi2013}), most examples in the 
literature use one of the different linear relations proposed.

In the present work, we adopted the following relations:

\begin{equation}\label{eq:c03}
\langle M_{V} \rangle = 0.866(\pm 0.085) + 0.214(\pm0.047) [Fe/H]
\end{equation} 
from \citet{Clementini2003}, and
\begin{equation}\label{eq:b03}
\langle M_{V} \rangle = 0.768(\pm0.072) + 0.177(\pm0.069) [Fe/H],
\end{equation} 
from \citet{Bono2003}\footnote{The zero-point of this equation, as well as for 
the FOBE equation~\ref{eq:fobe} presented in next section, has been modified
according with the shift of +0.05 to correct for the electron-conduction 
opacities \citep{Cassisi2007}.}.

The latter is valid only for metallicity lower than [Fe/H]=--1.6, which is 
appropriate for the six ISLAndS galaxies (where the metallicity of the 
old population is considered to be [Fe/H]=--1.8 or lower). Columns 3 and 4 
of Table~\ref{tab:distance_summary} show the true distance moduli obtained using 
the two relations. They are in excellent agreement with each other and with 
those derived previously using the PWR.\\

	\subsection{The FOBE method}\label{sec:fobe}

Another method that can be used to estimate the distance is based on 
the predicted period-luminosity-metallicity relation (PLMR) for pulsators 
located along the FOBE of the IS \citep[see][]{Caputo2000}:

\begin{eqnarray}\label{eq:fobe}
M_{V,FOBE}= -0.635 -2.255 \log(P_{FOBE}) - \nonumber\\
1.259 \log(M/M_{\odot}) +0.058 \log(Z)
\end{eqnarray}
which has an intrinsic dispersion of $\sigma_V$ = 0.07 mag.

This is considered a particularly robust technique for stellar systems 
with significant numbers of first-overtone RRL (RRc) stars, especially 
if the blue side of the IS is well populated. Thus, it can be applied 
safely to five of our six galaxies\footnote{And~XVI only has 5 RRc stars} 
(see Figure~\ref{fig:bailey}).
The distance modulus is derived by matching the observed distribution of 
RRc stars to eq.~\ref{eq:fobe}. That is, for a given metallicity and a 
mass corresponding to the typical effective temperature for RRL stars, 
we shift the relation until the FOBE matches the observed distribution 
of RRc stars. 

For the adopted metallicity listed in Table~\ref{tab:metallicities}, and 
using the evolutionary models from BaSTI \citep{Pietrinferni2004}, we 
obtain masses at log T$_{\rm eff}\approx$3.86 of M $\sim$ 0.7 M$_{\odot}$.
True distance moduli obtained for each galaxy using this method are shown 
in column 5 of Table~\ref{tab:distance_summary}, and are in good agreement 
with those described in the previous section.

	\subsection{The tip of the RGB}\label{sec:trgb}

It is well established that the TRGB is a good standard candle thanks to its 
weak dependence on age \citep{Salaris2002} and, in the $I$ band in 
particular, on the metallicity \citep[at least for relatively metal-poor systems,][]{DaCosta1990,Lee1993}. 
The TRGB is frequently used to obtain reliable 
distance estimates to galaxies of all morphological types, in the LG 
and beyond \citep[e.g.,][]{Rizzi2007, Bellazzini2011, Wu2014}.
However, determining the cutoff in the luminosity function at the bright 
end of the RGB is not straightforward in low-mass systems, because more 
than about 100 stars populating the top magnitude of the RGB are required 
to reliably derive the location of the tip 
\citep{Madore1995, Bellazzini2002, Bellazzini2008}. This condition
is fulfilled only in And~I ($>$200), And~II ($>$150), and nearly in 
And~III ($\sim$90). The low number of stars in the other three galaxies 
prevents us from deriving a reliable measurement of the apparent magnitude 
of the TRGB.

We applied the same method from \citet{Bernard2013} to determine the magnitude 
of the TRGB.  We convolved the $F814W$ luminosity functions with a Sobel kernel 
of the form [−1,−2,0,2,1]. From the filter response function, we obtain 
the center of the peak corresponding to the TRGB of each galaxy: 
$F814W_{0, AndI}$=20.45$\pm$0.09, $F814W_{0, AndI}$=20.05$\pm$0.12, 
and $F814W_{0, AndIII}$=20.25$\pm$0.19 mag, where the uncertainty is the 
Gaussian \textit{rms} width of the peak of the Sobel filter response.

The distances were obtained from the TRGB magnitudes using three calibrations:

\textit{i)} the empirical calibrations in the HST flight bands from \citet[][R07]{Rizzi2007}:
\begin{eqnarray}
M^{F814W}_{TRGB}=-4.06+0.15[(F555W-F814W)_0 \nonumber\\
-1.74] ~(\sigma=0.10)
\end{eqnarray}

\textit{ii)} the empirical calibration reported in \citet[][B11]{Bellazzini2011}, derived by 
\citet{Bellazzini2008} from the original calibration as a function of [Fe/H] obtained in 
\citet{Bellazzini2001} and revised in \citet{Bellazzini2004}:
\begin{eqnarray}
M^{F814W}_{TRGB} \approx M^I_{TRGB}=0.080(V-I)_0^2 \nonumber\\
- 0.194(V-I)_0 -3.93 ~(\sigma=0.12)
\end{eqnarray}

\textit{iii)} the theoretical calibration M$^{F814W}_{TRGB}$, as a function of the color 
($F475W$--$F814W$)$_0$, obtained in this work by fitting the BaSTI predictions 
\citet{Pietrinferni2004,Pietrinferni2006} for the TRGB brightness for 
an old ($\sim$12 Gyr) stellar population, a wide metallicity range and 
an alpha-enhanced heavy element distribution\footnote{We note that the zero-point of 
this theoretical calibration has been corrected in order to account for the impact 
on the TRGB brightness of more accurate conductive opacity evaluations. Following 
the results obtained by \citet{Cassisi2007} we have corrected the 
M$^{F814W}_{TRGB}$, by adding +0.08 mag.}: 

\begin{eqnarray}
M^{F814W}_{TRGB} = -4.11 + 0.07 [(F475W-F814W)_0-2.5)  \nonumber\\
+ 0.09[(F475-F814)_0-2.5]^2 ~(\sigma=0.02)~~~~
\end{eqnarray} \\

In the case of the \citet{Rizzi2007} calibration, we considered that 
$F555W$--$F814W$ $\sim$ $V$--$I$\footnote{We do not have $F555W$ 
magnitudes for the ISLAndS dSphs, but the ($F555W$--$F814W$) color 
is very close to ($V$--$I$)}. In fact, for both this calibration
and that of \citet{Bellazzini2011}, we use the following equation
to determine the expected ($V$--$I$)$_0$ color: 
($V$--$I$)$_{TRGB,0}$=0.581[Fe/H]$^2$+2.472[Fe/H]+4.013 
\citep{Bellazzini2001}. Since this last equation is based on 
\citet[ZW84]{ZinnWest1984} scale, in order to use it properly, we 
have to apply the conversion scales provided by 
\citet{Carretta2009}: [Fe/H]$_{ZW84}$=([Fe/H]$_{C09}$--0.160)/1.105. 
Columns 6, 7, and 8 in Table~\ref{tab:distance_summary} give the values 
of the true distance moduli calculated using the previous relations 
for And~I, And~II, and And~III. All three calibrations lead to distances 
that are in good agreement with each other and with the previously 
calculated RRL based distances. 

\begin{table*}
\renewcommand{\thetable}{\arabic{table}}
\centering
\caption{Summary of the different true distance moduli ($\mu_0$) obtained using several methods.}
\label{tab:distance_summary}
\begin{scriptsize}
\begin{tabular}{rccccccc} 
\hline
\hline
& \multicolumn{4}{c}{RRL stars} & \multicolumn{3}{c}{RGB stars} \\
 \cmidrule(lr){2-5}  \cmidrule(lr){6-8} 
Galaxy              &   PWR                  &     LMR$_{B03}$    &     LMR$_{C03}$    &     FOBE$^{*}$         & Tip$_{R07}$             &   Tip$_{B11}$           &   Tip$_{BaSTI}$ \\
\hline 
\textbf{And~I}      &  24.49$\pm$0.08(0.11)  &  24.54$\pm$0.16(0.10)  &  24.50$\pm$0.14(0.10)  &  24.49$\pm$0.10  &  24.52$\pm$0.11(0.09)  & 24.49$\pm$0.12(0.09)  &  24.56$\pm$0.04(0.09) \\
\textbf{And~II}     &  24.16$\pm$0.08(0.10)  &  24.15$\pm$0.16(0.09)  &  24.11$\pm$0.14(0.09)  &  23.95$\pm$0.10  &  24.10$\pm$0.11(0.12)  & 24.08$\pm$0.12(0.12)  &  24.17$\pm$0.04(0.12) \\
\textbf{And~III}    &  24.36$\pm$0.08(0.08)  &  24.44$\pm$0.16(0.08)  &  24.41$\pm$0.14(0.08)  &  24.48$\pm$0.10  &  24.35$\pm$0.11(0.19)  & 24.30$\pm$0.12(0.19)  &  24.37$\pm$0.04(0.19) \\
\textbf{And~XV}     &  24.42$\pm$0.08(0.09)  &  24.50$\pm$0.16(0.07)  &  24.47$\pm$0.14(0.07)  &  24.45$\pm$0.10  &  ---  & --- & --- \\
\textbf{And~XVI}    &  23.70$\pm$0.08(0.09)  &  23.73$\pm$0.16(0.07)  &  23.70$\pm$0.14(0.07)  &  23.75$\pm$0.10  &  ---  & --- & --- \\
\textbf{And~XXVIII} &  24.43$\pm$0.08(0.07)  &  24.45$\pm$0.16(0.08)  &  24.42$\pm$0.14(0.08)  &  24.41$\pm$0.10  &  ---  & --- & --- \\
\hline
\hline
\end{tabular}
\end{scriptsize}
\begin{tablenotes}
\begin{scriptsize}
\item Notes.- Systematic uncertainties of each estimation are given without parenthesis, while the random 
uncertainties are bracketed.
\item The uncertainties include the contribution from a possible metallicity dispersion of 0.3 dex.
\item $^{*}$ FOBE method is based in only one RRc star, for this reason they do not have standard deviation.
\end{scriptsize}
\end{tablenotes}
\end{table*}
    \subsection{On the consistency of the different methods}\label{sec:consistency}

As show in Table~\ref{tab:distance_summary}, all the distances obtained
from the different methods are in agreement within less than 1-$\sigma$ with 
each other. The inclusion of the TRGB method in this study was mainly for
checking the distances we obtained using the properties of the RRL stars
with those assessed with this method. In fact, it is worth mentioning that
since the TRGBs of these galaxies are not densely populated (we have $\gtrsim$ 200 
only for And~I), this technique is secondary in our study, but 
it serves to show that the metallicity we have assumed for the old population 
is robust. The good sampling of our LCs together with the large amount of RRL 
stars in all these galaxies (with exception of And~XVI), make them the best 
distance indicators that we have in these galaxies so far.

We adopt the distances obtained with the PWR as preferred 
because: \textit{i)} they are obtained with the RRL stars, \textit{ii)} the
PWR used for deriving them come from the most updated study 
\citep{Marconi2015}, and \textit{iii)} the systematic uncertainties are the 
smallest (see Table~\ref{tab:distance_summary}).

\begin{figure}
    \hspace{-0.5cm}
	\includegraphics[scale=0.47]{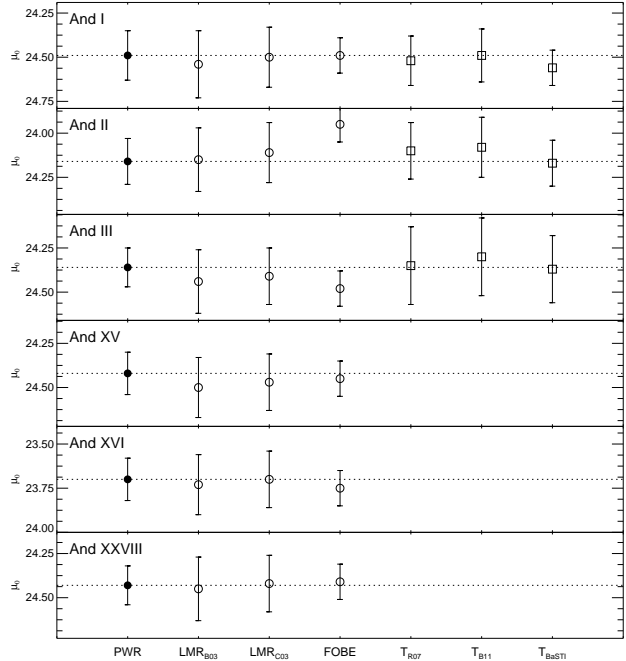}
	\caption{Summary of our derived distances. Circles report values based on 
    RRL stars while squares are based on the TRGB (provided only for the most 
    massive galaxies, for which the TRGB could be reliably estimated). The 
    filled circles and the dotted lines show the measurements based on the 
    PWR, which are the final adopted distances. Open symbols show 
    values obtained with the other methods, for comparison.}
	\label{fig:dist_1}
\end{figure}

\begin{figure}
    \hspace{-0.5cm}
	\includegraphics[scale=0.47]{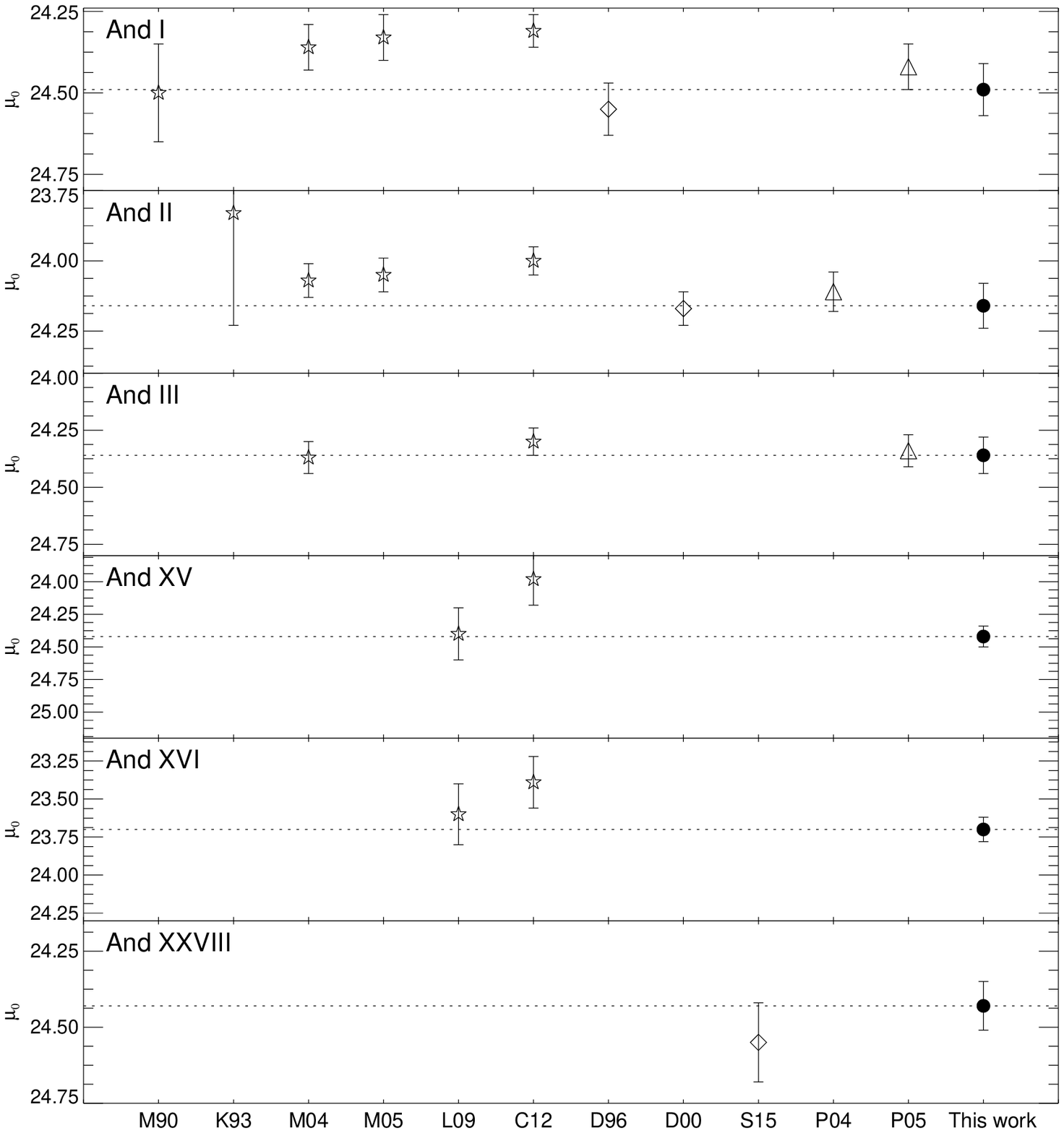}
	\caption{Comparison of our adopted distance moduli (based on the PWR, filled 
    circles and dotted lines) with the literature values (open symbols).
    In particular, we report values based on the TRGB
    \citep[stars:][]{Mould1990,Koenig1993,McConnachie2004,McConnachie2005,Letarte2009,Conn2012},
	HB luminosity \citep[diamond:][]{DaCosta1996, DaCosta2000, Slater2015}, and 
	RRL stars \citep[triangles:][]{Pritzl2004,Pritzl2005}.}
	\label{fig:dist_2}
\end{figure}

Figure~\ref{fig:dist_1} summarizes the distance determinations derived in this work. 
In particular, the filled circles together with the dotted line show the adopted 
final distance measurement coming from the PWR (\S~\ref{sec:wese}).
Open circles show the results from the RRL-based methods presented in previous 
sections (see Table~\ref{tab:distance_summary}), while the open squares show the 
TRGB distances. The plot shows that the agreement between the different methods 
presented here is remarkably good, as most of the derived distances agree within 
1$\sigma$. Taking as reference the PWR distance, some general trends can be noted 
between the results of the different methods adopted. The distance derived using 
the LMR with the \citet{Bono2003} calibration provides marginally larger distances 
with respect to both the \citet{Clementini2003} calibration (in agreement with the 
difference in the zero point), and also with the distance obtained from the PWR. 
The FOBE distance is larger than the PWR distance in three cases (And~III, XV, and
XVI), and shorter for And~II. Nevertheless, this method is the most sensitive to 
the sampling of the IS, and in particular the lack of RRL close to the blue edge 
of the IS introduces a bias toward larger distances. The TRGB technique could 
only be applied to the three most massive systems. Interestingly, in the case 
of And~II and And~III the derived distance seems to be, on average, marginally 
smaller, independent of the calibration adopted. We note that in the case of 
And~I the agreement between different indicators and methods is remarkably good. 
This is possibly linked to the fact that it presents the largest sample of RRL stars 
and the most populated TRGB region, thus suggesting that statistical fluctuations 
have a minimal effect.

	\subsection{Comparison with previous works}\label{sec:previous_distance}

Figure~\ref{fig:dist_2} displays a comparison with distance estimates
available in the literature and derived with different techniques:
RRL stars \citep[open triangles:][]{Pritzl2004,Pritzl2005}, the HB luminosity 
\citep[open diamonds:][]{DaCosta1996, DaCosta2000, Slater2015}, and the TRGB 
\citep[open stars:][]{Mould1990, Koenig1993,McConnachie2004,McConnachie2005,Letarte2009,Conn2012}. 
This figure shows an overall good agreement with our estimates, within the uncertainties. 
We note that the TRGB tends to provide closer distances than the 
RRL and the HB luminosity, though it is still compatible within 1.5-$\sigma$.
A couple of discrepant cases (And~XV and XVI from \citealt{Conn2012}) can be 
ascribed to the sparsely populated bright portion of the RGB in these galaxies.


\section{Other variables}\label{sec:other}

\subsection{Anomalous Cepheid stars} \label{sec:ac}

AC stars are variables stars in the core He-burning 
evolutionary phase at luminosity higher than RRL stars. Their periods 
range from $\sim$0.5 to $\sim$2.0 days and their masses are thought to 
be higher than 1 M$_{\odot}$. In order to have ACs with such masses, two
different channels are likely \citep{Bono1997a,Gallart2004,Cassisi2013}. 
They can be the progeny of coalesced binary stars, thus evolved blue 
straggler stars (BSS) tracing the old population 
\citep{Renzini1977,Hirshfeld1980,Sills2009}. Alternatively, 
they can be an evolutionary stage of single stars with mass between 
$\sim$1.2 and $\sim$2.2 M$_\odot$ and age between 1 and 6 Gyr 
\citep{Demarque1975,Norris1975,Castellani1995,Caputo1999,Dolphin2002,Fiorentino2006}. 
In both scenarios, they trace the existence of a metal poor (Z$<$0.0006) 
stellar population thus providing sound constraints to the metal 
enrichment of the galaxy.

Typically, purely-old (age$\ga$10 Gyr) nearby LG dwarf galaxies host a
few of them. However, they are very rare in GCs; so far, 
only one candidate has been confirmed in the metal-poor ([Fe/H] 
$\sim$ --2 dex) cluster NGC 5466 \citep{Zinn1976}, and a few others have been 
suggested \citep{Corwin1999,ArellanoFerro2008,Kuehn2011,Walker2017}. 
On the other hand, large samples of ACs have been collected in the LMC (141) and in the SMC (109) in the 
framework of the OGLE-IV project \citep{Soszynski2015}. In this context, 
it is worth mentioning that the work of \citet{Mateo1995} --updated by 
\citet{Fiorentino2012b} collecting data from nearby dwarf galaxies, for 
which SFH is provided and ACs have been found (see their Figure 7)-- 
noted a correlation between the frequency of ACs and the total luminosity 
of the host galaxy; the frequency of ACs decreases for increasing luminosity 
of the host galaxy. Another parameter that seems to impact the ACs frequency
is the SFH of the host system, with primarily old systems, or \emph{fast galaxies} 
\citep[as defined by][]{Gallart2015} having a lower specific frequency of ACs 
compared to systems containing an important amount of intermediate-age and 
young populations (\emph{slow galaxies}).

Figure~\ref{fig:cmds_acs} shows the presence of a few variable stars located 
1 to 2 mag above the HB. Given their pulsation properties, their position in 
the period-Wesenheit diagram (see e.g., Figure~1 in \citealt{Fiorentino2012b}), 
and the shape of their light curves, we classify them 
as ACs. In particular, a total of 15 ACs have been detected in the ACS field 
of And~II, III, XV,and XXVIII. No ACs have been detected in And~I, in 
agreement with \citet{Pritzl2005}, nor in And~XVI. The lack of ACs in 
And~I can be a hint of the fast chemical enrichment of this galaxy. This is 
supported by the high metallicity of this galaxy when compared with the 
remaining ISLAndS galaxies. On the other hand, And~XVI has the proper mean
low metallicity, but its low total mass is the most probable culprit for 
the lack of ACs. Interestingly, And~XVI shows active star formation until
about 6 Gyr ago, thus not extended enough to produce AC through the single 
star channel.

As none of the four ISLAndS galaxies where ACs have been detected shows 
evidence of star formation younger than 2 Gyr \citep{Skillman2017}, it is 
unlikely that ACs come from the evolution of a young metal-poor star.
The sequence of blue objects between the oldest MSTO and the HB are 
most likely blue straggler stars (BSS) descending from primordial 
binary stars of the old population in agreement with the SFHs 
obtained by \citet{Skillman2017}. Therefore we assume that 
the AC, here detected, are the progeny of coalesced binary stars 
(evolved BSS), thus tracing the old population 
\citep{Renzini1977,Hirshfeld1980,Sills2009}.

\begin{figure*}
	\includegraphics[width=1.0\textwidth]{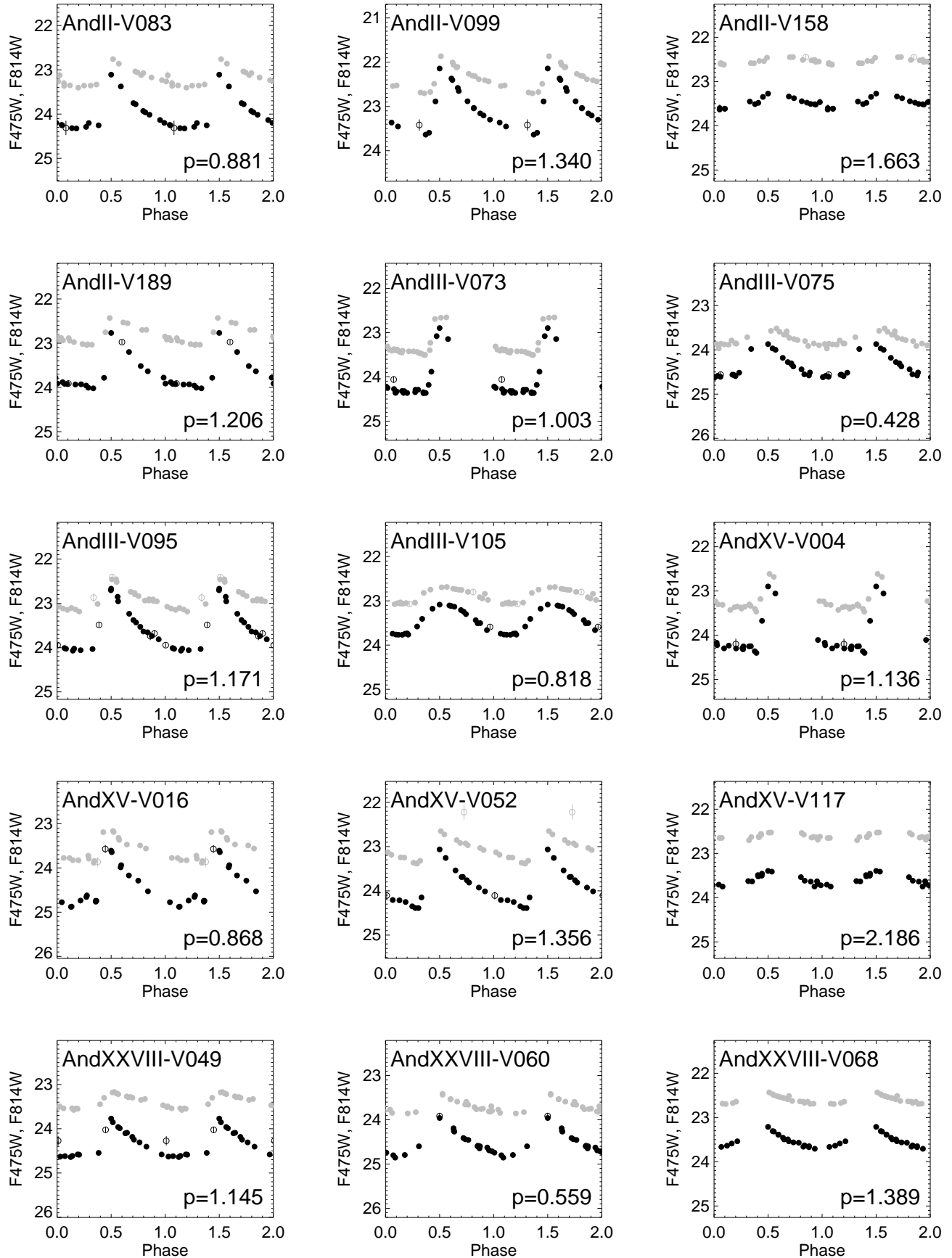}
	\caption{Light curves of member AC stars for four of the six ISLAndS 
    galaxies in the $F475W$ (black) and $F814W$ (gray) bands. Periods 
    (in days) are given in the lower-right corner, while the name of the 
    variable is displayed at the top of each panel. Open symbols show the 
    data for which the uncertainties are larger than 3-$\sigma$ above the
    mean error of a given star; these data were not used in periods and 
    mean magnitudes calculations.}
	\label{fig:ac_lcv}
\end{figure*}

From our sample, only four ACs were already known\footnote{Contrary to 
\citet[][theirV9]{Pritzl2005}, variable AndIII-V100 was classified as a 
RRL star as its location on the CMD is not compatible with an AC.}: 
AndII-V083  (V14 by \citealt{Pritzl2004}), AndIII-V073, AndIII-V075, 
and AndIII-V105 (V01, V07, and V06 by \citealt{Pritzl2005}). 
The LCs of all the  ACs detected are shown in Figure~\ref{fig:ac_lcv}. 
The different shapes are an indication of  different pulsational
modes. However, the classification of the pulsation mode of ACs is not
trivial and cannot be easily determined from the morphology of the 
LCs alone \citep{Marconi2004}.

\begin{figure}
    \hspace{-1.1cm}
    \includegraphics[scale=0.57]{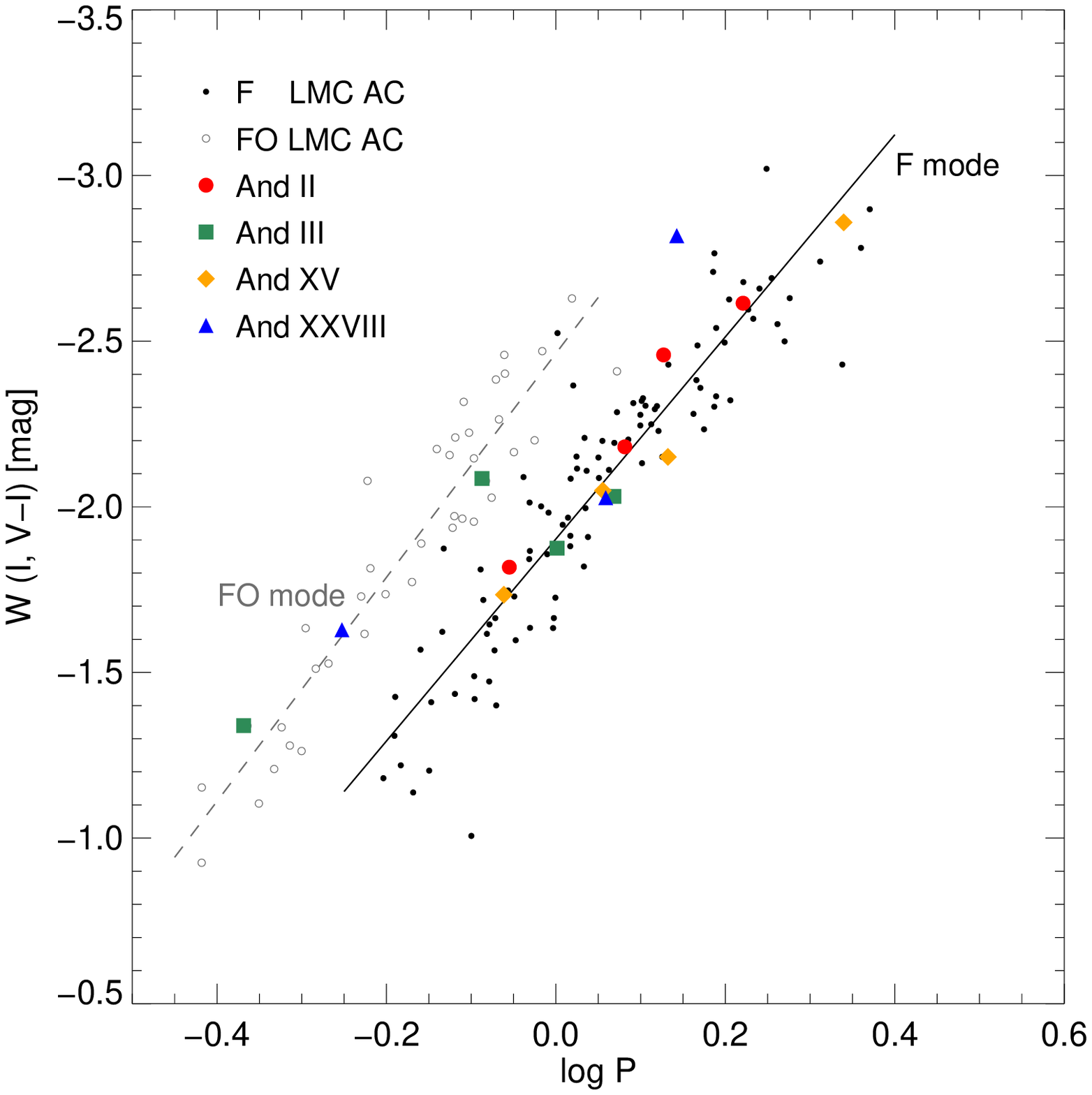}
    \caption{Period-Wesenheit diagram for ACs. 
    Black dots (F pulsators) and grey circles (FO pulsators) represent 
    the ACs of the LMC from the OGLE-IV release \citep{Soszynski2015}. 
    ACs discovered in our galaxies are represented by red circles (And~II), 
    green squares (And~III), orange diamonds (And~XV), and blue triangles 
    (And~XXVIII). The solid and dashed lines are the empirical 
    period-luminosity relations obtained by \citet{Soszynski2015} for ACs 
    in the LMC for the F and FO mode, respectively.}
    \label{fig:wes_ac}
\end{figure}

Figure~\ref{fig:wes_ac} shows the period-Wesenheit plane for the 
reddening-free index W($I$, $B$-$I$) --which reduces the scatter 
due to the interstellar  reddening and the intrinsic width of the 
IS -- for the ACs of the LMC (open symbols) published by the OGLE 
collaboration \citep{Soszynski2015}. This plot shows a clear 
separation between fundamental (F, black dots) and
first-overtone (FO, open circles) pulsators for ACs. 
Therefore, ACs are defined by different PL relations, and fundamental 
and first overtone pulsation can also be distinguished in this way. 
We therefore overplot the 15 ACs found in this work with the 
aim of checking  their nature and identifying their pulsation modes. 
The four ACs found in And~II are represented by red circles, the
four in And~III by green squares, the four in And~XV by orange diamonds 
and the three in And~XXVIII by blue triangles. First, Figure~\ref{fig:wes_ac} 
supports their classification as ACs.  We note that only one star 
(AndXXVIII-V068) is somewhat distant from the bulk of the F mode ACs of the 
LMC. From an inspection of the light curve of this star (see Figure~\ref{fig:ac_lcv}), 
the lack of phase points close to the maximum light is evident. 
Therefore, the measurement of the mean magnitude of this star may be biased 
to brighter magnitude, as the fit with templates tends to overestimate the 
amplitude. Figure~\ref{fig:wes_ac} indicates that the 
majority of the ACs (12) are pulsating in the F mode, and only three of 
them (AndIII-V075, AndIII-V105 and AndXXVIII-V060) are FO pulsators.

\subsection{Eclipsing binaries candidates}\label{sec:eb}

\begin{figure}
	\hspace{-0.4cm}
	\includegraphics[width=9cm]{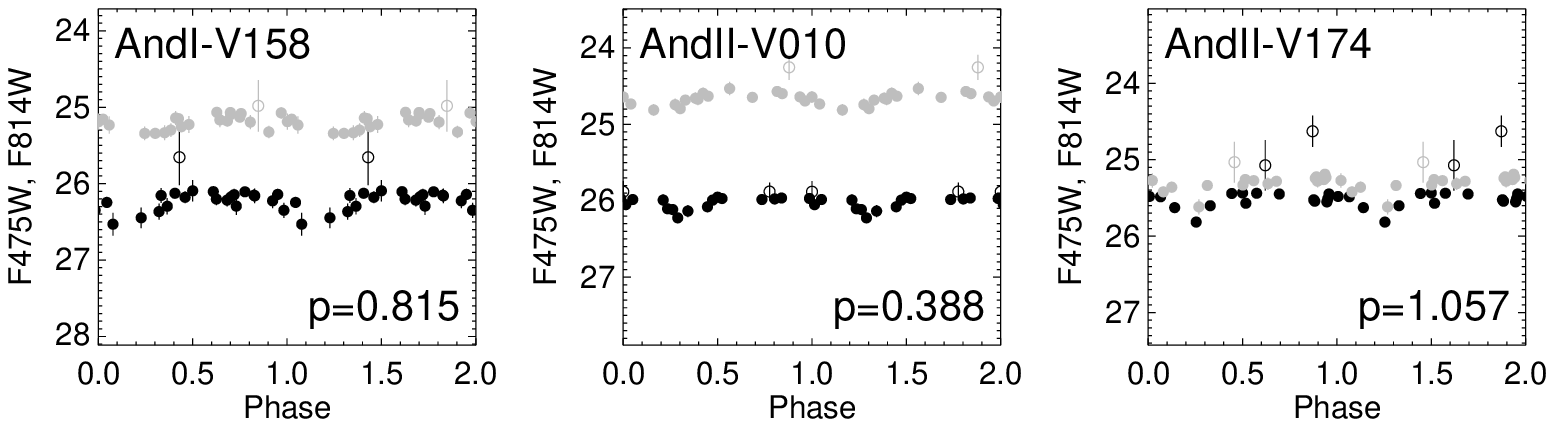}
	\caption{Light curves of the three EB candidates detected 
    in the field of And~I and And~II in the $F475W$ (black) and $F814W$ 
    (gray) bands. Periods (in days) are given in the lower-right corner, 
    while the name of the variable is displayed at the top of each panel. 
    Open symbols show the data for which the uncertainties are larger 
    than 3-$\sigma$ above the mean error of a given star; these data
	were not used in periods and mean magnitudes calculations.}
	\label{fig:eb_lcv}
\end{figure}

For the sake of completeness, we report the detection of three EBs, 
one in And~I and two in And~II. Figure~\ref{fig:eb_lcv} shows their 
LCs, which in all cases show a minimum. For the three candidates 
the minimum occurs at the same phase in the two bands. This feature, 
together with the flat bright part of the light curves, the periods, 
and their position in the CMD support the classification as EBs.

The fact that only such a small number of candidate EBs was detected is 
due to both the relatively small number of points per light curve taken, 
the non-optimal time sampling for this kind of variable, and the limited 
region of the CMD that was searched for variables.\\

\section{Summary and Final Remarks}\label{sec:conclusions}

In this paper we have analyzed multi-epoch HST data for six dSphs 
satellites of M31 in order to study their population of variable 
stars. The main findings of the current study are: 

$\bullet$ We have detected 895 variable stars in And~I, II, III, 
XV, XVI, and XXVIII: 678 of them are new discoveries. 
In particular, we classified 870 RRL stars, 15 ACs, 3 EB and 7 
field variable stars (5 of them probably belonging to the GSS of 
M31). Interestingly, no ACs were found in And~I despite being the 
second most massive dwarf in our sample, which we interpret as a 
hint of the fast chemical enrichment of  this galaxy. \\
$\bullet$ Pulsational properties (period, amplitude, mean magnitude) 
were derived for all detected variables. Moreover, we provide all 
the light curves and time series photometry. \\
$\bullet$ Using the properties of RRL stars, we derived new homogeneous 
distances to the six galaxies using three different methods: the 
period-Wesenheit relation, the metallicity-luminosity relation and the 
first overtone blue edge method. A fourth independent estimate was 
derived using the tip of the RGB for the three most populated systems. 
We find a satisfactory agreement both between different methods and 
with most of the estimates available in the literature. It is worth noting
that those values obtained using the RRL stars are more accurate and precise. 
For these reasons, we adopted as final distance moduli those which are obtained 
through the period-Wesenheit relation, which are the most precise values and 
based in the most updated relation for RRLs to date. \\
$\bullet$ We have shown that, similar to MW satellites, the mean
period of RRab variables of the six ISLAndS is close to 0.6 day, a value that
is typical of Oo-intermediate objects. On the other hand, the distribution of
RRL stars in the Bailey diagram is such that the majority of stars ($\sim$80\%)
are distributed close to the locus of Oo-I type. And~XXVIII appears to be a peculiar
object, both because the RRab stars do not follow any Oosterhoff relation in
particular, and because the fraction of RRc type stars is the largest among 
nearby galaxies.\\
$\bullet$ In spite of the slight difference in the HB morphology parameter
(R$_{HB}$), when we restrict the comparison between M31 and MW systems to the 
properties of RRL stars only, we do not find significant differences between 
the two groups of galaxies. In particular, based on a sample of 16 satellites 
of M31 and 15 of MW, we find a similar trend between the mean period and the 
mean metallicity. This suggests overall similar characteristics of the oldest 
($>$ 10 Gyr) population in the two systems in agreement with what is discussed 
by \citet{Monelli2017} using the global period distributions of 
thousands of RRL stars belonging to faint and bright satellites of M31 and the MW.

To date, none of the known Local Group dwarf galaxies has a complete census 
of their entire population of variable stars.  However, we are at the dawn of 
a new era for variability studies. Current and future surveys are about to 
bring an unprecedented amount of information on the variable stars 
populating the surroundings of the MW and of the Local Group. \textit{Gaia} 
will bring the discovery of thousands of new RRL stars (G$\la$20.7 mag) in 
the MW Halo (\citealt{Clementini2016}, DR1) and, in particular, will help us 
to complete the census of RRL stars in some MW satellites \citep{Antoja2015}. 
Additionally, the LSST will produce a flood of data that will enable the
discovery of a highly complete sample of RRL stars out to hundreds of kpc.

Finally, regarding RRL stars in the M31 satellites, the large number 
of dwarf galaxies discovered in this system during the last few years 
remain largely unexplored. The advent of the wide-field imaging capability 
of telescopes, such as WFIRST (with a field of view roughly 100 times greater 
than that of HST), could substantially accelerate these studies and offer us the 
possibility to understand the similarities and differences between the two 
systems of the Local Group: MW and M31 satellites.

\acknowledgments

The authors thank the anonymous referee for the useful comments
which helped to improve the manuscript.
This research made extensive use of NASA's Astrophysics Data System 
Bibliographic Services and the NASA/IPAC Extragalactic Database (NED), 
which is operated by the Jet Propulsion Laboratory, California Institute 
of Technology, under contract with the National Aeronautics and Space 
Administration.
Support for this work was provided by NASA through grants GO-13028 and 
GO-13749 from the Space Telescope Science Institute, which is operated 
by AURA, Inc., under NASA contract NAS5-26555.
This work has also been supported by the Spanish Ministry of Economy and 
Competitiveness (MINECO) under the grant (project reference AYA2014-56795-P).
EJB acknowledges support from the CNES postdoctoral fellowship program.
GF has been supported by the Futuro in Ricerca 2013 (grant RBFR13J716).
Support for D.R.W. is provided by NASA through Hubble Fellowship grants 
HST-HF-51331.01 awarded by the Space Telescope Science Institute. 



\vspace{5mm}
\facilities{HST(ACS, WFC3)}

\software{IDL, DAOPHOT/ALLFRAME}



\clearpage
\appendix

\section{Observing logs for ISLAnds galaxies}\label{sec:observing_logs}

This work is based on observations obtained with the ACS and WFC3 onboard the HST. 
These data were collected in different runs for each galaxy over about 2 and 5.3 
consecutive days between 2013 October 4 to 2015 September 6 as part of a
large HST proposal (GO-13028 and GO-13749, P.I.: E. Skillman).
The observing sequence consisted of alternating $\sim$1100 s exposures 
in F475W and F814W for an optimal sampling of the light curves. 
The complete observing logs for And~I, II, III, XV, XVI, and XXVIII are given in 
Tables~\ref{tab:observing_log_andi}, \ref{tab:observing_log_andii},
\ref{tab:observing_log_andiii}, \ref{tab:observing_log_andxv},
\ref{tab:observing_log_andxvi}, and \ref{tab:observing_log_andxxviii}, 
respectively. These tables collect the name of the image (column 1) as it appears in 
the HST archive(\url{http://archive.stsci.edu/}), the date (column 2) and the
UT start of each exposure (column 3) the filter used (column 4), and the 
exposure time (column 5).

\begin{table*}
\centering
\caption{Observing log for And~I.} 
\label{tab:observing_log_andi}
\begin{tabular}{ccccc} 
\hline
Image Name & Date         & UT Start      & Filter & Exp. time \\
  name     &  (YY-MM-DD)  & (hh:mm:ss)    &  name  &    (s)    \\ 
\hline
\multicolumn{5}{c}{\emph{ACS (RA=00:45:42.8, Dec=+38:02:22.8)}} \\
\hline
jcnb01lyq\_flc.fits       & 2015-09-01 & 09:56:05 & F475W  & 1264 \\ 
jcnb01m1q\_flc.fits       & 2015-09-01 & 10:20:06 & F814W  & 1002 \\ 
jcnb01m3q\_flc.fits       & 2015-09-01 & 11:24:32 & F814W  & 1086 \\ 
jcnb01m7q\_flc.fits       & 2015-09-01 & 11:45:36 & F475W  & 1372 \\ 
jcnb02mtq\_flc.fits       & 2015-09-01 & 16:17:43 & F475W  & 1264 \\ 
... & ... & ... & ... & ...  \\
\hline
\multicolumn{5}{c}{\emph{WFC3 (RA=00:45:14.2, Dec=+38:00:03.0)}} \\
\hline
icnb01lzq\_flc.fits       & 2015-09-01 & 09:55:10 & F475W  &1308 \\
icnb01m0q\_flc.fits       & 2015-09-01 & 10:19:26 & F814W  &1046 \\
icnb01m4q\_flc.fits       & 2015-09-01 & 11:24:27 & F814W  &1103 \\
icnb01m8q\_flc.fits       & 2015-09-01 & 11:45:24 & F475W  &1389 \\
icnb02muq\_flc.fits       & 2015-09-01 & 16:16:48 & F475W  &1308 \\
... & ... & ... & ... & ...  \\
\hline
\end{tabular}
\begin{tablenotes}
\begin{scriptsize}
\item This table is a portion of its entirely form which will be available in the online journal.
\end{scriptsize}
\end{tablenotes}
\end{table*}
\begin{table*}
\centering
\caption{Observing log for And~II.} 
\label{tab:observing_log_andii}
\begin{tabular}{ccccc} 
\hline
Image Name & Date         & UT Start      & Filter & Exp. time \\
  name     &  (YY-MM-DD)  & (hh:mm:ss)    &  name  &    (s)    \\ 
\hline
\multicolumn{5}{c}{\emph{ACS (RA=01:16:23.8, Dec=+33:26:05.5)}} \\
\hline
jc1d01wfq\_flc.fits       & 2013-10-04 & 03:50:09 & F475W  &	  1280 \\ 
jc1d01whq\_flc.fits       & 2013-10-04 & 04:14:26 & F814W  &	   987 \\ 
jc1d01x5q\_flc.fits       & 2013-10-04 & 05:21:50 & F814W  &	  1100 \\ 
jc1d01x9q\_flc.fits       & 2013-10-04 & 05:43:08 & F475W  &	  1359 \\ 
jc1d02ycq\_flc.fits       & 2013-10-04 & 10:12:02 & F475W  &	  1280 \\ 
... & ... & ... & ... & ...  \\
\hline
\multicolumn{5}{c}{\emph{WFC3 (RA=01:16:04.4, Dec=+33:21:31.7)}} \\
\hline
ic1d01wgq\_flc.fits       & 2013-10-04 & 03:49:14 & F475W  &1350 \\
ic1d01wiq\_flc.fits       & 2013-10-04 & 04:14:12 & F814W  &1122 \\
ic1d01x6q\_flc.fits       & 2013-10-04 & 05:21:45 & F814W  &1200 \\
ic1d01xbq\_flc.fits       & 2013-10-04 & 05:44:19 & F475W  &1409 \\
ic1d02ydq\_flc.fits       & 2013-10-04 & 10:11:07 & F475W  &1350 \\
... & ... & ... & ... & ...  \\
\hline
\end{tabular}
\begin{tablenotes}
\begin{scriptsize}
\item This table is a portion of its entirely form which will be available in the online journal.
\end{scriptsize}
\end{tablenotes}
\end{table*}
\begin{table*}
\centering
\caption{Observing log for And~III.} 
\label{tab:observing_log_andiii}
\begin{tabular}{ccccc} 
\hline
Image Name & Date         & UT Start      & Filter & Exp. time \\
  name     &  (YY-MM-DD)  & (hh:mm:ss)    &  name  &    (s)    \\ 
\hline
\multicolumn{5}{c}{\emph{ACS (RA=00:35:30.7, Dec=+36:30:14.2)}} \\
\hline
jcnb12c4q\_flc.fits  	 & 2014-11-24 & 05:33:55 & F475W  &	  1264 \\ 
jcnb12c7q\_flc.fits  	 & 2014-11-24 & 05:57:56 & F814W  &	  1002 \\ 
jcnb12c9q\_flc.fits  	 & 2014-11-24 & 06:56:08 & F814W  &	  1086 \\ 
jcnb12cdq\_flc.fits  	 & 2014-11-24 & 07:17:12 & F475W  &	  1372 \\ 
jcnb13cpq\_flc.fits  	 & 2014-11-24 & 10:20:31 & F475W  &	  1264 \\ 
... & ... & ... & ... & ...  \\
\hline
\multicolumn{5}{c}{\emph{WFC3 (RA= 00:35:51.5, Dec=+36:25:48.5)}} \\
\hline
icnb12c5q\_flc.fits       & 2014-11-24 & 05:33:00 & F475W  &1308 \\  
icnb12c6q\_flc.fits       & 2014-11-24 & 05:57:16 & F814W  &1046 \\  
icnb12caq\_flc.fits       & 2014-11-24 & 06:56:03 & F814W  &1103 \\  
icnb12ceq\_flc.fits       & 2014-11-24 & 07:17:00 & F475W  &1389 \\  
icnb13cqq\_flc.fits       & 2014-11-24 & 10:19:36 & F475W  &1308 \\  
... & ... & ... & ... & ...  \\
\hline
\end{tabular}
\begin{tablenotes}
\begin{scriptsize}
\item This table is a portion of its entirely form which will be available in the online journal.
\end{scriptsize}
\end{tablenotes}
\end{table*}
\begin{table*}
\centering
\caption{Observing log for And~XV.} 
\label{tab:observing_log_andxv}
\begin{tabular}{ccccc} 
\hline
Image Name & Date         & UT Start      & Filter & Exp. time \\
  name     &  (YY-MM-DD)  & (hh:mm:ss)    &  name  &    (s)    \\ 
\hline
\multicolumn{5}{c}{\emph{ACS (RA=01:14:18.7,	Dec=+38:07:03.0)}} \\
\hline
jcnb23w3q\_flc.fits       & 2014-09-17 & 11:23:38 & F475W  &  1264 \\
jcnb23w6q\_flc.fits       & 2014-09-17 & 11:47:39 & F814W  &  1002 \\
jcnb23w8q\_flc.fits       & 2014-09-17 & 12:53:44 & F814W  &  1086 \\
jcnb23wcq\_flc.fits       & 2014-09-17 & 13:14:48 & F475W  &  1372 \\
jcnb24azq\_flc.fits       & 2014-09-18 & 16:05:20 & F475W  &  1264 \\
... & ... & ... & ... & ...  \\
\hline
\multicolumn{5}{c}{\emph{WFC3 (RA=01:13:50.3, Dec=+38:04:37.3)}} \\
\hline
icnb23w4q\_flc.fits       & 2014-09-17 & 11:22:43 & F475W  &1308 \\ 
icnb23w5q\_flc.fits       & 2014-09-17 & 11:46:59 & F814W  &1046 \\ 
icnb23w9q\_flc.fits       & 2014-09-17 & 12:53:39 & F814W  &1103 \\ 
icnb23wdq\_flc.fits       & 2014-09-17 & 13:14:36 & F475W  &1389 \\ 
icnb24b0q\_flc.fits       & 2014-09-18 & 16:04:25 & F475W  &1308 \\ 
... & ... & ... & ... & ...  \\
\hline
\end{tabular}
\begin{tablenotes}
\begin{scriptsize}
\item This table is a portion of its entirely form which will be available in the online journal.
\end{scriptsize}
\end{tablenotes}
\end{table*}
\begin{table*}
\centering
\caption{Observing log for And~XVI.} 
\label{tab:observing_log_andxvi}
\begin{tabular}{ccccc} 
\hline
Image Name & Date         & UT Start      & Filter & Exp. time \\
  name     &  (YY-MM-DD)  & (hh:mm:ss)    &  name  &    (s)    \\ 
\hline
\multicolumn{5}{c}{\emph{ACS (RA=00:59:32.3, Dec=+32:23:38.9)}} \\
\hline
jc1d09upq\_1.fits         & 2013-11-20 & 12:46:13 &  F475W &  1280 \\
jc1d09urq\_1.fits         & 2013-11-20 & 13:10:30 &  F814W &   987 \\
jc1d09uuq\_1.fits         & 2013-11-20 & 14:13:37 &  F814W &  1100 \\
jc1d09uyq\_1.fits         & 2013-11-20 & 14:34:55 &  F475W &  1359 \\
jc1d10wdq\_1.fits         & 2013-11-20 & 23:55:40 &  F475W &  1280 \\
... & ... & ... & ... & ...  \\
\hline
\multicolumn{5}{c}{\emph{WFC3 (RA=00:59:48.6, Dec=+32:18:37.2)}} \\
\hline
ic1d09uqq\_flc.fits       & 2013-11-20 & 12:45:18 & F475W  &1350 \\
ic1d09usq\_flc.fits       & 2013-11-20 & 13:10:16 & F814W  &1122 \\
ic1d09uvq\_flc.fits       & 2013-11-20 & 14:13:32 & F814W  &1200 \\
ic1d09v0q\_flc.fits       & 2013-11-20 & 14:36:06 & F475W  &1409 \\
ic1d10weq\_flc.fits       & 2013-11-20 & 23:54:45 & F475W  &1350 \\
... & ... & ... & ... & ...  \\
\hline
\end{tabular}
\begin{tablenotes}
\begin{scriptsize}
\item This table is a portion of its entirely form which will be available in the online journal.
\end{scriptsize}
\end{tablenotes}
\end{table*}
\begin{table*}
\centering
\caption{Observing log for And~XXVIII.} 
\label{tab:observing_log_andxxviii}
\begin{tabular}{ccccc} 
\hline
Image Name & Date         & UT Start      & Filter & Exp. time \\
  name     &  (YY-MM-DD)  & (hh:mm:ss)    &  name  &    (s)    \\ 
\hline
\multicolumn{5}{c}{\emph{ACS (RA=22:32:41.2, Dec=+31:12:58.2)}} \\
\hline
jcnb31psq\_flc.fits	 & 2015-01-20 & 23:57:41 & F475W  &	 1264 \\
jcnb31pvq\_flc.fits	 & 2015-01-21 & 00:21:42 & F814W  &	 1002 \\
jcnb31qoq\_flc.fits	 & 2015-01-21 & 01:19:28 & F814W  &	 1086 \\
jcnb31qsq\_flc.fits	 & 2015-01-21 & 01:40:32 & F475W  &	 1372 \\
jcnb32rnq\_flc.fits	 & 2015-01-21 & 04:44:14 & F475W  &	 1264 \\
jcnb32rqq\_flc.fits	 & 2015-01-21 & 05:08:15 & F814W  &	 1002 \\
... & ... & ... & ... & ...  \\
\hline
\multicolumn{5}{c}{\emph{WFC3 (RA=22:33:09.6, Dec=+31:13:31.0)}} \\
\hline
icnb31ptq\_flc.fits       & 2015-01-20 & 23:56:46 &  F475W & 1308 \\ 
icnb31puq\_flc.fits       & 2015-01-21 & 00:21:02 &  F814W & 1046 \\ 
icnb31qpq\_flc.fits       & 2015-01-21 & 01:19:23 &  F814W & 1103 \\ 
icnb31qtq\_flc.fits       & 2015-01-21 & 01:40:20 &  F475W & 1389 \\ 
icnb32roq\_flc.fits       & 2015-01-21 & 04:43:19 &  F475W & 1308 \\ 
... & ... & ... & ... & ...  \\
\hline
\end{tabular}
\begin{tablenotes}
\begin{scriptsize}
\item This table is a portion of its entirely form which will be available in the online journal.
\end{scriptsize}
\end{tablenotes}
\end{table*}
\clearpage

\section{Time series of Variable stars in ISLAndS galaxies}\label{sec:time_series}

The individual F475W and F814W measurements for all of the variables found
in each galaxy of this work are listed in Tables~\ref{tab:photometry_andi}, 
\ref{tab:photometry_andii}, \ref{tab:photometry_andiii}, 
\ref{tab:photometry_andxv}, \ref{tab:photometry_andxvi}, and
\ref{tab:photometry_andxxviii}, respectively.

\begin{table*}
\centering
\caption{Photometry of the variable stars in And~I dSph.} 
\label{tab:photometry_andi}
\begin{scriptsize}
\begin{tabular}{cccccc}
\hline
\hline 
MHJD$^*$    &    $F475W$    &    $\sigma_{F475W}$    &    MHJD$^*$    &    $F814W$    &    $\sigma_{F814W}$  \\
\hline
\multicolumn{6}{c}{AndI-V001} \\
\hline
   57266.425781  &   25.254  &    0.089  &   57266.441406  &   24.713  &    0.090 \\
   57266.500000  &   25.523  &    0.056  &   57266.484375  &   24.417  &    0.119 \\
   57266.691406  &   25.801  &    0.041  &   57266.707031  &   24.953  &    0.067 \\
   57266.765625  &   25.917  &    0.064  &   57266.750000  &   25.028  &    0.093 \\
   57267.417969  &   25.175  &    0.036  &   57267.433594  &   24.364  &    0.043 \\
   57267.496094  &   25.118  &    0.036  &   57267.480469  &   24.549  &    0.064 \\
   57267.683594  &   25.733  &    0.062  &   57267.699219  &   24.876  &    0.054 \\
   57267.761719  &   25.863  &    0.052  &   57267.746094  &   25.007  &    0.127 \\
   57268.281250  &   25.768  &    0.069  &   57268.296875  &   24.734  &    0.053 \\
   57268.355469  &   25.832  &    0.063  &   57268.339844  &   24.781  &    0.087 \\
   57268.890625  &   25.811  &    0.054  &   57268.824219  &   24.701  &    0.095 \\
   57269.417969  &   25.788  &    0.064  &   57268.875000  &   24.767  &    0.080 \\
   57269.339844  &   25.706  &    0.058  &   57269.355469  &   24.708  &    0.079 \\
   57269.539062  &   25.700  &    0.072  &   57269.402344  &   24.811  &    0.062 \\
   57268.808594  &   25.720  &    0.052  &   57269.554688  &   24.983  &    0.062 \\
   57269.617188  &   26.001  &    0.065  &   57269.601562  &   24.958  &    0.111 \\
   57270.464844  &   25.608  &    0.063  &   57270.480469  &   24.784  &    0.089 \\
   57270.542969  &   25.681  &    0.088  &   57270.527344  &   24.797  &    0.082 \\
   57270.664062  &   25.856  &    0.061  &   57270.679688  &   24.860  &    0.062 \\
   57270.742188  &   26.026  &    0.062  &   57270.726562  &   25.066  &    0.104 \\
   57271.660156  &   25.765  &    0.058  &   57271.675781  &   24.759  &    0.055 \\
   57271.738281  &   25.904  &    0.033  &   57271.718750  &   24.856  &    0.078 \\
\hline
\hline
\end{tabular}
\end{scriptsize}
\begin{tablenotes}
\begin{scriptsize}
\item $^*$ Modified Heliocentric Julian Date of mid-exposure: HJD - 2,400,000
\item (This table is a portion of its entirely form which will be available in the online journal.)
\end{scriptsize}
\end{tablenotes}
\end{table*}

\begin{table*}
\centering
\caption{Photometry of the variable stars in And~II dSph.} 
\label{tab:photometry_andii}
\begin{scriptsize}
\begin{tabular}{cccccc}
\hline
\hline 
MHJD$^*$    &    $F475W$    &    $\sigma_{F475W}$    &    MHJD$^*$    &    $F814W$    &    $\sigma_{F814W}$  \\
\hline
\multicolumn{6}{c}{AndII-V001} \\
\hline
   56569.171875  &   24.838  &    0.031  &   56569.187500  &   24.319  &    0.047 \\
   56569.253906  &   24.822  &    0.051  &   56569.234375  &   24.196  &    0.047 \\
   56569.437500  &   25.382  &    0.041  &   56569.453125  &   24.654  &    0.061 \\
   56569.519531  &   24.785  &    0.069  &   56569.500000  &   24.326  &    0.059 \\
   56570.101562  &   25.326  &    0.042  &   56570.117188  &   24.509  &    0.058 \\
   56570.183594  &   24.775  &    0.048  &   56570.164062  &   24.268  &    0.047 \\
   56570.500000  &   24.833  &    0.036  &   56570.515625  &   24.399  &    0.051 \\
   56570.582031  &   24.849  &    0.050  &   56570.562500  &   24.332  &    0.036 \\
   56570.699219  &   25.400  &    0.042  &   56570.714844  &   24.593  &    0.051 \\
   56570.781250  &   25.339  &    0.062  &   56570.761719  &   24.677  &    0.084 \\
   56570.898438  &   24.797  &    0.043  &   56570.914062  &   24.285  &    0.051 \\
   56570.980469  &   25.153  &    0.048  &   56570.960938  &   24.457  &    0.049 \\
   56571.097656  &   25.369  &    0.083  &   56571.113281  &   24.569  &    0.079 \\
   56571.179688  &   24.772  &    0.052  &   56571.160156  &   24.300  &    0.062 \\
   56571.562500  &   24.790  &    0.064  &   56571.578125  &   24.393  &    0.055 \\
   56571.644531  &   25.160  &    0.055  &   56571.625000  &   24.396  &    0.056 \\
   56571.695312  &   25.362  &    0.067  &   56571.710938  &   24.548  &    0.058 \\
\hline
\hline
\end{tabular}
\end{scriptsize}
\begin{tablenotes}
\begin{scriptsize}	
\item $^*$ Modified Heliocentric Julian Date of mid-exposure: HJD - 2,400,000
\item (This table is a portion of its entirely form which will be available in the online journal.)
\end{scriptsize}
\end{tablenotes}
\end{table*}

\begin{table*}
\centering
\caption{Photometry of the variable stars in And~III dSph.} 
\label{tab:photometry_andiii}
\begin{scriptsize}
\begin{tabular}{cccccc} 
\hline
\hline
MHJD$^*$    &    $F475W$    &    $\sigma_{F475W}$    &    MHJD$^*$    &    $F814W$    &    $\sigma_{F814W}$  \\
\hline
\multicolumn{6}{c}{AndIII-V001} \\
\hline
   56985.244721  &   25.181  &    0.034  &   56985.259883  &   24.593  &    0.036 \\
   56985.317071  &   25.254  &    0.021  &   56985.300786  &   24.459  &    0.035 \\
   56985.443750  &   25.538  &    0.047  &   56985.458912  &   24.765  &    0.038 \\
   56985.516204  &   25.525  &    0.048  &   56985.499919  &   24.701  &    0.062 \\
   56985.642767  &   25.035  &    0.040  &   56985.657929  &   24.488  &    0.048 \\
   56985.721413  &   24.881  &    0.149  &   56985.705128  &   24.540  &    0.038 \\
   56986.239807  &   25.663  &    0.046  &   56986.254969  &   24.688  &    0.059 \\
   56986.312643  &   25.630  &    0.029  &   56986.296358  &   24.256  &    0.173 \\
   56987.168353  &   24.976  &    0.086  &   56987.183515  &   24.517  &    0.053 \\
   56987.241710  &   25.183  &    0.023  &   56987.225425  &   24.404  &    0.044 \\
   56987.499975  &   25.635  &    0.047  &   56987.515137  &   24.716  &    0.046 \\
   56987.573493  &   25.028  &    0.034  &   56987.557209  &   24.472  &    0.038 \\
   56988.229862  &   25.751  &    0.036  &   56988.245024  &   24.743  &    0.034 \\
   56988.303346  &   25.523  &    0.029  &   56988.287061  &   24.663  &    0.027 \\
   56988.428810  &   25.330  &    0.035  &   56988.443972  &   24.534  &    0.043 \\
   56988.502386  &   25.340  &    0.038  &   56988.486102  &   24.496  &    0.048 \\
   56989.158280  &   25.062  &    0.021  &   56989.173442  &   24.382  &    0.047 \\
   56989.232169  &   25.009  &    0.072  &   56989.215885  &   24.380  &    0.050 \\
   56989.357228  &   25.337  &    0.045  &   56989.372390  &   24.719  &    0.030 \\
   56989.431186  &   25.660  &    0.044  &   56989.414902  &   24.680  &    0.059 \\
   56989.630215  &   25.272  &    0.038  &   56989.613930  &   24.464  &    0.046 \\
   56989.556175  &   25.420  &    0.063  &   56989.571337  &   24.501  &    0.065 \\
\hline
\hline
\end{tabular}
\end{scriptsize}
\begin{tablenotes}
\begin{scriptsize}	
\item $^*$ Modified Heliocentric Julian Date of mid-exposure: HJD - 2,400,000
\item (This table is a portion of its entirely form which will be available in the online journal.)
\end{scriptsize}
\end{tablenotes}
\end{table*}

\begin{table*}
\centering
\caption{Photometry of the variable stars in And~XV dSph.} 
\label{tab:photometry_andxv}
\begin{scriptsize}
\begin{tabular}{cccccc}
\hline
\hline 
MHJD$^*$    &    $F475W$    &    $\sigma_{F475W}$    &    MHJD$^*$    &    $F814W$    &    $\sigma_{F814W}$  \\
\hline
\multicolumn{6}{c}{AndXV-V001} \\
\hline
   56917.486432  &   24.553  &    0.033  &   56917.501594  &   24.281  &    0.036 \\
   56917.564258  &   25.095  &    0.039  &   56917.547973  &   24.378  &    0.030 \\
   56918.682087  &   25.631  &    0.055  &   56918.697250  &   24.707  &    0.048 \\
   56918.758837  &   25.772  &    0.053  &   56918.742552  &   24.872  &    0.067 \\
   56918.881259  &   25.826  &    0.056  &   56918.896422  &   24.939  &    0.064 \\
   56918.957951  &   25.254  &    0.098  &   56918.941666  &   24.810  &    0.057 \\
   56919.476714  &   24.797  &    0.034  &   56919.491877  &   24.240  &    0.038 \\
   56919.555350  &   24.883  &    0.033  &   56919.539065  &   24.342  &    0.035 \\
   56919.678374  &   25.536  &    0.037  &   56919.693537  &   24.710  &    0.061 \\
   56919.754476  &   25.851  &    0.040  &   56919.738191  &   24.777  &    0.052 \\
   56919.877581  &   25.733  &    0.064  &   56919.892743  &   24.902  &    0.074 \\
   56919.953659  &   25.653  &    0.045  &   56919.937374  &   25.007  &    0.056 \\
   56920.541683  &   24.732  &    0.057  &   56920.556846  &   24.369  &    0.063 \\
   56920.617634  &   25.256  &    0.039  &   56920.601349  &   24.489  &    0.055 \\
   56920.740890  &   25.675  &    0.043  &   56920.756052  &   24.853  &    0.056 \\
   56920.816829  &   25.835  &    0.069  &   56920.800544  &   24.877  &    0.057 \\
   56920.883382  &   25.712  &    0.046  &   56920.867293  &   24.903  &    0.054 \\
\hline
\hline
\end{tabular}
\end{scriptsize}
\begin{tablenotes}
\begin{scriptsize}	
\item $^*$ Modified Heliocentric Julian Date of mid-exposure: HJD - 2,400,000
\item (This table is a portion of its entirely form which will be available in the online journal.)
\end{scriptsize}
\end{tablenotes}
\end{table*}

\begin{table*}
\centering
\caption{Photometry of the variable stars in And~XVI dSph.} 
\label{tab:photometry_andxvi}
\begin{scriptsize}
\begin{tabular}{cccccc} 
\hline
\hline
MHJD$^*$    &    $F475W$    &    $\sigma_{F475W}$    &    MHJD$^*$    &    $F814W$    &    $\sigma_{F814W}$  \\
\hline
\multicolumn{6}{c}{AndXVI-V001} \\
\hline
   56616.545139  &   25.496  &    0.025  &   56616.560301  &   24.366  &    0.153 \\
   56616.621076  &   25.675  &    0.034  &   56616.604792  &   24.668  &    0.042 \\
   56617.010037  &   25.030  &    0.024  &   56617.025199  &   24.462  &    0.056 \\
   56617.085986  &   25.392  &    0.022  &   56617.069701  &   24.566  &    0.029 \\
   56617.408499  &   25.885  &    0.050  &   56617.423673  &   24.846  &    0.054 \\
   56617.484483  &   25.923  &    0.063  &   56617.468199  &   24.823  &    0.057 \\
   56617.674172  &   25.619  &    0.043  &   56617.689334  &   24.660  &    0.054 \\
   56617.752522  &   25.330  &    0.034  &   56617.736660  &   24.650  &    0.044 \\
   56618.006245  &   25.789  &    0.046  &   56618.021407  &   24.766  &    0.069 \\
   56618.082218  &   25.945  &    0.049  &   56618.065933  &   24.916  &    0.052 \\
   56618.471143  &   25.801  &    0.047  &   56618.485872  &   24.788  &    0.057 \\
   56618.547139  &   25.841  &    0.067  &   56618.530854  &   24.764  &    0.066 \\
   56618.598771  &   25.825  &    0.048  &   56618.615056  &   24.860  &    0.053 \\
\hline
\hline
\end{tabular}
\end{scriptsize}
\begin{tablenotes}
\begin{scriptsize}
\item $^*$ Modified Heliocentric Julian Date of mid-exposure: HJD - 2,400,000
\item (This table is a portion of its entirely form which will be available in the online journal.)
\end{scriptsize}
\end{tablenotes}
\end{table*}

\begin{table*}
\centering
\caption{Photometry of the variable stars in And~XXVIII dSph.} 
\label{tab:photometry_andxxviii}
\begin{scriptsize}
\begin{tabular}{cccccc} 
\hline
\hline
MHJD$^*$    &    $F475W$    &    $\sigma_{F475W}$    &    MHJD$^*$    &    $F814W$    &    $\sigma_{F814W}$  \\
\hline
\multicolumn{6}{c}{AndXXVIII-V001} \\
\hline
   57043.010749  &   25.951  &    0.110  &   57043.025911  &   24.779  &    0.060 \\
   57043.082797  &   25.813  &    0.054  &   57043.066513  &   24.845  &    0.031 \\
   57043.209741  &   25.053  &    0.062  &   57043.224903  &   24.318  &    0.052 \\
   57043.281893  &   24.992  &    0.038  &   57043.265608  &   24.327  &    0.039 \\
   57044.204663  &   25.908  &    0.053  &   57044.219825  &   24.515  &    0.075 \\
   57044.277290  &   25.903  &    0.043  &   57044.261005  &   24.683  &    0.175 \\
   57044.337289  &   25.878  &    0.077  &   57044.352451  &   24.875  &    0.046 \\
   57044.412127  &   26.123  &    0.084  &   57044.395842  &   24.798  &    0.046 \\
   57045.007156  &   25.852  &    0.091  &   57044.949321  &   24.892  &    0.067 \\
   57044.934159  &   25.758  &    0.102  &   57044.990872  &   24.887  &    0.070 \\
   57045.066797  &   26.149  &    0.065  &   57045.081959  &   24.925  &    0.075 \\
   57045.139852  &   25.151  &    0.055  &   57045.123567  &   24.522  &    0.075 \\
   57045.928885  &    ---    &     ---   &   57045.944046  &    ---    &     ---  \\
   57046.002310  &    ---    &     ---   &   57045.986025  &    ---    &     ---  \\
   57046.061499  &   25.748  &    0.054  &   57046.076661  &   24.674  &    0.054 \\
   57046.134994  &   25.818  &    0.042  &   57046.118709  &   24.739  &    0.063 \\
   57046.923494  &   25.931  &    0.072  &   57046.938656  &   24.971  &    0.069 \\
   57046.997347  &   25.977  &    0.071  &   57046.981063  &   24.940  &    0.067 \\
   57047.056108  &   25.098  &    0.043  &   57047.071270  &   24.255  &    0.076 \\
   57047.130008  &   25.142  &    0.042  &   57047.113724  &   24.387  &    0.040 \\
\hline
\hline
\end{tabular}
\end{scriptsize}
\begin{tablenotes}
\begin{scriptsize}
\item $^*$ Modified Heliocentric Julian Date of mid-exposure: HJD - 2,400,000
\item (This table is a portion of its entirely form which will be available in the online journal.)
\end{scriptsize}
\end{tablenotes}
\end{table*}
\clearpage

\section{Comparison with the literature}\label{sec:comp_pritzl}

\begin{figure*}
	\epsscale{0.80}
	\plotone{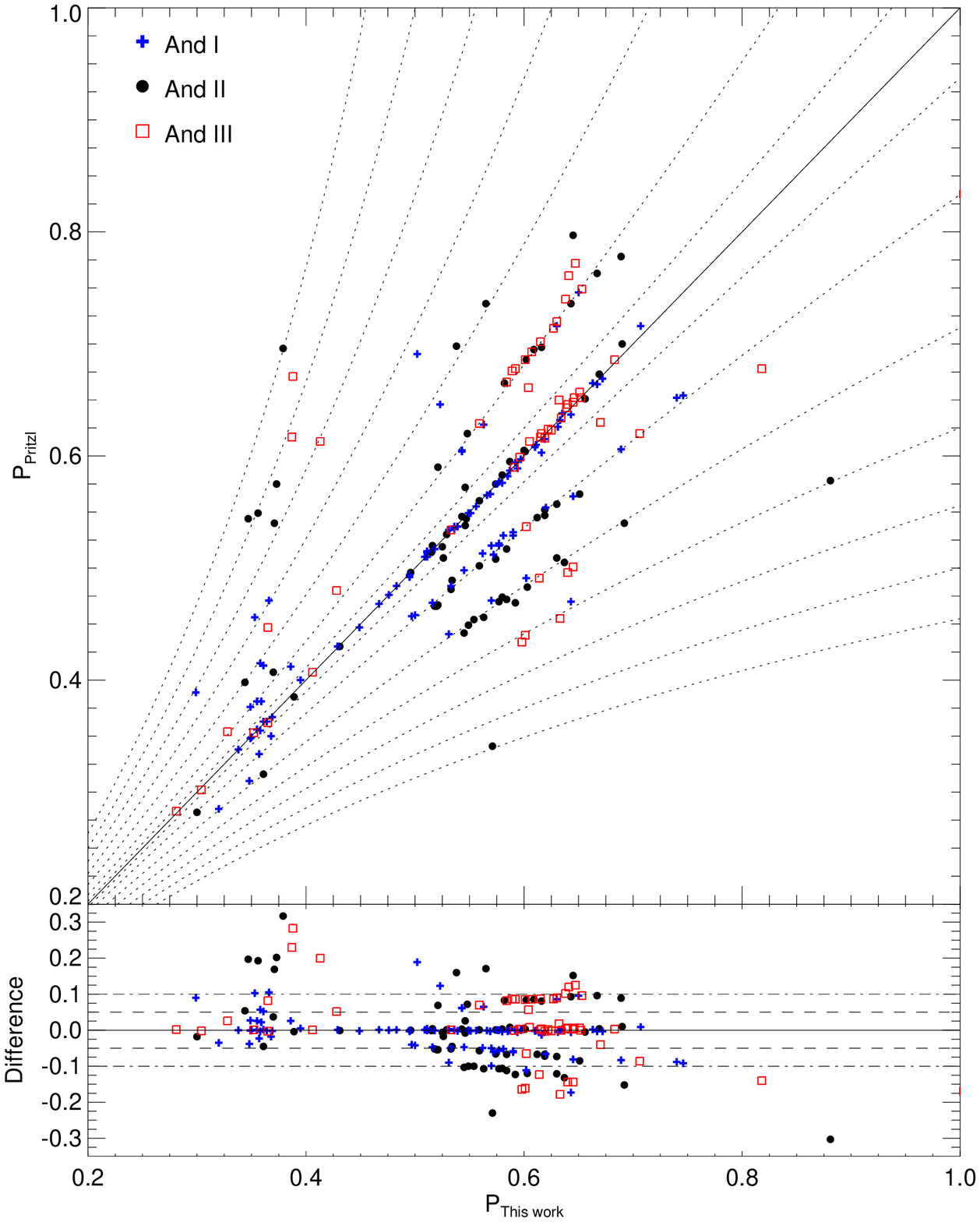}
	\caption{Current period versus the period (top) and period difference (bottom) found by 
	\citet{Pritzl2004,Pritzl2005} for the 94, 69 and 54 stars matched in And~I (black circles), 
	 And~II (blue plusses) and And~III (red open squares), respectively. The dotted curves in the 
     top panel are the aliasing lines (see text). We have taken the HST orbital period (96 minutes) as 
	 $f_{sampling}$ and $|N|$=[1,3,6,9,15,18] for obtaining these curves (aliasing lines).  
	 Note how the outliers follow the aliasing lines in most cases. This indicates the high probability 
	 that the offsets are due to aliasing.}
	\label{fig:comp_pritzl}
\end{figure*}

Figure~\ref{fig:comp_pritzl} compares the periods of variable stars in common between 
our work and the published period for the three galaxies already studied in the literature 
\citep[And~I, And~II and And~III][]{Pritzl2004,Pritzl2005}, according to the 
labelled symbols.
Pritzl's data for each galaxy consist in two data-sets separated by 4-5 days. 
In each data-set, they started observing first all the $F555W$ and then all the 
$F450W$ images. The cadence of the data depends on the individual case. For And~I, 
the strategy was 3$\times F555W$ (4$\times F555W$) and 6$\times F450W$ 
(6$\times F450W$) in the first (second) set, with 1 image per orbit. For And~II, 
the data were collected as 3$\times F555W$ (4$\times F555W$) and 7$\times F450W$ 
(8$\times F450W$) in the first (second) set, with 2 images per orbit. For And~III 
the strategy was similar to that for And~II, but collecting 4$\times F555W$ and 
8$\times F450W$ in each data-set.

A total of 94, 69 and 54 stars (out of 100, 73 and 56 in their catalogues) 
were recovered for And~I, And~II and And~III, respectively. The cross-identifications made
between our catalogues and Pritzl's are shown in Tables~\ref{tab:pritzl_andi}, 
\ref{tab:pritzl_andii}, \ref{tab:pritzl_andiii}. The small fraction of stars that were not 
recovered either appear as non variable in our data or fall in the ACS gap (see the column 
``Notes'' of Tables~\ref{tab:pritzl_andi}, \ref{tab:pritzl_andii}, \ref{tab:pritzl_andiii}). 
We note that the matching was complicated by the fact that the coordinates listed in the 
Pritzl catalogs were significantly offset, with different offsets for each WFPC2 chip, in 
particular in the case of And III.

\begin{table*}
\centering
\caption{Cross-identification with the Pritzl et al. catalog of variable stars in And~I.} 
\label{tab:pritzl_andi}
\begin{scriptsize}
\centering
\begin{tabular}{ccccl}
\hline
ID$_{Pritzl}$ & Period$_{Pritzl}$ & ID$_{This~work}$ & Period$_{This~work}$ & Notes \\  
\hline
V1   &  1.630  &    ---         &   ---   &  near to the BSS region; affected by near saturated field star; not variable in our data \\
V2   &  0.348  &  AndI-V185  &  0.349  &  \\
V3   &  0.412  &  AndI-V182  &  0.386  &  \\
V4   &  9.999  &  AndI-V172  &  0.607  &  \\
V5   &  0.654  &  AndI-V154  &  0.746  &  \\
V6   &  0.430  &  AndI-V186  &  0.429  &  \\
...  & ... & ... & ... & ... \\
V44  &  0.748  &     ---        &   ---   &  minor variation in F814; not variable in our data \\
V45  &  0.772  &     ---        &   ---   &  not variable in our data \\
.....  & ... & ... & ... & ... \\
V89  &  0.523  &      ---       &   ---   &  in ACS gap \\
...  & ... & ... & ... & ... \\
V98  &  0.625  &      ---       &   ---   &  in ACS gap \\
V99  &  0.716  &  AndI-V237  &  0.630  &  \\
V100 &  0.782  &      ---       &   ---   &  not variable in our data \\
\hline
\end{tabular}
\end{scriptsize}
\begin{tablenotes}
\begin{scriptsize}
\item The full table is available as Supporting Information with the online version of the paper.
\end{scriptsize}
\end{tablenotes}
\end{table*}
\begin{table*}
\centering
\caption{Cross-identification with the Pritzl et al. catalog of variable stars in And~II.} 
\label{tab:pritzl_andii}
\begin{scriptsize}
\begin{tabular}{ccccl}
\hline
ID$_{Pritzl}$ & Period$_{Pritzl}$ & ID$_{This~work}$ & Period$_{This~work}$ & Notes \\  
\hline
V01  &  0.407  &  AndII-V080  &  0.370 &  \\  
V02  &  0.546  &  AndII-V071  &  0.543 &  \\ 
V03  &  0.520  &  AndII-V098  &  0.516 &  \\ 
V04  &  0.540  &  AndII-V064  &  0.692 &  \\ 
V05  &  0.583  &  AndII-V081  &  0.580 &  \\  
... & ... & ... & ... & ... \\
V16  &  0.346  &      ---        &   ---  &  not variable in our data \\ 
... & ... & ... & ... & ... \\
V37  &  0.751  &      ---        &   ---  & not variable in our data \\  
... & ... & ... & ... & ... \\
V43  &  0.490  &      ---        &   ---  &  in ACS gap \\
... & ... & ... & ... & ... \\
V70  &  0.707  &      ---        &   ---  &  not variable in our data \\
V71  &  0.474  &  AndII-V153  &  0.580 &  \\  
V72  &  0.469  &  AndII-V087  &  0.592 &  \\ 
V73  &  0.698  &  AndII-V118  &  0.538 &  \\ 
\hline
\end{tabular}
\end{scriptsize}
\begin{tablenotes}
\begin{scriptsize}
\item The full table is available as Supporting Information with the online version of the paper.
\end{scriptsize}
\end{tablenotes}
\end{table*}
\begin{table*}
\centering
\caption{Cross-identification with the Pritzl et al. catalog of variable stars in And~III.} 
\label{tab:pritzl_andiii}
\begin{scriptsize}
\begin{tabular}{ccccl}
\hline
ID$_{Pritzl}$ & Period$_{Pritzl}$ & ID$_{This~work}$ & Period$_{This~work}$ & Notes \\  
\hline
V01  &  0.834  &  AndIII-V073  &  1.003 & \\ 
V02  &  0.590  &  AndIII-V069  &  0.591 & \\ 
V03  &  0.773  &     ---          &   ---  &  in ACS gap \\
V04  &  0.629  &  AndIII-V065  &  0.559 & \\ 
V05  &  0.650  &  AndIII-V067  &  0.632 & \\ 
V06  &  0.678  &  AndIII-V105  &  0.818 & \\ 
V07  &  0.480  &  AndIII-V075  &  0.428 & \\ 
V08  &  1.510  &      ---         &   ---  &  not variable in our data \\
... & ... & ... & ... & ... \\
V53  &  0.534  &  AndIII-V039  &  0.533 & \\ 
V54  &  0.623  &  AndIII-V041  &  0.625 & \\ 
V55  &  0.599  &  AndIII-V031  &  0.596 & \\ 
V56  &  0.496  &  AndIII-V046  &  0.640 & \\  
\hline
\end{tabular}
\end{scriptsize}
\begin{tablenotes}
\begin{scriptsize}
\item The full table is available as Supporting Information with the online version of the paper.
\end{scriptsize}
\end{tablenotes}
\end{table*}

The comparison discloses a general good agreement (53\% of the stars within a 
difference of 0.05 days and 80\% within 0.1 days) though with a few outliers for 
which the period is significantly discrepant (20\% have a difference larger than 0.1 day).
However, this effect can be easily explained taking into account that HST observations 
may suffer from a strong aliasing introduced by the orbital time. By having more 
epochs and scheduling them to avoid redundant periods, our program strongly 
suppressed the effects of aliasing.

The aliasing effect is a very common \textit{error} in signal treatment, as it happens whenever
the original periodic signal (in our case, the light curve of the variable stars) 
is reconstructed using a discrete sampling. When a periodic signal of frequency 
$f_{true}$ is sampled with a frequency $f_{sampling}$, the resulting number of 
cycles per sample is $f_{true}/f_{sampling}$ (normalized frequency), and the 
samples are indistinguishable from those of another sinusoid (or periodic signal), 
called an \textit{alias}, whose normalized frequency differs from $f_{true}/f_{sampling}$
by an integer \citep[see e.g.,][]{VanderPlas2017}. 
Then, we can express the aliases of frequency as
$f_{alias}=|f_{true}-N f_{sampling}|$, being N an integer. In this case, the calculation 
of the period can be affected by the HST orbital cadence of 96 minutes.

The curves overplotted on Figure~\ref{fig:comp_pritzl} represent how the true period 
is affected by a cadence of 96 minutes. Interestingly, most if not all the discrepant 
points are explained by this aliasing effect. Taking into account the limited 
number of phase points in previous studies, and the optimized strategy of our 
observations, we suggest that has resulted in more precise period determinations. 
Nevertheless, this comparison also supports the quality of the previous analysis, 
given the observational material available.

\clearpage

\section{Pulsation Properties of Variable stars in ISLAndS galaxies}\label{sec:pulsation_properties}

The properties of the variable stars found in this work for 
And~I, II, III, XV, XVI, and XXVIII are detailed in Tables~\ref{tab:variables_andi},
\ref{tab:variables_andii}, \ref{tab:variables_andiii}, \ref{tab:variables_andxv},
\ref{tab:variables_andxvi}, and \ref{tab:variables_andxxviii}, respectively.
The first columns give the identification number and the next two list the 
equatorial coordinates (J2000.0). Column 4 give the period of the variable in days, while 
columns 5 to 14 list the intensity-averaged magnitudes and amplitude in the 
filters $F475$, $F814$, $B$, $V$, and $I$, respectively. Last column displays the variable type.

\begin{table*}
\caption{Parameters of the variable stars in And~I dSph.} 
\label{tab:variables_andi}
\begin{scriptsize}
\hspace{-100pt}
\centering
\begin{tabular}{ccccccccccccccc}
\hline
\hline
ID & RA & DEC & Period & $\langle F475W\rangle$ & A$_{F475W}$ & $\langle F814W\rangle$ & A$_{F814W}$ & $\langle B\rangle$ & A$_{B}$ & $\langle V\rangle$ & A$_{V}$ & $\langle I\rangle$ & A$_{I}$ & Type \\
name & (J2000) & (J2000) & (current) & & & & & & & & & & & \\ 
\hline 
AndI-V001  &  0:45:09.233  &  +37:58:47.19  &  0.569  &  25.532  &  1.038  &  24.692  &  0.574  &  25.657  &  1.137  &  25.266  &  0.915  &  24.677  &  0.580  &  RRab    \\
AndI-V002  &  0:45:09.646  &  +37:59:48.86  &  0.567  &  25.505  &  0.590  &  24.737  &  0.358  &  25.613  &  0.669  &  25.251  &  0.436  &  24.719  &  0.360  &  RRab    \\
AndI-V003  &  0:45:09.819  &  +37:59:32.31  &  0.296  &  25.329  &  0.336  &  24.831  &  0.091  &  25.398  &  0.392  &  25.159  &  0.207  &  24.819  &  0.097  &  RRc     \\
AndI-V004  &  0:45:10.116  &  +37:58:44.43  &  0.598  &  25.322  &  1.390  &  24.607  &  0.699  &  25.415  &  1.605  &  25.103  &  0.949  &  24.599  &  0.669  &  RRab    \\
AndI-V005  &  0:45:10.429  &  +37:58:56.47  &  0.585  &  25.557  &  0.807  &  24.733  &  0.608  &  25.680  &  0.864  &  25.276  &  0.660  &  24.722  &  0.576  &  RRab    \\
AndI-V006  &  0:45:11.526  &  +37:58:45.53  &  0.349  &  25.442  &  1.054  &  24.806  &  0.508  &  25.531  &  1.120  &  25.228  &  0.875  &  24.790  &  0.518  &  RRd     \\
AndI-V007  &  0:45:12.020  &  +38:00:25.42  &  0.352  &  25.404  &  0.543  &  24.762  &  0.254  &  25.488  &  0.603  &  25.196  &  0.436  &  24.748  &  0.250  &  RRd     \\
AndI-V008  &  0:45:12.286  &  +37:58:45.04  &  0.791  &  25.338  &  0.304  &  24.450  &  0.243  &  25.478  &  0.321  &  25.023  &  0.263  &  24.421  &  0.273  &  RRab    \\
AndI-V009  &  0:45:13.494  &  +38:00:57.92  &  0.353  &  25.364  &  0.479  &  24.714  &  0.210  &  25.455  &  0.545  &  25.142  &  0.354  &  24.699  &  0.214  &  RRc     \\
AndI-V010  &  0:45:13.931  &  +37:59:21.84  &  0.581  &  25.273  &  1.393  &  24.614  &  0.658  &  25.357  &  1.550  &  25.096  &  0.999  &  24.611  &  0.617  &  RRab    \\
AndI-V011  &  0:45:14.260  &  +37:58:40.41  &  0.479  &  25.476  &  1.252  &  24.849  &  0.947  &  25.605  &  1.300  &  25.230  &  1.078  &  24.836  &  0.904  &  RRab    \\
AndI-V012  &  0:45:14.735  &  +38:00:35.23  &  0.623  &  25.172  &  1.365  &  24.434  &  0.656  &  25.302  &  1.483  &  24.921  &  1.069  &  24.421  &  0.647  &  RRab    \\
AndI-V013  &  0:45:15.057  &  +38:00:50.11  &  0.515  &  25.423  &  1.213  &  24.688  &  0.731  &  25.522  &  1.350  &  25.194  &  0.845  &  24.681  &  0.734  &  RRab    \\
AndI-V014  &  0:45:15.796  &  +37:59:31.21  &  0.359  &  25.326  &  0.484  &  24.685  &  0.235  &  25.420  &  0.530  &  25.106  &  0.381  &  24.670  &  0.234  &  RRc     \\
AndI-V015  &  0:45:16.130  &  +38:01:24.28  &  0.703  &  25.345  &  0.422  &  24.430  &  0.208  &  25.486  &  0.463  &  25.037  &  0.377  &  24.416  &  0.214  &  RRab    \\
AndI-V016  &  0:45:16.173  &  +37:59:27.10  &  0.644  &  25.458  &  0.576  &  24.553  &  0.329  &  25.606  &  0.634  &  25.139  &  0.478  &  24.538  &  0.338  &  RRab    \\
AndI-V017  &  0:45:16.178  &  +37:58:57.24  &  0.640  &  25.508  &  0.420  &  24.633  &  0.124  &  25.640  &  0.489  &  25.213  &  0.317  &  24.622  &  0.141  &  RRab    \\
AndI-V018  &  0:45:16.711  &  +37:59:41.10  &  0.301  &  25.213  &  0.212  &  24.726  &  0.135  &  25.290  &  0.222  &  25.033  &  0.174  &  24.718  &  0.139  &  RRc     \\
AndI-V019  &  0:45:16.813  &  +37:59:09.94  &  0.564  &  25.381  &  1.011  &  24.679  &  0.677  &  25.475  &  1.142  &  25.167  &  0.722  &  24.666  &  0.582  &  RRab    \\
AndI-V020  &  0:45:17.035  &  +38:00:02.48  &  0.574  &  25.453  &  1.113  &  24.692  &  0.536  &  25.560  &  1.257  &  25.216  &  0.836  &  24.685  &  0.525  &  RRab    \\
\hline
\hline
\end{tabular}
\end{scriptsize}
\begin{tablenotes}
\begin{scriptsize}	
\item Stars from ``AndI-V001'' to ``AndI-V0038'' were detected in the WFC3 field, while stars from ``AndI-V039'' to ``AndI-V314'' were detected in the ACS field.
\item Full version are available as Supporting Information with the online version of the paper.
\end{scriptsize}
\end{tablenotes}
\end{table*}
\begin{table*}
\caption{Parameters of the variable stars in And~II dSph.} 
\label{tab:variables_andii}
\begin{scriptsize}
\hspace{-100pt}
\centering
\begin{tabular}{ccccccccccccccc} 
\hline
\hline
ID & RA & DEC & Period & $\langle F475W\rangle$ & A$_{F475W}$ & $\langle F814W\rangle$ & A$_{F814W}$ & $\langle B\rangle$ & A$_{B}$ & $\langle V\rangle$ & A$_{V}$ & $\langle I\rangle$ & A$_{I}$ & Type \\
name & (J2000) & (J2000) & (current) & & & & & & & & & & & \\ 
\hline 
AndII-V001  &  1:15:56.325  &  +33:21:19.95  &  0.332  &  25.061  &  0.686  &  24.441  &  0.336  &  25.149  &  0.747  &  24.848  &  0.563  &  24.425  &  0.330  &  RRc   \\
AndII-V002  &  1:15:57.224  &  +33:21:32.52  &  0.601  &  25.064  &  1.016  &  24.272  &  0.362  &  25.310  &  0.948  &  24.785  &  0.685  &  24.239  &  0.325  &  RRab  \\
AndII-V003  &  1:15:58.620  &  +33:21:29.91  &  0.625  &  25.115  &  0.766  &  24.243  &  0.375  &  25.257  &  0.886  &  24.811  &  0.613  &  24.231  &  0.383  &  RRab  \\
AndII-V004  &  1:15:58.951  &  +33:21:14.99  &  0.663  &  25.124  &  0.339  &  24.234  &  0.158  &  25.258  &  0.378  &  24.828  &  0.294  &  24.221  &  0.180  &  RRab  \\
AndII-V005  &  1:15:58.986  &  +33:21:05.31  &  0.769  &  24.971  &  0.561  &  24.102  &  0.297  &  25.109  &  0.620  &  24.667  &  0.470  &  24.088  &  0.308  &  RRab  \\
AndII-V006  &  1:15:59.033  &  +33:21:08.26  &  0.590  &  24.957  &  1.231  &  24.204  &  0.557  &  25.064  &  1.356  &  24.700  &  1.008  &  24.140  &  0.573  &  RRab  \\
AndII-V007  &  1:15:59.056  &  +33:20:45.97  &  0.606  &  24.927  &  1.551  &  24.194  &  0.791  &  25.033  &  1.692  &  24.670  &  1.306  &  24.180  &  0.792  &  RRab  \\
AndII-V008  &  1:15:59.377  &  +33:21:57.11  &  0.622  &  25.128  &  0.762  &  24.267  &  0.380  &  25.261  &  0.836  &  24.845  &  0.644  &  24.271  &  0.348  &  RRab  \\
AndII-V009  &  1:15:59.439  &  +33:21:26.87  &  0.345  &  24.990  &  0.655  &  24.378  &  0.339  &  25.080  &  0.705  &  24.768  &  0.554  &  24.363  &  0.336  &  RRc   \\
AndII-V010  &  1:16:00.025  &  +33:20:38.76  &  0.388  &  26.014  &  0.174  &  24.680  &  0.252  &  26.276  &  0.219  &  25.551  &  0.161  &  24.699  &  0.236  &  EB    \\
AndII-V011  &  1:16:01.326  &  +33:22:38.74  &  0.612  &  25.053  &  0.540  &  24.218  &  0.319  &  25.184  &  0.570  &  24.763  &  0.501  &  24.207  &  0.332  &  RRab  \\
AndII-V012  &  1:16:01.426  &  +33:20:39.01  &  0.640  &  24.976  &  0.540  &  24.125  &  0.267  &  25.108  &  0.593  &  24.686  &  0.431  &  24.109  &  0.273  &  RRab  \\
AndII-V013  &  1:16:02.447  &  +33:20:24.48  &  0.357  &  25.041  &  0.452  &  24.378  &  0.291  &  25.140  &  0.507  &  24.804  &  0.358  &  24.356  &  0.296  &  RRd   \\
AndII-V014  &  1:16:02.573  &  +33:22:07.27  &  0.621  &  25.124  &  0.521  &  24.366  &  0.259  &  25.245  &  0.578  &  24.844  &  0.417  &  24.337  &  0.239  &  RRab  \\
AndII-V015  &  1:16:02.657  &  +33:23:07.14  &  0.570  &  25.157  &  1.160  &  24.493  &  0.400  &  25.266  &  1.317  &  24.919  &  0.885  &  24.478  &  0.392  &  RRab  \\
AndII-V016  &  1:16:03.585  &  +33:22:15.65  &  0.556  &  24.914  &  1.441  &  24.419  &  0.683  &  24.993  &  1.554  &  24.757  &  1.123  &  24.412  &  0.642  &  RRab  \\
AndII-V017  &  1:16:04.235  &  +33:20:58.80  &  0.347  &  24.825  &  0.194  &  24.293  &  0.061  &  24.897  &  0.217  &  24.647  &  0.153  &  24.267  &  0.080  &  RRc   \\
AndII-V018  &  1:16:04.463  &  +33:22:25.54  &  0.572  &  25.082  &  1.090  &  24.397  &  0.599  &  25.176  &  1.185  &  24.857  &  0.895  &  24.457  &  0.388  &  RRab  \\
AndII-V019  &  1:16:04.471  &  +33:22:06.01  &  0.641  &  25.065  &  0.845  &  24.331  &  0.402  &  25.171  &  0.952  &  24.815  &  0.698  &  24.316  &  0.395  &  RRab  \\
AndII-V020  &  1:16:05.220  &  +33:19:59.10  &  0.751  &  25.048  &  0.546  &  24.139  &  0.296  &  25.187  &  0.600  &  24.743  &  0.459  &  24.126  &  0.303  &  RRab  \\
\hline
\hline
\end{tabular}
\end{scriptsize}
\begin{tablenotes}
\begin{scriptsize}
\item Stars from ``AndII-V001'' to ``AndII-V0035'' were detected in the WFC3 field, while stars from ``AndII-V036'' to ``AndII-V260'' were detected in the ACS field.
\item Full version are available as Supporting Information with the online version of the paper.
\end{scriptsize}
\end{tablenotes}
\end{table*}
\begin{table*}
\caption{Parameters of the variable stars in And~III dSph.} 
\label{tab:variables_andiii}
\begin{scriptsize}
\hspace{-100pt}
\centering
\begin{tabular}{ccccccccccccccc} 
\hline
\hline
ID & RA & DEC & Period & $\langle F475W\rangle$ & A$_{F475W}$ & $\langle F814W\rangle$ & A$_{F814W}$ & $\langle B\rangle$ & A$_{B}$ & $\langle V\rangle$ & A$_{V}$ & $\langle I\rangle$ & A$_{I}$ & Type \\
name & (J2000) & (J2000) & (current) & & & & & & & & & & & \\ 
\hline 
AndIII-V001  &  0:35:22.101  &  +36:29:14.19  &  0.400  &  25.357  &  0.568  &  24.585  &  0.299  &  25.486  &  0.644  &  25.079  &  0.481  &  24.570  &  0.303  &  RRd   \\  
AndIII-V002  &  0:35:22.192  &  +36:29:30.74  &  0.645  &  25.410  &  0.710  &  24.493  &  0.467  &  25.551  &  0.781  &  25.107  &  0.591  &  24.490  &  0.445  &  RRab  \\  
AndIII-V003  &  0:35:22.642  &  +36:31:02.97  &  0.655  &  25.325  &  0.973  &  24.477  &  0.506  &  25.461  &  1.071  &  25.030  &  0.730  &  24.464  &  0.510  &  RRab  \\  
AndIII-V004  &  0:35:22.827  &  +36:31:49.08  &  0.626  &  25.462  &  0.470  &  24.587  &  0.269  &  25.599  &  0.518  &  25.161  &  0.392  &  24.573  &  0.273  &  RRab  \\  
AndIII-V005  &  0:35:23.254  &  +36:30:04.01  &  0.399  &  25.257  &  0.545  &  24.510  &  0.335  &  25.376  &  0.595  &  24.991  &  0.450  &  24.492  &  0.330  &  RRc   \\  
AndIII-V006  &  0:35:23.687  &  +36:31:51.12  &  0.627  &  25.420  &  0.752  &  24.562  &  0.387  &  25.557  &  0.814  &  25.123  &  0.672  &  24.548  &  0.394  &  RRab  \\  
AndIII-V007  &  0:35:23.985  &  +36:31:11.84  &  0.706  &  25.167  &  0.765  &  24.290  &  0.381  &  25.315  &  0.819  &  24.849  &  0.681  &  24.275  &  0.387  &  RRab  \\  
AndIII-V008  &  0:35:24.109  &  +36:31:14.42  &  0.653  &  25.390  &  0.430  &  24.465  &  0.280  &  25.540  &  0.439  &  25.065  &  0.423  &  24.452  &  0.288  &  RRab  \\  
AndIII-V009  &  0:35:24.208  &  +36:31:05.28  &  0.375  &  25.363  &  0.462  &  24.611  &  0.230  &  25.474  &  0.523  &  25.115  &  0.372  &  24.594  &  0.232  &  RRd   \\  
AndIII-V010  &  0:35:24.240  &  +36:30:03.13  &  0.606  &  25.311  &  0.825  &  24.476  &  0.406  &  25.439  &  0.906  &  25.026  &  0.630  &  24.466  &  0.439  &  RRab  \\  
AndIII-V011  &  0:35:24.529  &  +36:30:22.74  &  0.607  &  25.313  &  0.915  &  24.484  &  0.512  &  25.446  &  0.976  &  25.019  &  0.764  &  24.469  &  0.521  &  RRab  \\  
AndIII-V012  &  0:35:25.360  &  +36:29:24.55  &  0.601  &  25.407  &  0.779  &  24.489  &  0.429  &  25.553  &  0.862  &  25.099  &  0.652  &  24.474  &  0.442  &  RRab  \\ 
AndIII-V013  &  0:35:25.445  &  +36:30:56.36  &  0.650  &  25.384  &  0.263  &  24.496  &  0.153  &  25.525  &  0.284  &  25.074  &  0.225  &  24.473  &  0.143  &  RRab  \\  
AndIII-V014  &  0:35:25.888  &  +36:29:46.85  &  0.646  &  25.276  &  0.519  &  24.372  &  0.396  &  25.417  &  0.568  &  24.967  &  0.460  &  24.370  &  0.329  &  RRab  \\ 
AndIII-V015  &  0:35:26.072  &  +36:31:35.04  &  0.661  &  25.302  &  0.666  &  24.436  &  0.301  &  25.432  &  0.742  &  25.012  &  0.594  &  24.425  &  0.324  &  RRab  \\  
AndIII-V016  &  0:35:26.112  &  +36:29:53.61  &  0.614  &  25.343  &  0.949  &  24.510  &  0.453  &  25.463  &  1.073  &  25.053  &  0.776  &  24.504  &  0.498  &  RRab  \\ 
AndIII-V017  &  0:35:26.231  &  +36:30:26.37  &  0.406  &  25.235  &  0.499  &  24.481  &  0.200  &  25.349  &  0.557  &  24.979  &  0.391  &  24.465  &  0.207  &  RRd   \\  
AndIII-V018  &  0:35:26.302  &  +36:30:44.48  &  0.413  &  25.269  &  0.490  &  24.497  &  0.288  &  25.383  &  0.528  &  25.000  &  0.424  &  24.479  &  0.295  &  RRd   \\  
AndIII-V019  &  0:35:26.384  &  +36:30:24.06  &  0.328  &  25.353  &  0.622  &  24.718  &  0.302  &  25.453  &  0.646  &  25.117  &  0.569  &  24.700  &  0.299  &  RRc   \\  
AndIII-V020  &  0:35:26.533  &  +36:30:51.13  &  0.406  &  25.239  &  0.422  &  24.483  &  0.210  &  25.355  &  0.471  &  24.978  &  0.341  &  24.465  &  0.209  &  RRd   \\  
\hline
\hline
\end{tabular}
\end{scriptsize}
\begin{tablenotes}
\begin{scriptsize}
\item Stars from ``AndIII-V001'' to ``AndIII-V114'' were detected in the ACS field, while stars from ``AndIII-V115'' to ``AndIII-V118'' were detected in the WFC3 field.
\item Full version are available as Supporting Information with the online version of the paper.
\end{scriptsize}
\end{tablenotes}
\end{table*}
\begin{table*}
\caption{Parameters of the variable stars in And~XV dSph.} 
\label{tab:variables_andxv}
\begin{scriptsize}
\hspace{-100pt}
\centering
\begin{tabular}{ccccccccccccccc}
\hline
\hline
ID & RA & DEC & Period & $\langle F475W\rangle$ & A$_{F475W}$ & $\langle F814W\rangle$ & A$_{F814W}$ & $\langle B\rangle$ & A$_{B}$ & $\langle V\rangle$ & A$_{V}$ & $\langle I\rangle$ & A$_{I}$ & Type \\
name & (J2000) & (J2000) & (current) & & & & & & & & & & & \\ 
\hline 
AndXV-V001  &  1:14:10.037  &  +38:06:34.23  &  0.503  &  25.366  &  1.443  &  24.657  &  0.761  &  25.469  &  1.552  &  25.128  &  1.215  &  24.643  &  0.760  &  RRab  \\
AndXV-V002  &  1:14:10.426  &  +38:06:37.79  &  0.618  &  25.364  &  0.861  &  24.523  &  0.458  &  25.494  &  0.948  &  25.071  &  0.707  &  24.508  &  0.465  &  RRab  \\
AndXV-V003  &  1:14:10.438  &  +38:06:38.32  &  0.587  &  25.312  &  1.015  &  24.541  &  0.633  &  25.429  &  1.089  &  25.047  &  0.876  &  24.524  &  0.640  &  RRab  \\
AndXV-V004  &  1:14:11.397  &  +38:06:27.19  &  1.136  &  23.779  &  1.574  &  23.104  &  0.828  &  23.875  &  1.697  &  23.558  &  1.319  &  23.093  &  0.819  &  AC    \\
AndXV-V005  &  1:14:11.724  &  +38:06:47.28  &  0.547  &  25.335  &  1.217  &  24.546  &  0.808  &  25.440  &  1.359  &  25.083  &  0.996  &  24.530  &  0.833  &  RRab  \\
AndXV-V006  &  1:14:11.949  &  +38:06:51.41  &  0.377  &  25.235  &  0.602  &  24.599  &  0.433  &  25.329  &  0.665  &  25.014  &  0.488  &  24.581  &  0.434  &  RRc   \\
AndXV-V007  &  1:14:12.105  &  +38:07:17.08  &  0.629  &  25.320  &  0.671  &  24.470  &  0.385  &  25.449  &  0.736  &  25.032  &  0.565  &  24.455  &  0.393  &  RRab  \\
AndXV-V008  &  1:14:12.198  &  +38:06:54.71  &  0.329  &  25.246  &  0.605  &  24.632  &  0.393  &  25.332  &  0.664  &  25.038  &  0.482  &  24.610  &  0.390  &  RRc   \\
AndXV-V009  &  1:14:12.555  &  +38:06:08.23  &  0.365  &  25.305  &  0.493  &  24.659  &  0.240  &  25.404  &  0.526  &  25.074  &  0.416  &  24.641  &  0.239  &  RRc   \\
AndXV-V010  &  1:14:13.229  &  +38:07:23.41  &  0.576  &  25.356  &  0.452  &  24.537  &  0.351  &  25.475  &  0.482  &  25.083  &  0.404  &  24.521  &  0.355  &  RRab  \\
AndXV-V011  &  1:14:13.380  &  +38:06:34.38  &  0.609  &  25.360  &  1.069  &  24.522  &  0.558  &  25.490  &  1.164  &  25.074  &  0.905  &  24.513  &  0.544  &  RRab  \\
AndXV-V012  &  1:14:13.441  &  +38:07:03.68  &  0.543  &  25.256  &  0.949  &  24.537  &  0.415  &  25.360  &  1.032  &  25.006  &  0.722  &  24.521  &  0.422  &  RRab  \\
AndXV-V013  &  1:14:13.592  &  +38:07:44.73  &  0.608  &  25.244  &  1.430  &  24.481  &  0.668  &  25.356  &  1.613  &  24.983  &  1.140  &  24.468  &  0.672  &  RRab  \\
AndXV-V014  &  1:14:13.642  &  +38:06:04.53  &  0.518  &  25.359  &  1.305  &  24.685  &  0.669  &  25.457  &  1.428  &  25.135  &  1.039  &  24.673  &  0.665  &  RRab  \\
AndXV-V015  &  1:14:13.936  &  +38:06:04.72  &  0.621  &  25.385  &  0.631  &  24.559  &  0.347  &  25.512  &  0.698  &  25.101  &  0.515  &  24.544  &  0.354  &  RRab  \\
AndXV-V016  &  1:14:13.944  &  +38:07:35.31  &  0.868  &  24.331  &  1.406  &  23.554  &  0.776  &  24.436  &  1.550  &  24.084  &  1.146  &  23.536  &  0.734  &  AC    \\
AndXV-V017  &  1:14:14.008  &  +38:08:00.67  &  0.677  &  25.226  &  0.821  &  24.406  &  0.427  &  25.356  &  0.919  &  24.942  &  0.649  &  24.393  &  0.440  &  RRab  \\
AndXV-V018  &  1:14:14.094  &  +38:07:06.15  &  0.601  &  25.308  &  0.946  &  24.479  &  0.518  &  25.434  &  1.042  &  25.022  &  0.730  &  24.466  &  0.534  &  RRab  \\
AndXV-V019  &  1:14:14.115  &  +38:07:40.49  &  0.719  &  25.313  &  0.801  &  24.418  &  0.351  &  25.451  &  0.890  &  25.020  &  0.597  &  24.406  &  0.358  &  RRab  \\
AndXV-V020  &  1:14:14.371  &  +38:07:08.30  &  0.644  &  25.491  &  0.593  &  24.562  &  0.330  &  25.638  &  0.652  &  25.167  &  0.477  &  24.553  &  0.328  &  RRab  \\
\hline
\hline
\end{tabular}
\end{scriptsize}
\begin{tablenotes}
\begin{scriptsize}	
\item All stars were detected in the ACS field.
\item Full version are available as Supporting Information with the online version of the paper.
\end{scriptsize}
\end{tablenotes}
\end{table*}
\begin{table*}
\caption{Parameters of the variable stars in And~XVI dSph.} 
\label{tab:variables_andxvi}
\begin{scriptsize}
\hspace{-100pt}
\centering
\begin{tabular}{ccccccccccccccc}
\hline
\hline
ID & RA & DEC & Period & $\langle F475W\rangle$ & A$_{F475W}$ & $\langle F814W\rangle$ & A$_{F814W}$ & $\langle B\rangle$ & A$_{B}$ & $\langle V\rangle$ & A$_{V}$ & $\langle I\rangle$ & A$_{I}$ & Type \\
name & (J2000) & (J2000) & (current) & & & & & & & & & & & \\ 
\hline 
AndXVI-V001  &  0:59:24.386  &  +32:22:33.16  &  0.622  &  25.459  &  1.142  &  24.642  &  0.555  &  25.584  &  1.281  &  25.173  &  0.912  &  24.627  &  0.571  &  M31 RRL \\
AndXVI-V002  &  0:59:25.335  &  +32:22:16.10  &  0.358  &  24.563  &  0.624  &  23.877  &  0.402  &  24.671  &  0.692  &  24.306  &  0.484  &  23.857  &  0.394  &  RRc     \\
AndXVI-V003  &  0:59:27.972  &  +32:22:57.58  &  0.389  &  24.560  &  0.552  &  23.791  &  0.374  &  24.667  &  0.612  &  24.313  &  0.445  &  23.774  &  0.375  &  RRc     \\
AndXVI-V004  &  0:59:29.434  &  +32:22:25.90  &  0.350  &  24.580  &  0.520  &  23.900  &  0.346  &  24.682  &  0.547  &  24.346  &  0.467  &  23.882  &  0.347  &  RRc     \\
AndXVI-V005  &  0:59:30.846  &  +32:22:14.01  &  0.617  &  24.603  &  0.922  &  23.756  &  0.581  &  24.734  &  0.992  &  24.335  &  0.759  &  23.741  &  0.586  &  RRab    \\
AndXVI-V006  &  0:59:34.271  &  +32:21:59.44  &  0.640  &  24.594  &  1.199  &  23.746  &  0.657  &  24.725  &  1.296  &  24.280  &  0.894  &  23.730  &  0.671  &  RRab    \\
AndXVI-V007  &  0:59:36.072  &  +32:23:16.35  &  0.392  &  24.606  &  0.456  &  23.877  &  0.227  &  24.715  &  0.500  &  24.352  &  0.388  &  23.858  &  0.231  &  RRc     \\
AndXVI-V008  &  0:59:37.515  &  +32:22:10.10  &  0.289  &  24.669  &  0.298  &  24.163  &  0.198  &  24.738  &  0.312  &  24.495  &  0.264  &  24.147  &  0.195  &  RRc     \\
AndXVI-V009  &  0:59:38.101  &  +32:23:15.78  &  0.651  &  24.610  &  0.663  &  23.785  &  0.424  &  24.737  &  0.712  &  24.324  &  0.589  &  23.769  &  0.430  &  RRab    \\
\hline
\hline
\end{tabular}
\end{scriptsize}
\begin{tablenotes}
\begin{scriptsize}	
\item All stars were detected in the ACS field.
\end{scriptsize}
\end{tablenotes}
\end{table*}
\begin{table*}
\caption{Parameters of the variable stars in And~XXVIII dSph.} 
\label{tab:variables_andxxviii}
\begin{scriptsize}
\hspace{-100pt}
\centering
\begin{tabular}{ccccccccccccccc}
\hline
\hline
ID & RA & DEC & Period & $\langle F475W\rangle$ & A$_{F475W}$ & $\langle F814W\rangle$ & A$_{F814W}$ & $\langle B\rangle$ & A$_{B}$ & $\langle V\rangle$ & A$_{V}$ & $\langle I\rangle$ & A$_{I}$ & Type \\
name & (J2000) & (J2000) & (current) & & & & & & & & & & & \\ 
\hline 
AndXXVIII-V001  &  22:32:32.098  &  +31:13:11.52  &  0.642  &  25.549  &  1.142  &  24.648  &  0.601  &  25.694  &  1.259  &  25.229  &  0.960  &  24.639  &  0.620  & RRab  \\
AndXXVIII-V002  &  22:32:33.432  &  +31:13:07.56  &  0.608  &  25.392  &  0.697  &  24.659  &  0.228  &  25.508  &  0.590  &  25.141  &  0.392  &  24.615  &  0.203  & RRL?  \\
AndXXVIII-V003  &  22:32:35.539  &  +31:12:15.41  &  0.407  &  25.414  &  0.460  &  24.603  &  0.217  &  25.536  &  0.517  &  25.131  &  0.366  &  24.587  &  0.222  & RRc   \\
AndXXVIII-V004  &  22:32:35.594  &  +31:12:32.34  &  0.366  &  25.490  &  0.501  &  24.730  &  0.272  &  25.607  &  0.538  &  25.215  &  0.435  &  24.713  &  0.275  & RRc   \\
AndXXVIII-V005  &  22:32:36.365  &  +31:12:22.37  &  0.565  &  25.529  &  0.974  &  24.647  &  0.566  &  25.653  &  1.102  &  25.244  &  0.769  &  24.635  &  0.582  & RRab  \\
AndXXVIII-V006  &  22:32:36.682  &  +31:14:05.83  &  0.540  &  25.493  &  1.222  &  24.639  &  0.771  &  25.621  &  1.344  &  25.198  &  1.016  &  24.629  &  0.785  & RRab  \\
AndXXVIII-V007  &  22:32:36.703  &  +31:13:45.74  &  0.681  &  25.363  &  0.720  &  24.456  &  0.345  &  25.502  &  0.802  &  25.055  &  0.606  &  24.445  &  0.355  & RRab  \\
AndXXVIII-V008  &  22:32:36.958  &  +31:13:14.25  &  0.341  &  25.411  &  0.541  &  24.559  &  0.246  &  25.542  &  0.592  &  25.123  &  0.451  &  24.546  &  0.250  & RRd   \\
AndXXVIII-V009  &  22:32:37.075  &  +31:12:45.87  &  0.362  &  25.453  &  0.480  &  24.715  &  0.396  &  25.567  &  0.508  &  25.196  &  0.459  &  24.698  &  0.403  & RRd   \\
AndXXVIII-V010  &  22:32:37.332  &  +31:12:31.50  &  0.510  &  25.471  &  1.147  &  24.682  &  0.678  &  25.596  &  1.263  &  25.195  &  0.936  &  24.668  &  0.683  & RRab  \\
AndXXVIII-V011  &  22:32:37.507  &  +31:12:31.46  &  0.366  &  25.465  &  0.493  &  24.722  &  0.334  &  25.572  &  0.554  &  25.213  &  0.390  &  24.706  &  0.331  & RRd   \\
AndXXVIII-V012  &  22:32:37.510  &  +31:11:55.04  &  0.385  &  25.482  &  0.393  &  24.683  &  0.227  &  25.603  &  0.421  &  25.208  &  0.339  &  24.667  &  0.226  & RRd   \\
AndXXVIII-V013  &  22:32:37.975  &  +31:13:40.09  &  0.366  &  25.394  &  0.541  &  24.613  &  0.341  &  25.508  &  0.583  &  25.134  &  0.478  &  24.596  &  0.346  & RRc   \\
AndXXVIII-V014  &  22:32:37.980  &  +31:14:01.26  &  0.369  &  25.466  &  0.572  &  24.647  &  0.324  &  25.587  &  0.626  &  25.200  &  0.492  &  24.631  &  0.321  & RRd   \\
AndXXVIII-V015  &  22:32:38.287  &  +31:14:04.83  &  0.646  &  25.412  &  0.790  &  24.488  &  0.301  &  25.565  &  0.910  &  25.092  &  0.524  &  24.403  &  0.483  & RRab  \\
AndXXVIII-V016  &  22:32:38.614  &  +31:13:12.52  &  0.397  &  25.178  &  0.574  &  24.534  &  0.257  &  25.271  &  0.625  &  24.960  &  0.491  &  24.517  &  0.257  & RRc   \\
AndXXVIII-V017  &  22:32:38.635  &  +31:13:34.34  &  0.558  &  25.351  &  1.367  &  24.601  &  0.771  &  25.461  &  1.476  &  25.100  &  1.138  &  24.585  &  0.778  & RRab  \\
AndXXVIII-V018  &  22:32:38.690  &  +31:14:14.99  &  0.412  &  25.353  &  0.569  &  24.603  &  0.251  &  25.463  &  0.634  &  25.097  &  0.454  &  24.588  &  0.257  & RRc   \\
AndXXVIII-V019  &  22:32:38.837  &  +31:13:12.22  &  0.524  &  25.455  &  1.076  &  24.678  &  0.608  &  25.571  &  1.156  &  25.170  &  0.809  &  24.662  &  0.710  & RRab  \\
AndXXVIII-V020  &  22:32:39.029  &  +31:12:51.00  &  0.651  &  25.427  &  0.432  &  24.543  &  0.189  &  25.566  &  0.493  &  25.125  &  0.370  &  24.528  &  0.199  & RRab  \\
\hline
\hline
\end{tabular}
\end{scriptsize}
\begin{tablenotes}
\begin{scriptsize}	
\item All stars were detected in the ACS field.
\item Full version are available as Supporting Information with the online version of the paper.
\end{scriptsize}
\end{tablenotes}
\end{table*}




\bibliographystyle{apj}



\end{document}